\documentstyle[aps,twocolumn]{revtex}
\input psfig

\begin{document}

\twocolumn[\hsize\textwidth\columnwidth\hsize\csname
@twocolumnfalse\endcsname

\draft

\title{Disorder-Induced Critical Phenomena in Hysteresis:\\
A Numerical Scaling Analysis}
\author{Olga Perkovi\'{c}, Karin A. Dahmen, and James P. Sethna}
\address{Laboratory of Atomic and Solid--State Physics,\\
Cornell University,\\
Ithaca,  NY 14853--2501.}
\maketitle

\begin{abstract}
Experimental systems with a first order phase transition will often
exhibit hysteresis when out of equilibrium. If defects are present, the
hysteresis loop can have different shapes: with small disorder the
hysteresis loop has a macroscopic jump, while for large disorder the
hysteresis loop is smooth. The transition between these two shapes is
critical, with diverging length scales and power laws. We simulate such
a system with the zero temperature random field Ising model, in $2$,
$3$, $4$, $5$, $7$, and $9$ dimensions, with systems of up to $1000^3$
spins, and find the critical exponents from scaling collapses of several
measurements. The numerical results agree well with the analytical
predictions from a renormalization group calculation
\cite{Dahmen1}.
\end{abstract}


]



\narrowtext

\section{Introduction}

The increased interest in real materials in condensed matter physics has
brought disordered systems into the spotlight. Dirt changes the free
energy landscape of a system, and can introduce metastable states with
large energy barriers\ \cite{barrier}. This can lead to extremely slow
relaxation towards the equilibrium state. On long length scales and
practical time scales, a system driven by an external field will move
from one metastable local free-energy minimum to the next. The
equilibrium, global free energy minimum and the thermal fluctuations
that drive the system toward it, are in this case irrelevant. The state
of the system will instead depend on its history.

The motion from one local minima to the next is a collective process
involving many local (magnetic) domains in a local region - {\it an
avalanche}. In magnetic materials, as the external magnetic field $H$ is
changed continuously, these avalanches lead to the magnetic noise: the
Barkhausen effect\cite{Jiles,McClure}. This effect can be picked up as
voltage pulses in a coil surrounding the magnet. The distribution of
pulse (avalanche) sizes is found\
\cite{McClure,Barkhausen,Urbach,SOC_example} to follow a power law with
a cutoff after a few decades, and was interpreted by some\
\cite{SOC_example} to be an example of self-organized criticality\
\cite{SOC}. (In SOC, a system organizes itself into a critical state
without the need to tune an external parameter.) Other systems can
exhibit avalanches as well. Several examples where disorder may play a
part are: superconducting vortex line avalanches\ \cite{Field},
resistance avalanches in superconducting films\ \cite{Wu}, and capillary
condensation of helium in Nuclepore\ \cite{Nuclepore}.

The history dependence of the state of the system leads to hysteresis.
Experiments with magnetic tapes\ \cite{Berger} have shown that the shape
of the hysteresis curve changes with the annealing temperature. The
hysteresis curve goes from smooth to discontinuous as the annealing
temperature is increased. This transition can be explained in terms of a
{\it plain old critical point} with two tunable parameters: the
annealing temperature and the external field. At the critical
temperature and field, the correlation length diverges, and the
distribution of pulse (avalanche) sizes follows a power law.

We have argued earlier\ \cite{Perkovic} that the Barkhausen noise
experiments can be quantitatively explained by a model\ \cite{Sethna}
with two tunable parameters (external field and disorder), which
exhibits {\it universal}, non-equilibrium collective behavior. The model
is athermal and incorporates collective behavior through nearest
neighbor interactions. The role of {\it dirt} or disorder, as we call
it, is played by random fields. This paper presents the results and
conclusions of a large scale simulation of that model: the
non-equilibrium zero-temperature Random Field Ising Model (RFIM), with a
deterministic dynamics. The results compare very well to our $\epsilon$
expansion\cite{Dahmen1,Dahmen2}, and to experiments in Barkhausen
noise\cite{Perkovic}. A more detailed comparison to experimental systems
is in process\cite{Dahmen3}.

The paper is divided as follows. Section II quickly reviews the model.
Section III explains the simulation method that we use. Section IV
explains the data analysis and shows results for the simulation in $2$,
$3$, $4$, and $5$ dimensions, as well as in mean field. Section V gives
a comparison between the simulation and the $\epsilon$ expansion
exponents, and a comparison between the shape of the magnetization
curves in $5$, $7$, and $9$ dimensions, and the predicted shape from the
$\epsilon$ expansion. Section VI summarizes the results. This is
followed by three appendices that cover derivations that were omitted in
the text for continuity.



\narrowtext

\section{The Model}

To model the long-range, far from equilibrium, collective behavior
mentioned in the previous section, we define\cite{Sethna} spins $s_i$ on
a hypercubic lattice, which can take two values: $s_i = \pm 1$. The
spins interact ferromagnetically with their nearest neighbors with a
strength $J_{ij}$, and are sitting in a uniform magnetic field $H$
(which is directed along the spins). Dirt is simulated by a random field
$h_i$, associated with each site of the lattice, which is given by a
gaussian distribution function $\rho (h_i)$:
\begin{equation}
\rho (h_i) = {1 \over {\sqrt {2\pi}} R}\ e^{-{h_i}^2 \over 2R^2}
\label{model_equ1}
\end{equation}
of width proportional to $R$ which we call the disorder parameter, or
just disorder. The hamiltonian is then
\begin{equation}
{\cal H} = - \sum_{<i,j>} J_{ij} s_i s_j - \sum_{i} (H + h_i) s_i
\label{model_equ2}
\end{equation}
For the analytic calculation, as well as the simulation, we have set the
interaction between the spins to be independent of the spins and equal
to one for nearest neighbors, $J_{ij}=J=1$, and zero otherwise.

The dynamics is deterministic, and is defined such that a spin $s_i$
will flip only when its local effective field $h^{ef\!f}_i$:
\begin{equation}
h^{ef\!f}_i = J \sum_{j} s_j + H + h_i
\label{model_equ3}
\end{equation}
changes sign. All the spins start pointing down ($s_i=-1$ for all $i$).
As the field is adiabatically increased, a spin will flip. Due to the
nearest neighbor interaction, a flipped spin will push a neighbor to
flip, which in turn might push another neighbor, and so on, thereby
generating an avalanche of spin flips. During each avalanche, the
external field is kept constant. For large disorders, the distribution
of random fields is wide, and spins will tend to flip independently of
each other. Only small avalanches will exist, and the magnetization
curve will be smooth. On the other hand, a small disorder implies a
narrow random field distribution which allows larger avalanches to
occur. As the disorder is lowered, at the disorder $R=R_c$ and field
$H=H_c$, an infinite avalanche in the thermodynamic system will occur
for the first time, and the magnetization curve will show a
discontinuity. Near $R_c$ and $H_c$, we find critical scaling behavior
and avalanches of all sizes. Therefore, the system has two tunable
parameters: the external field $H$ and the disorder $R$. We found from
the mean field calculation\ \cite{Dahmen1,Dahmen2} and the simulation
that a discontinuity in the magnetization exists for disorders $R \le
R_c$, at the field $H_c(R) \ge H_c(R_c)$, but that only at $(R_c, H_c)$,
do we have critical behavior. For finite size systems of length $L$, the
transition occurs at the disorder $R_c^{ef\!f}(L)$ near which avalanches
first begin to span the system in one of the {\it d} dimensions
(spanning avalanches). The effective critical disorder $R_c^{ef\!f}(L)$
is larger than $R_c$, and $R_c^{ef\!f}(L) \rightarrow R_c$ as $L
\rightarrow \infty$.

%



\narrowtext

\section{Algorithm}

There are several methods that can be used to simulate the above model.
The simplest but most time and space (memory) consuming method starts by
assigning a random field to each spin on the hypercubic lattice. At the
beginning of the simulation, all the spins are pointing down. The
external field $H$ is then increased by small increments, starting from
a large negative value. After each increase of the field, all the spins
are checked to find if one of them should flip (a spin flips when its
effective field changes sign). If a spin flips, its neighbors are
checked, and so on until no spins are left that can flip. Then, the
external field is further increased, and the process repeated. Since the
external magnetic field is increased in equal increments, a large amount
of time is spent searching the lattice for spins that can flip. The
increments have to be big enough to avoid searching the lattice when
there are no spins that can flip, but small enough so that two or more
spins far apart don't flip at the same field. This is the method used
experimentally, but it is suited only for ``that kind of'' massively
parallel computing.

A variation on the above method, removes the searching through the
lattice that is done even if there are no spins that can flip. It
involves looking at all the spins, finding the next one that will flip
and {\it then} increasing the external field so that it does. The
average searching time for a flip is decreased, but is still very large.
Far from the critical point, where spins will tend to flip independently
of each other, the time for searching scales like $N^2$ where $N$ is the
number of spins in the system.

The search time can be further decreased if the random fields are
initially ordered in a list. The first spin that will flip is the one on
``top'' of the list. The external field is increased until the effective
field of the top spin become zero, and the spin flips. We then check its
nearest neighbors, and so on, while keeping the external field constant.
When no spins are left to flip, the external field needs to be increased
again. The change in the external field $\Delta H$, necessary to flip
the next spin, is found by looking for the spin whose random field $h_i$
satisfies:
\begin{equation}
h_i \ge - (H_{old}+\Delta H) - (2{n_{\uparrow}}-z)J
\label{num_equ1}
\end{equation}
where $H_{old}$ is the field at which the previous spins have flipped,
$z$ is the coordination number, and ${n_{\uparrow}}$ is the number of
nearest neighbors pointing up ($s_j = +1 $) for spin $s_i$. In general,
there will be a minimum of $z+1$ spins to check from the list, since
${n_{\uparrow}}$ can have the integer value between zero and $z$. The
spin for which equation (\ref{num_equ1}) is satisfied for the smallest
$\Delta H$, and for which the number of up neighbors is
${n_{\uparrow}}$, will flip. In general, more than $z+1$ spins will need
to be checked because a spin can satisfy equation (\ref{num_equ1}) for
some value of ${n_{\uparrow}}$ but might not have that number of up
neighbors, or the spin might have already flipped. This algorithm
decreases the searching time since not all the spins need to be checked
to find the next spin that will flip. Our early simulation work
\cite{Sethna,Dahmen2} used this method. In practice, about half of the
time was spent for the $N\ log_2 N$ initial sorting of the list of
random field numbers, where $N$ is the total number of spins in the
system. The big drawback of this method (as for the ones mentioned
above) is the huge amount of storage space needed to store the random
fields, the positions of each spin, and the values of the spins. This
becomes particularly important when larger size systems are simulated.

The results in this paper use a more sophisticated algorithm which
removes the need for a large storage space. It revolves around the idea
that the change $\Delta H$ in the external field, between two
avalanches, follows a probability distribution since the random fields
$h_i$ are given by a Gaussian distribution. The increments $\Delta H$ in
the external field should be chosen according to that distribution. The
probability distribution itself is not known explicitly, but its
integral from $0$ to some finite $\Delta H$ is. It is the probability,
$P_{all}^{none}(\Delta H)$, that {\it no} spin will flip in the whole
system during a field change less than $\Delta H$. It is given by:
\begin{equation}
P_{all}^{none}(\Delta H) = \Pi_{n_{\uparrow}}\ P_{n_{\uparrow}}^{none}(\Delta H)
\label{simul_equ1}
\end{equation}
where the product is over $n_{\uparrow}=0,1,...,z$, and
$P_{n_{\uparrow}}^{none}(\Delta H)$ is the probability for a down spin with
$n_{\uparrow}$ up nearest neighbors not to flip when the external field
changes by less than $\Delta H$:
\begin{eqnarray}
P_{n_{\uparrow}}^{none}(\Delta H)\ =\ \nonumber \\
\biggl(1 - {\int_{0}^{H_{local}(n_{\uparrow})}
\rho (f)\ df - \int_{0}^{H_{local}^{new}(n_{\uparrow})} \rho (f)\ df \over
P_{n_{\uparrow}}^{noflip} \Bigl(H_{local}(n_{\uparrow})\Bigr)}
\biggr)^{N_{n_{\uparrow}}}
\label{simul_equ2}
\end{eqnarray}
The function $\rho (f)$ is the random field distribution function, and
$H_{local}(n_{\uparrow})$ and $H_{local}^{new}(n_{\uparrow})$ are
defined respectively as:
\begin{equation}
H_{local}(n_{\uparrow}) = - H - (2n_{\uparrow} - z)J
\label{simul_equ4}
\end{equation}
and
\begin{equation}
H_{local}^{new} (n_{\uparrow}) = -(H+\Delta H) - (2n_{\uparrow} - z)J.
\label{simul_equ5}
\end{equation}
$P_{n_{\uparrow}}^{noflip}(H_{local})$ gives the probability that a spin
with $n_{\uparrow}$ up nearest neighbors has not flipped {\it before}
the field has reached the external magnetic field value $H$:
\begin{equation}
P_{n_{\uparrow}}^{noflip}\Bigl(H_{local}(n_{\uparrow})\Bigr) =
{1 \over 2} + \int_{0}^{H_{local}(n_{\uparrow})}
\rho (f)\ df,
\label{simul_equ6}
\end{equation}
and $N_{n_{\uparrow}}$ is the number of {\it down} spins that have
$n_{\uparrow}$ up neighbors.

A field increment $\Delta H$ that has the required probability
distribution is found by choosing a uniform random number between zero
and one and solving for $\Delta H$ from equation\ (\ref{simul_equ1}), by
setting the probability $P_{all}^{none}(\Delta H)$ equal to the value of
the random number. Once the increment $\Delta H$ is known, we can find
the next spin that will flip. We first calculate\ \cite{note1} the
probability $P^{flip}(n_{\uparrow})$ for a down spin with $n_{\uparrow}$
up neighbors to flip at the new field $H + \Delta H$:
\begin{equation}
P^{flip}(n_{\uparrow}) = {R_{n_{\uparrow}} \over R_{tot}}
\label{simul_equ61}
\end{equation}
where
\begin{equation}
R_{n_{\uparrow}} = {N_{n{\uparrow}}\
\rho \Bigl(H_{local}^{new}(n_{\uparrow})\Bigr) \over
P_{n_{\uparrow}}^{noflip} \Bigl(H_{local}^{new}(n_{\uparrow})\Bigr)}
\label{simul_equ62}
\end{equation}
is the rate at which down spins with $n_{\uparrow}$ up neighbors would
flip, and $R_{tot}$ is the sum of the rates $R_{n_{\uparrow}}$ for all
$n_{\uparrow}$. The spin that flips will have $k$ up neighbors, which is
found by satisfying the following inequality:
\begin{equation}
\Sigma_{n_{\uparrow}=0}^{k}\ P^{flip}(n_{\uparrow}) > C >
\Sigma_{n_{\uparrow}=0}^{k-1}\ P^{flip}(n_{\uparrow})
\label{simul_equ63}
\end{equation}
where the cutoff $C$ is a random number between $0$ and $1$. Once $k$ is
known, a spin is then randomly picked from the list of down spins with
$k$ up neighbors.

After the first spin has flipped, its neighbors are checked. The
probability for one of the neighbors, with ($n_{\uparrow} + 1$) up
nearest neighbors, to flip at $H + \Delta H$, given that it has not yet
flipped, is:
\begin{equation}
P_{next}(n_{\uparrow}, H+\Delta H) = 1 - {{{1 \over 2}\ +\
\int_{0}^{H_{local}^{new} (n_{\uparrow}+1)}
\rho (f)\ df} \over {{1 \over 2}\ +\
\int_{0}^{H_{local}^{new}(n_{\uparrow})} \rho (f)\ df}}
\label{simul_equ7}
\end{equation}
When all the neighbors have been checked, the size of the avalanche is
stored, as well as all the other measurements. The external magnetic
field $H$ is then incremented again by finding the next $\Delta H$,
starting back with equation\ (\ref{simul_equ1}).

The important characteristic of this method is that the random fields
are not assigned to the spins at the beginning of the simulation, which
for large system sizes decreases memory requirements tremendously
(asymptotically, we use one bit per spin). This method has allowed us to
simulate system sizes of up to $30000^2$, $1000^3$, $80^4$, and $50^5$
spins. The majority of the data analysis was performed on systems of
sizes $7000^2$, $320^3$, $80^4$, and $30^5$. The SP1 and SP2
supercomputers at the Cornell Theory Center, and IBM RS6000 model 560
and J30 workstations were used for the simulation. Using this new
algorithm, close to the critical disorder, one run (for a particular
random field configuration) for a $320^3$ system took more than $1$ CPU
hour on a SP1 node at the Cornell Theory Center, while it took close to
$37$ CPU hours for a $800^3$ system on an IBM RS6000 model 560
workstation. Far above the critical disorder $R_c$, the simulation time
increases substantially: $40\%$ above the critical disorder, for the
$320^3$ system, the simulation time was six times longer than for the
simulation at $10\%$ above $R_c$.




\narrowtext

\section{The Simulation Results}

The following measurements were obtained from the simulation as a function
of disorder R: \par
$\bullet$ the magnetization $M(H,R)$ as a function of the \par external
field $H$. \par
$\bullet$ the avalanche size distribution integrated over the \par
field $H$: $D_{int}(S,R)$. \par
$\bullet$ the avalanche correlation function integrated over the \par
field $H$:
$G_{int}(x,R)$. \par
$\bullet$ the number of spanning avalanches $N(L,R)$ as a \par function of the
system length $L$, integrated over the \par field $H$. \par
$\bullet$ the discontinuity in the magnetization $\Delta M (L,R)$ as \par a
function of the system length $L$. \par
$\bullet$ the second $\langle S^2 \rangle_{int}(L,R)$, third
$\langle S^3 \rangle_{int}(L,R)$, and \par
fourth $\langle S^4 \rangle_{int}(L,R)$
moments of the avalanche size \par
distribution as a function of the system length $L$, \par
integrated over the field $H$. \par

{\noindent In addition, we have measured:} \par
$\bullet$ the avalanche size distribution $D(S,H,R)$ as a \par function of the
field $H$ and disorder $R$. \par
$\bullet$ the distribution of avalanche times $D_{t}^{(int)}(S,t)$
as a \par function of the avalanche size $S$, at
$R=R_c$, integrated \par over the field $H$. \par

The data obtained from the simulation was used to find and describe the
critical transition. It was analyzed using {\bf scaling collapses}. The
mean field calculation\cite{Sethna,Dahmen1} for our model shows
that near the critical point, the magnetization curve has the scaling
form
\begin{equation}
M(H,R) - M_c(H_c,R_c) \sim |r|^\beta\ {\cal M}_{\pm}(h/|r|^{\beta\delta})
\label{model_equ4}
\end{equation}
where $M_c$ is the critical magnetization (the magnetization at $H_c$,
for $R=R_c$), $r=(R_c-R)/R$ and $h=(H-H_c)$ are the reduced disorder and
reduced field respectively, and ${\cal M}_{\pm}$ is a universal scaling
function ($\pm$ refers to the sign of $r$). Both $r$ and $h$ are small.
The critical exponent $\beta$ gives the scaling for the magnetization at
the critical field $H_c$ ($h=0$). Its mean field value is $1/2$, and the
mean field value of $\beta\delta$ is $3/2$. (Appendix A gives a short
review on why scaling and scaling functions occur near a critical point,
and why they have the form they do).

The significance of scaling for experimental and numerical data is as
follows\cite{Goldenfeld}. If the magnetization data, for example, is
plotted against the field $H$, there would be one data curve for each
disorder $R$ (fig.\ \ref{mf_mofhfig}a). While if we plot $|r|^{-\beta}
M(H,R)$ against $h/|r|^{\beta\delta}$, all the curves close to $R_c$ and
$H_c$ will {\bf collapse} (fig. \ref{mf_mofhfig}b) onto either one of
two curves: one for $r<0$ (${\cal M}_{-}$), and one for $r>0$ (${\cal
M}_{+}$). The functions ${\cal M}_{\pm}$ depend only on the combination
$h/|r|^{\beta\delta}$ and not on the field $H$ and disorder $R$
separately, and are therefore {\it universal}. Usually, the exponents
are unknown and scaling or data collapses are used to obtain them: the
exponents are varied until all the curves lie on top of each other. This
method is useful for analyzing numerical as well as experimental data,
and is often preferred to ``data fitting'', as we will show.

Numerical simulations and experiments are done on finite size systems.
Often the properties of the system will depend on the linear size $L$.
Functions that depend on the system's length are analyzed using {\bf
finite size collapses}\cite{Goldenfeld,Barber}. An example is the number $N$ of
spanning avalanches: $N(L,R) \sim L^{\theta}\ {\cal N}(L^{1/\nu} |r|)$ (to
be explained later). If $N$ is plotted against $R$, there would be one
data curve for each length $L$. The exponents $\theta$ and $\nu$ are
obtained by plotting $L^{-\theta}N(L,R)$ against $L^{1/\nu} |r|$ onto one
curve (the collapse), and extracting the exponents.

The scaling forms we use for the collapses do not include corrections
that are present when the data is {\it not taken} in the limit $R
\rightarrow R_c$ and $L \rightarrow \infty$ (see appendix A for
corrections that exist in those limits). On the other hand, finite size
effects close to $R_c$ become important. It is thus necessary to
extrapolate to $R \rightarrow R_c$ and $L \rightarrow \infty$ to obtain
the correct exponents. We have done a mean field simulation to test our
extrapolation method. The mean field exponents can be calculated
analytically\ \cite{Sethna,Dahmen1}, but it is useful to check that the
numerical results from the mean field simulation, for disorders away
from $R_c$ and for finite sizes, extrapolate to the analytical values at
$R=R_c$ and $1/L=0$. We will see that this indeed occurs, and we will
use the same extrapolation method in $3$, $4$, and $5$ dimensions.

The mean field simulation was done with the same code, but with some
changes. In mean field, the interactions between spins are infinite in
range (each spin interacts when every spin in the system with the same
interaction). This means that distances and positions are not relevant,
and therefore we don't need to keep track of the spins and their
neighbors; we just need to know the total number of flipped spins, and
the value of the external field $H$. The following section will show the
results of the mean field simulation and explain the extrapolation
method. Then, we will turn to results in $3$, $4$, and $5$ dimensions.
And finally, we will cover the more subtle scaling behavior in two
dimensions.



\narrowtext

\subsection{Mean Field Simulation}

The mean field simulation shows how well the results for the critical
exponents, obtained close to $R_c$ and for finite size systems (finite
number of spins), extrapolate to the calculated values for a system in
the thermodynamic limit, at the critical disorder. Thus, we will omit in
this section some details that are only relevant for understanding the
non-mean field simulation results. We start with the curves for the
magnetization as a function of the field for different values of the
disorder, which we find are not useful for extracting critical
exponents. We then go on to measurements of spin avalanche sizes and
their moments. Avalanches that span the system from one ``side'' to
another will also be mentioned although since in mean field there are no
``sides'', we will define what we mean by a mean field spanning
avalanche. Since distances are irrelevant in mean field, we do not have
any correlation measurements, but we can still apply what we learn from
other collapses in mean field to the correlation measurement data in
$2$, $3$, $4$, and $5$ dimensions.

\begin{figure}
\centerline{
\psfig{figure=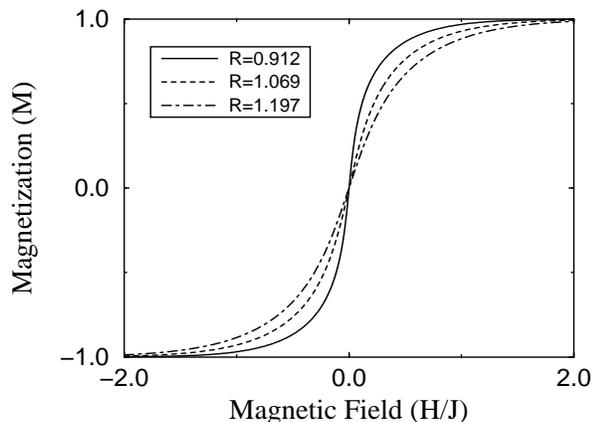,width=3truein} 
}
{\caption[Mean field magnetization curves]
        {{\bf Mean field magnetization} curves for $10^6$ spins.
        The critical disorder is $R_c= 0.79788456$.
        The curves are averages of $6$ to $10$ different initial realizations
        of the random field distribution. \label{mf_mofhfiga}}}
\end{figure}

\begin{figure}
\centerline{
\psfig{figure=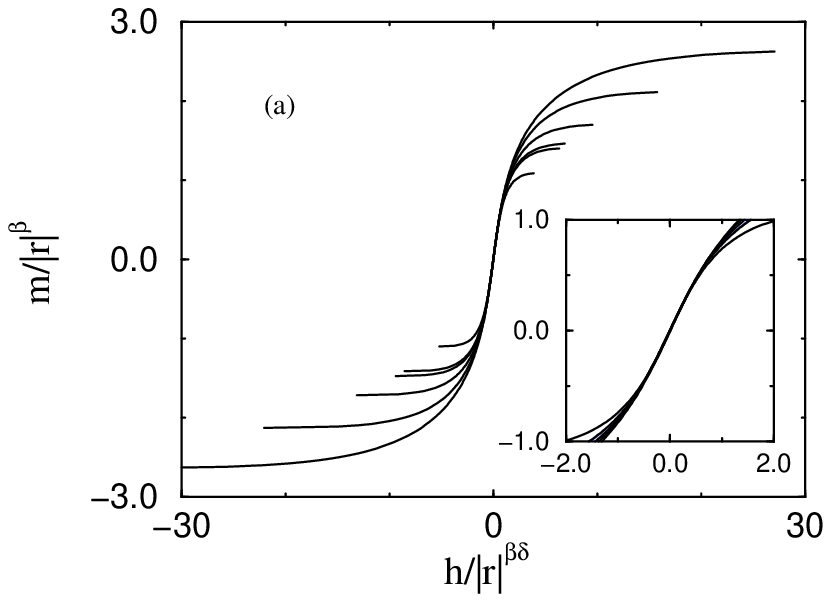,width=3truein} 
}
\nobreak
\centerline{
\psfig{figure=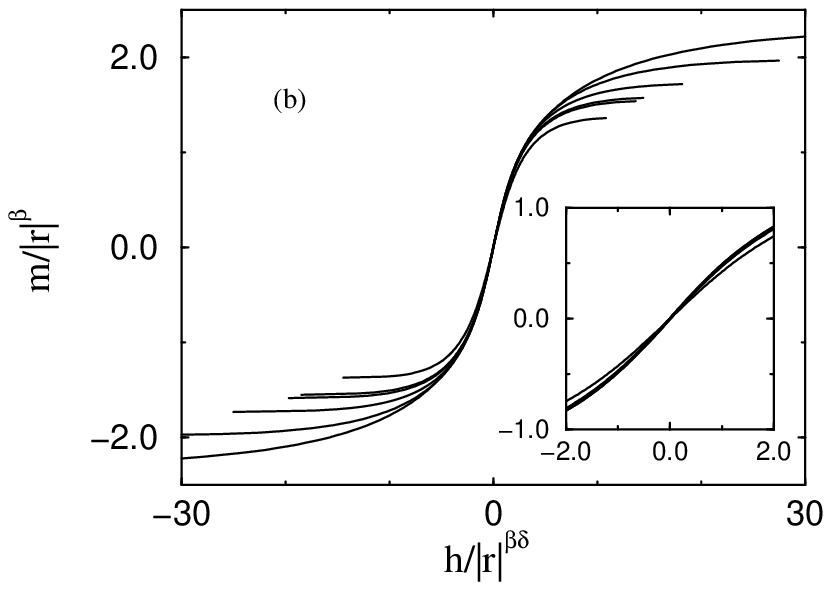,width=3truein} 
}
\caption[Scaling collapse of mean field magnetization curves]
        {(a) {\bf Scaling collapse for the mean field magnetization curves} at
        disorders $R=0.912$, $0.974$, $1.069$, $1.165$, $1.197$, and $1.460$.
        (These values of disorder were chosen relative to $R_c=0.79788456$,
        to match some of the values we measured in $3$ dimensions (see figure\
        \protect\ref{3d_MofH_fig}). The value of the critical disorder $R_c$ in
        $3$ dimensions has since been modified, and there is no correspondence
        anymore.) The
        exponents are $\beta=1/2$ and $\beta\delta=3/2$. $m$ is defined as
        $M-M_c$, and in mean field both $M_c$ and $H_c$ are zero. The inset
        shows a closeup of the critical region.
        (b) Scaling collapse of the same curves as in (a) but with the
        (wrong) exponents
        $\beta=0.4$ and $\beta\delta =1.65$. The two collapses are very
        similar. The inset is a closeup. \label{mf_mofhfig}}
\end{figure}

Figure\ \ref{mf_mofhfiga} shows the magnetization curves, and figure
\ref{mf_mofhfig}a shows a scaling collapse for a $10^6$ mean field spin
system and $r<0$ ($R>R_c$). As mentioned earlier, near the critical
point ($R_c = {\sqrt {2/\pi}}$ for $J=1$, in mean field), the
magnetization scales like\cite{Sethna,Dahmen1}
\begin{equation}
M(H,R) - M_c(H_c,R_c) \sim |r|^\beta\ {\cal M}_{\pm}(h/|r|^{\beta\delta})
\label{mf_mofh_equ1}
\end{equation}
where $\pm$ refers to the sign of the reduced disorder $r=(R_c-R)/R$,
and $h=(H-H_c)$. The mean field critical exponents are $\beta = 1/2$ and
$\beta\delta = 3/2$. Notice in figure\ \ref{mf_mofhfig}a that the
scaling region around $M_c=0$ and $H_c=0$ is very small; figure\
\ref{mf_mofhfig}b shows that a substantially different set of critical
exponents leads to a similar looking collapse. In general, the critical
field $H_c$ and the critical magnetization $M_c$ are not zero as in mean
field, and $M_c$ is not well determined numerically. In dimensions that
we simulate ($2$ through $5$), the critical region is not only small but
it is also poorly defined, which does not sufficiently constrain the
values of the exponents. This makes the magnetization function $M(H,R)$
a poor choice for extracting critical exponents.

\begin{figure}
\centerline{
\psfig{figure=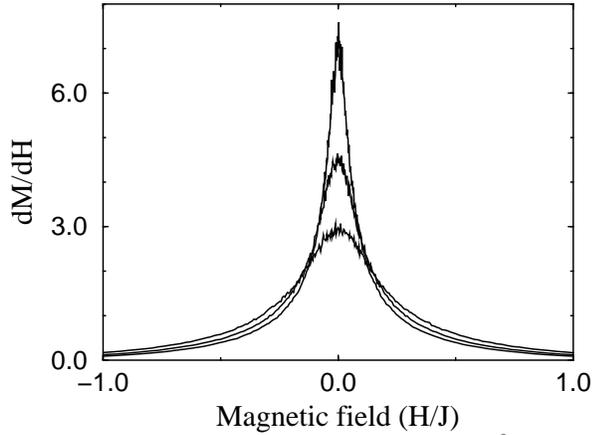,width=3truein} 
}
\caption[Mean field $dM/dH$ curves]{{\bf Mean field $dM/dH$ curves}
        for $10^6$ spins and disorders $R=0.912$,
        $0.974$, and $1.069$ (from largest to smallest peak).
        The original data is the same as in figure\ \protect\ref{mf_mofhfiga}.
        The critical disorder is $R_c = 0.79788456$. \label{mf_dmdh_figa}}
\end{figure}

\begin{figure}
\centerline{
\psfig{figure=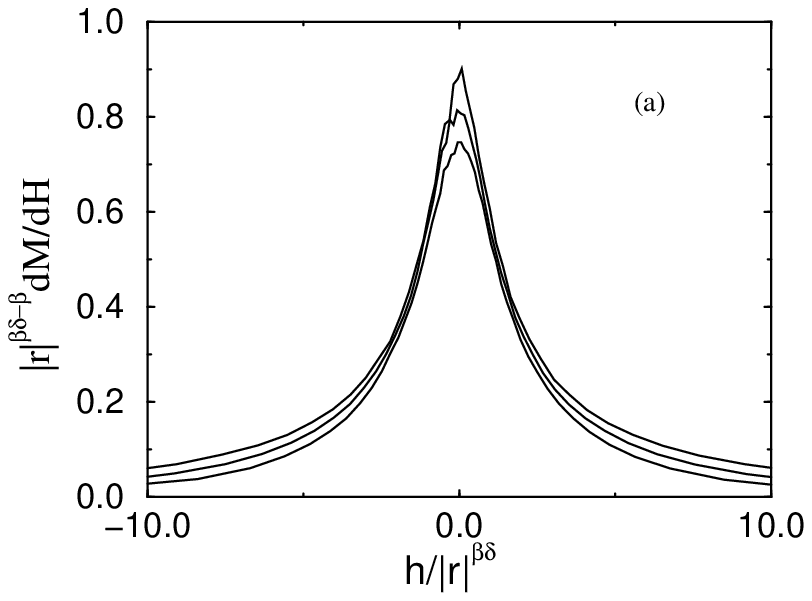,width=3truein} 
}
\nobreak
\centerline{
\psfig{figure=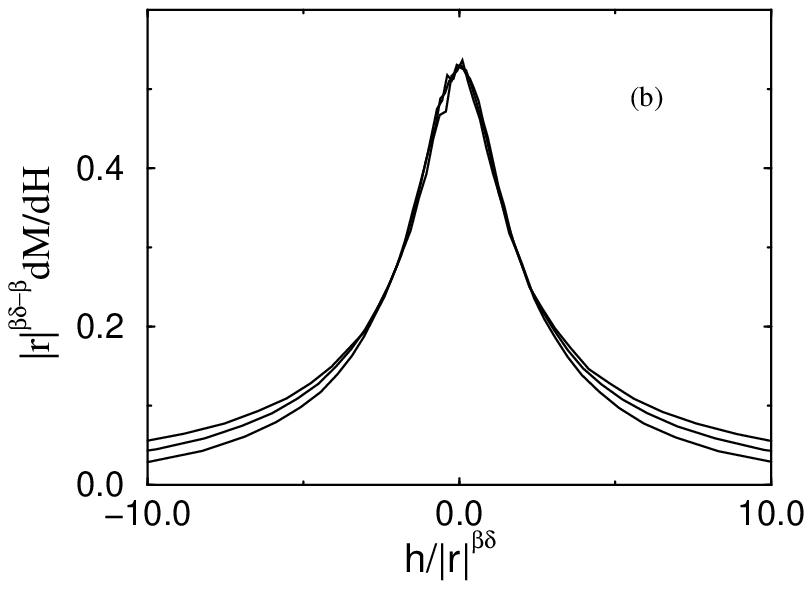,width=3truein} 
}
\caption[Scaling collapse of mean field $dM/dH$ curves]
        {(a) {\bf Scaling collapse of mean field $dM/dH$ curves} from figure
        \protect\ref{mf_dmdh_figa}. The exponents are
        $\beta=1/2$ and $\beta\delta=3/2$ (mean field values). The curves are
        smoothed over $5$ data points (using a running average)
        to show the collapse better. The curves
        do not collapse well for large and small $h/r^{\beta\delta}$,
        unless we get very close to the critical disorder (see figure
        \protect\ref{mf_dmdh_figd}).
        (b) Scaling collapse of data in (a) but with exponents
        $\beta=0.4$ and $\beta\delta=1.65$. The collapse is better, although
        the exponents are wrong. \label{mf_dmdh_fig}}
\end{figure}

\begin{figure}
\centerline{
\psfig{figure=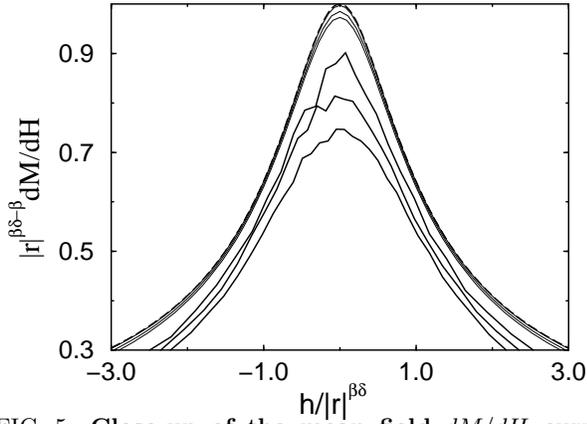,width=3truein} 
}
\caption[Close--up of mean field $dM/dH$ curves collapse]
        {{\bf Close-up of the mean field $dM/dH$ curves collapse}
        in figure \protect\ref{mf_dmdh_fig}a.
        Also plotted are
        three curves (thin lines) calculated using the mean field
        analytic solution to $M(H)$ (see text).
        These are for $R=0.80$, $0.81$, and $0.82$. We see that the
        scaling collapse, at the mean field exponents, of the $dM/dH$ curves
        converges to the expected mean field scaling function
        (thick dashed line), as $R \rightarrow R_c$. \label{mf_dmdh_figd}}
\end{figure}

The critical magnetization $M_c$ can be removed from the scaling form if
we look at the first derivative of the magnetization with respect to the
field instead. $dM/dH$ scales like:
\begin{equation}
{dM \over dH} (H,R) \sim |r|^{\beta-\beta\delta}\
\dot{\cal M}_{\pm}(h/|r|^{\beta\delta})
\label{dmdh_mf_equ1}
\end{equation}
where $\dot{\cal M}_{\pm}$ denotes the derivative of the scaling
function ${\cal M}_{\pm}$ with respect to its argument
$h/|r|^{\beta\delta}$. The $dM/dH$ mean field curves and collapses are
shown in figure~\ref{mf_dmdh_figa} and figures~\ref{mf_dmdh_fig}(a--b).
Notice that the incorrect exponents $\beta=0.4$ and $\beta\delta=1.65$
give a better collapse (fig.\ \ref{mf_dmdh_fig}b). Figure\
\ref{mf_dmdh_figd} shows a close up of figure\ \ref{mf_dmdh_fig}a,
alongside with three (thin) curves for disorders: $0.80,0.81,$ and
$0.82$. These are not measured in the simulation (the finite number of
mean field spins we use give rise to finite size effects near $R_c$ as
we will see soon); instead they are numerically calculated from the mean
field implicit equation for the magnetization\ \cite{Sethna,Dahmen1}:
\begin{equation}
M(H) = 1 - 2 \int_{-\infty}^{-J^{*}M(H)-H} \rho(f)\ df
\label{mofh_implicit}
\end{equation}
where $J^{*}$ denotes the coupling of one spin to {\it all} the other
spins in the system, and $\rho(f)$ is the random field distribution
function.

The scaling collapse converges to the expected scaling
function (dashed thick line) as we get closer to the critical disorder.
The expected scaling form is also obtained from an analytic expression
derived in mean field\ \cite{Sethna,Dahmen1}. It is given by the
smallest real root $g(y)$ of the cubic equation:
\begin{equation}
g^3 + {12 \over \pi} g - {12 \sqrt{2} \over \pi^{3/2} R_c} y = 0.
\label{g_eqn}
\end{equation}

We again find that the critical exponents and $R_c$, obtained from the
$dM/dH$ curves, are ill-determined. In finite dimensions, that is even
more true since we have another parameter to fit: $H_c$. For dimensions
$3$, $4$, and $5$, we extract $\beta$, $\beta\delta$, $H_c$, and $R_c$
by other means and simply show the resulting collapse of the $M(H)$ and
$dM/dH$ curves as a check.

\begin{figure}
\centerline{
\psfig{figure=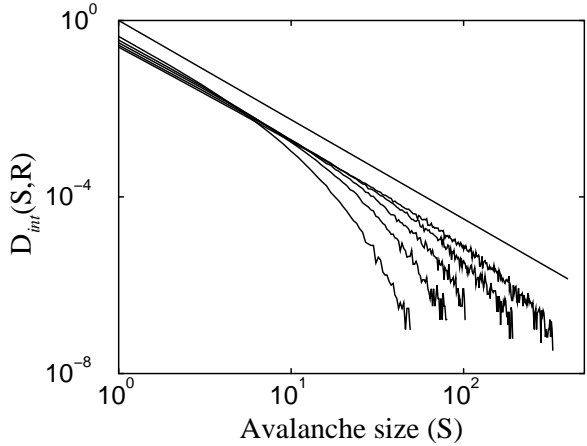,width=3truein} 
}
\caption[Mean field integrated avalanche size distribution curves]
        {{\bf Mean field integrated avalanche size distribution curves} for
        $10^6$ spins and disorders $R=0.912$, $0.974$, $1.069$, $1.197$, and
        $1.460$
        (from right to left). The straight line is the slope of the power law
        behavior in mean field: $\tau + \sigma\beta\delta=9/4$.
        \label{mf_aval_collapfigaa}}
\end{figure}

As mentioned earlier, the spins flip in avalanches of varying sizes. The
distribution of ${\it all}$ the avalanches that occur at a disorders $R$
while the external field $H$ is raised adiabatically from $-\infty$ to
$+\infty$ is plotted in figure\ \ref{mf_aval_collapfigaa}. The curves in
this plot are normalized by the number of spins in the system, and
therefore represent the probability {\it per spin} for an avalanche of
size $S$ to occur in the hysteresis loop, at disorder $R$. The curves
can be normalized to one if they are divided by the total number of
avalanches in the loop, and multiplied by the number of spins in the
system.

Often in experiments, the $\it binned$ avalanche size distribution,
which contains only avalanches that occur in a small range of fields
around a particular value of the field $H$, is measured instead. The
scaling form for this distribution\cite{note2} is\
\cite{Sethna,Dahmen1}:
\begin{equation}
D(S,R,H) \sim S^{-\tau}\ {\bar {\cal D}_{\pm}}
(S^\sigma |r|, h/|r|^{\beta\delta})
\label{int_aval0}
\end{equation}
where $S$ is the size of the avalanche and is large, and $r$ and $h$ are
small. In mean field, $\sigma=1/2$ and $\tau=3/2$. The scaling form for
the integrated avalanche size distribution is obtained by integrating
the above form over all fields:
\begin{equation}
D_{\it int}(S,R) \sim \int S^{-\tau}\ {\bar {\cal D}_{\pm}} (S^\sigma |r|,
h/|r|^{\beta\delta})\ dh
\label{int_aval1}
\end{equation}

With the change of variable $u=h/|r|^{\beta\delta}$, equation\
(\ref{int_aval1}) becomes:
\begin{equation}
D_{\it int}(S,R) \sim S^{-\tau}\ |r|^{\beta\delta} \int {\bar {\cal D}_{\pm}}
(S^\sigma |r|,u)\ du
\label{int_aval2}
\end{equation}
The integral in equation\ (\ref{int_aval2}) is a function of $S^\sigma
|r|$ only,so we can write it as:
\begin{equation}
(S^\sigma |r|)^{-\beta\delta}\ {\bar {\cal D}_{\pm}}^{(int)}
(S^\sigma |r|)
\label{int_aval2a}
\end{equation}
to obtain the scaling form for the integrated avalanche size
distribution:
\begin{equation}
D_{\it int} (S,R) \sim S^{-(\tau + \sigma\beta\delta)}\
{\bar {\cal D}_{\pm}}^{(int)} (S^\sigma |r|)
\label{int_aval3}
\end{equation}
To obtain equation\ (\ref{int_aval2a}), we have assumed that the
integral in (\ref{int_aval2}) converges. This is usually safe to do
since the distribution curves near the critical point drop off
exponentially for large arguments. The same kind of argument can be made
for the integrals of other measurements as well.

\begin{figure}
\centerline{
\psfig{figure=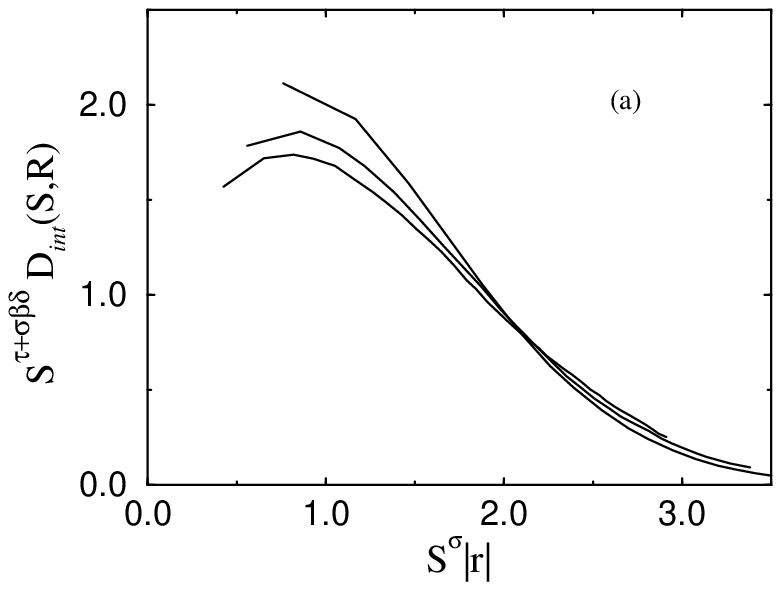,width=3truein} 
}
\nobreak
\centerline{
\psfig{figure=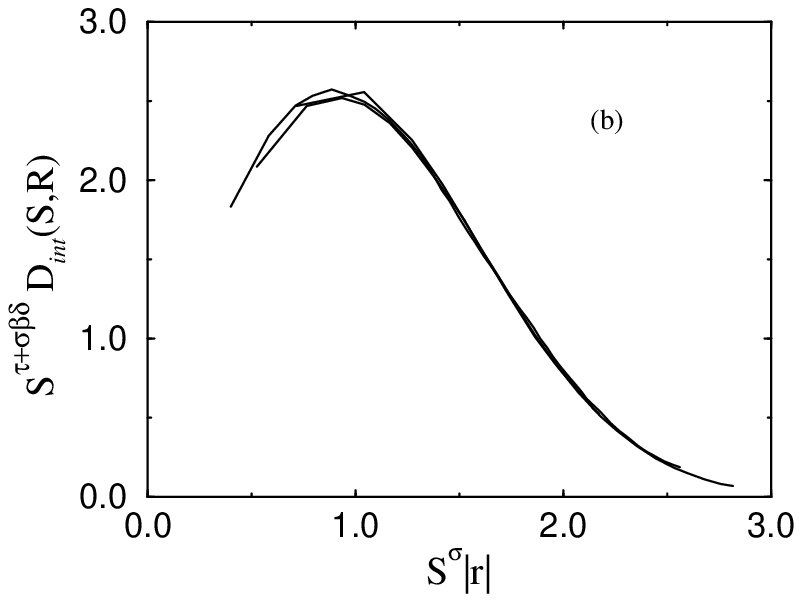,width=3truein} 
}
\caption[Scaling collapse of the mean field integrated avalanche
        size distribution curves]
        {(a) {\bf Scaling collapse of three integrated avalanche
        size distribution curves in mean field}, for disorders:
        $1.069$, $1.197$, and $1.460$.
        The curves are smoothed
        over $5$ data points before they are collapsed. The collapse
        is done using the mean field values of the exponents $\sigma$ and
        $\tau + \sigma\beta\delta$ ($1/2$ and $9/4$ respectively),
        and $r = (R_c-R)/R$.
        (b) Same curves and scaling form as in (a), but with the exponents
        $\sigma=0.44$ and $\tau + \sigma\beta\delta = 2.4$. The collapse is
        better for the incorrect exponents! We use this ``best'' collapse
        to extract exponents for figures\ \protect\ref{mf_aval_expfig}a
        and \protect\ref{mf_aval_expfig}b, and then extrapolate to $R=R_c$
        to obtain the correct mean field exponents. \label{mf_aval_collapfiga}}
\end{figure}

Figures\ \ref{mf_aval_collapfiga}a and \ref{mf_aval_collapfiga}b show
two collapses with different critical exponents of the curves from
figure\ \ref{mf_aval_collapfigaa}, using the scaling form in equation\
(\ref{int_aval3}). Notice that the collapse with the incorrect exponents
$\tau + \sigma\beta\delta =2.4$ and $\sigma = 0.44$ is better than the
collapse with the mean field exponents $\tau + \sigma\beta\delta = 9/4$
and $\sigma = 1/2$. Although the distribution curves in figures\
\ref{mf_aval_collapfiga}a and \ref{mf_aval_collapfiga}b have disorders
that are far from the critical disorder $R_c=0.79788456$, the curves
collapse but with the {\it wrong} exponents.

\begin{figure}
\centerline{
\psfig{figure=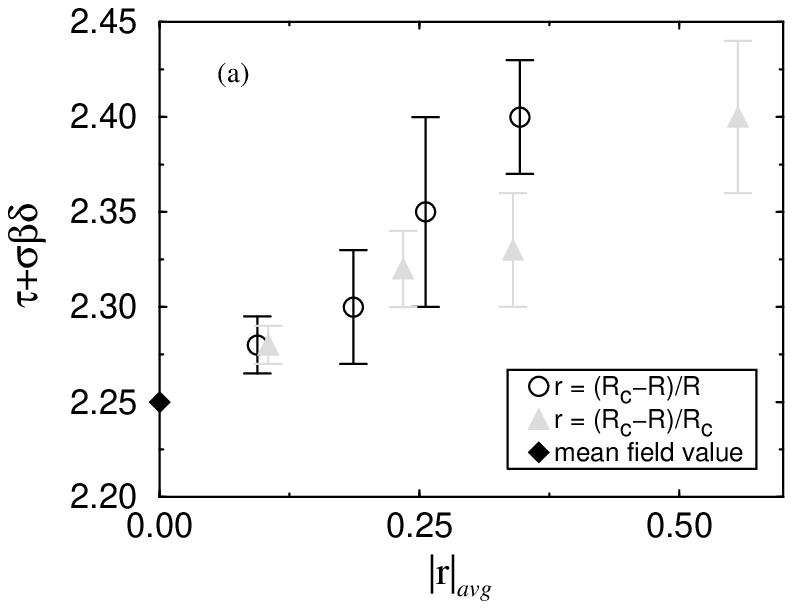,width=3truein} 
}
\nobreak
\centerline{
\psfig{figure=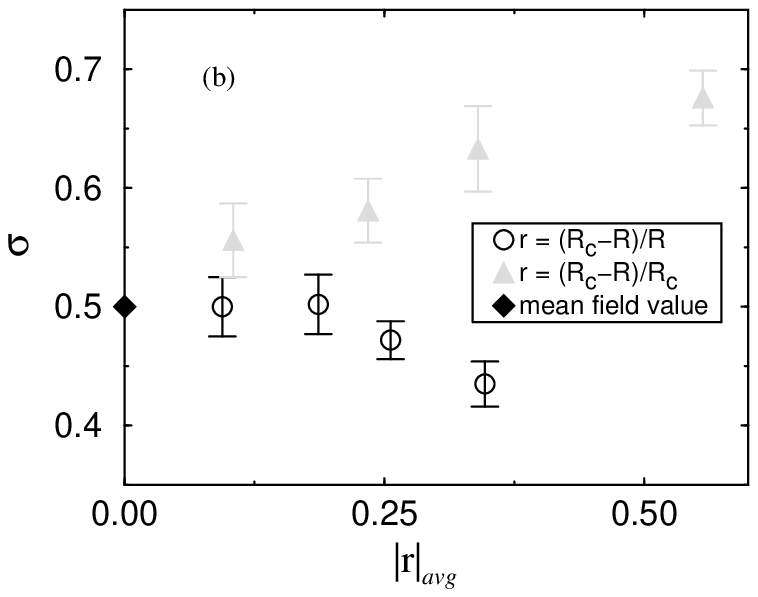,width=3truein} 
}
\caption[Mean field exponents $\tau + \sigma\beta\delta$ and $\sigma$,
        from the integrated avalanche size distribution]
        {(a) $\tau + \sigma \beta \delta$ from collapses of {\bf mean field
        integrated avalanche size distribution curves}
        for $10^6$ spins. The two points closest
        to $|r|_{\it avg}=0$ are for a system of $10^7$ spins.
        $|r|_{\it avg}$ is the average reduced disorder $|r|$
        for the three curves collapsed together (see text).
        (b) $\sigma$ from collapses of integrated avalanche size
        distribution curves for the system in (a). Again, the two closest
        points to $|r|_{\it avg}=0$ are for a system of $10^7$ spins. The mean
        field values are calculated analytically. \label{mf_aval_expfig}}
\end{figure}

It is surprising that these curves collapse at all since the scaling
form is correct only for $R$ close to $R_c$. Corrections to scaling
become important away from the critical point, but it seems that the
scaling form has enough ``freedom'' that collapses are possible even far
from $R_c$. In the limit of $R \rightarrow R_c$, we expect that the
exponents obtained from such collapses will converge to the mean field
value, and that the extrapolation will remove the question of scaling
corrections. To test this, we have collapsed three curves at a time, and
plotted the values of the exponents extracted from such collapses
against the average of the reduced disorder $|r|$ for the three curves,
which we call $|r|_{\it avg}$ (figures\ \ref{mf_aval_expfig}a and
\ref{mf_aval_expfig}b). In these figures, notice two things. First, the
linear extrapolation to $|r|_{\it avg}=0$ agrees quite well with the
mean field exponent values, and second, the points obtained by doing
collapses using $r=(R_c-R)/R$ either converge faster to the mean field
exponents or do as well as the points obtained from collapses done with
$r=(R_c-R)/R_c$. This is true for all the extrapolations that we have
done in mean field. Other models (see for example\ \cite{Robbins})
exhibit this behavior, and experimentalists seem to have known about
this for a while\ \cite{Souletie}.

In dimensions $2$ to $5$, we obtain the exponents
$\tau+\sigma\beta\delta$ and $\sigma$ in the limit $R \rightarrow R_c$,
using the above linear extrapolation method. For other collapses, if the
two extrapolation results differ substantially, we ``bias'' our result
towards the $r=(R_c-R)/R$ extrapolated value of the exponent.

Notice in figures\ \ref{mf_aval_collapfiga}a and
\ref{mf_aval_collapfiga}b that the scaling function ${\bar {\cal
D}_{-}^{(int)}}$ has a ``bump'' (the $-$ sign indicates that the
collapse is for curves with $R > R_c$). Although we will come back to
this point when we talk about the results in $5$ and lower dimensions,
it is interesting to know what the shape of the scaling function ${\bar
{\cal D}}_{-}^{(int)}$ is. In appendix B, we calculate the mean field
scaling function for $r<0$ (equations\ \ref{apA1_eq14} and
\ref{apA1_eq15}):
\begin{eqnarray}
{\bar {\cal D}}_{-}^{(int)}(-r S^{\sigma})\  =\  {e^{-(-r S^{\sigma})^2 \over 2}
\over \pi {\sqrt 2}}\ \times \nonumber \\
\int_0^{\infty}
e^{\Bigl(-(-r) S^{\sigma}\ u - {u^2 \over 2}\Bigr)}\
\Bigl(-r S^{\sigma} + u\Bigr)\ {du \over {\sqrt u}}
\label{int_aval3b}
\end{eqnarray}
where $\sigma=1/2$.

\begin{figure}
\centerline{
\psfig{figure=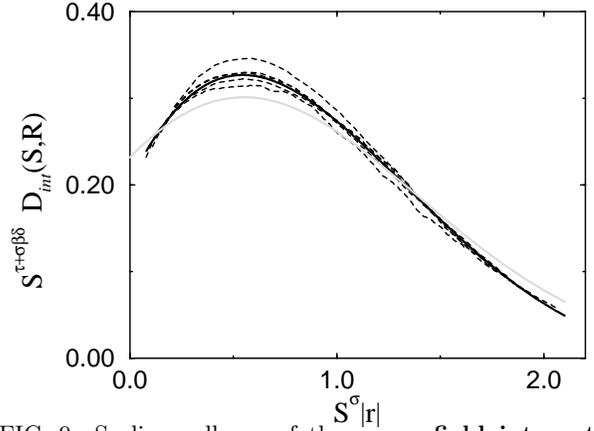,width=3truein} 
}
\caption[Mean field scaling function for the integrated avalanche size
        distribution]
        {Scaling collapse of the {\bf mean field integrated avalanche size
        distribution curves} (dashed lines),
        for $S=10^6$ spins and $R=0.912,0.974$, and
        $S=10^7$ spins and $R=0.854, 0.878, 0.912$. The critical exponents
        are: $\tau+\sigma\beta\delta =9/4$ and $\sigma =1/2$. The thick black
        line is the best fit to the data using a function that is the product
        of a polynomial and an exponential (eqn.\ (\protect\ref{int_aval3a})).
        The thick grey line is the ``real''
        mean field scaling function (see text).\label{mf_aval_fit}}
\end{figure}

A closed analytic form can not be obtained, but we can find the behavior
of this function for small and large arguments $-rS^\sigma$. For small
arguments $X=-rS^\sigma$, the scaling function is a polynomial in $X$
(\ref{apA1_eq17}), while for large arguments, the scaling function is
given by the product of an exponential decay in $X^2$ and the square
root of $X$ (\ref{apA1_eq18}). We can then try to fit our data (the
scaling collapse) with a function that will incorporate a polynomial
and an exponential decay (as an approximation to the real function). We
obtain:
\begin{eqnarray}
e^{-{X^2 \over 2}}\ &
(0.204 + 0.482 X - 0.391 X^2 + \nonumber \\
 & 0.204 X^3 - 0.048 X^4)
\label{int_aval3a}
\end{eqnarray}
This form has the expected exponential behavior at large $X$, but the
wrong pre-factor. On the other hand, for small $X$, the above function
is analytic. A better approach might be to use a parametric
representation\ \cite{Schofield}, which we have not yet tried.

Equation (\ref{int_aval3a}) can be compared with the curve obtained by
numerically integrating the scaling function ${\bar {\cal
D}}_{-}^{(int)}$ in equation (\ref{int_aval3b}). Figure\
\ref{mf_aval_fit} shows the fit in black (equation (\ref{int_aval3a}))
to the collapsed data, for curves (dashed lines) of different disorder,
and system size $S=10^6$ and $S=10^7$ spins. The grey curve is the
``real'' scaling function obtained from the numerical integration of
equation (\ref{int_aval3b}). Notice that the scaling collapse (done with
the mean field values of the exponents $\tau+\sigma\beta\delta$ and
$\sigma$) of even a system of $10^7$ spins and within $7\%$ of $R_c$
(ie. $R=0.854$) (this is the curve with the smallest peak in the graph)
is not close to the ``real'' scaling function (the thick grey curve).
The error is within $5\%$ for this curve (within $10\%$ for the fit).
However, as $R \rightarrow R_c$, the avalanche size distribution curves
seem to be approaching the ``real'' scaling function (grey curve). It is
important to keep in mind when analyzing experimental or numerical data
as we will in $5$ and lower dimensions, that the scaling collapse most
likely does not give the limiting curve one would obtain for $1/L
\rightarrow 0$ and $R \rightarrow R_c$, even for what seems like a large
size, and close to the critical disorder.

\begin{figure}
\centerline{
\psfig{figure=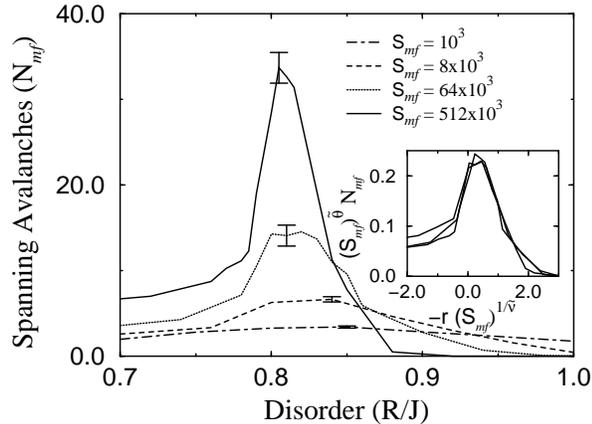,width=3truein} 
}
\caption[Spanning avalanches in mean field]
        {{\bf Number of mean field spanning avalanches}
        $N_{mf}$ as a function
        of the disorder $R$. Curves at sizes $125$ and
        $343$ are not plotted. All the
        error bars are not shown for clarity. The ones that are shown are
        representative for the peaks. The error bars are smaller for
        larger disorders.
        About 26 points are used for each curve; each point being an average
        between $250$ (for size $512000$) and $2500$ (for size $1000$)
        random field configurations. The inset shows the collapse
        of the three largest size curves using the mean field (calculated)
        exponents ${\tilde \theta}= 3/8$ and $1/{\tilde \nu}=1/4$.
        \label{span_aval_mfa}}
\end{figure}

The avalanches in the avalanche size distribution are finite, by which
we mean that they don't span the system. We have mentioned earlier that
due to the finite size of a system, close to the critical disorder
$R_c$, the largest avalanche or avalanches will span the system from one
side to another. We will talk about spanning avalanches in more details
later, but for now we just need to know that the number $N$ of spanning
avalanches scales as $N(L,R) \sim L^\theta\ {\cal N}_{\pm}(L^{1/\nu}
|r|)$ where ${\cal N}_{\pm}$ is a scaling function ($\pm$ indicates the
sign of $r$), $L$ is the linear size of the system, $\theta$ is the
exponent that arises from the existence of more than one spanning
avalanche, and $\nu$ is the correlation length ($\xi$) exponent: $\xi
\sim |r|^{-\nu}$.

\begin{figure}
\centerline{
\psfig{figure=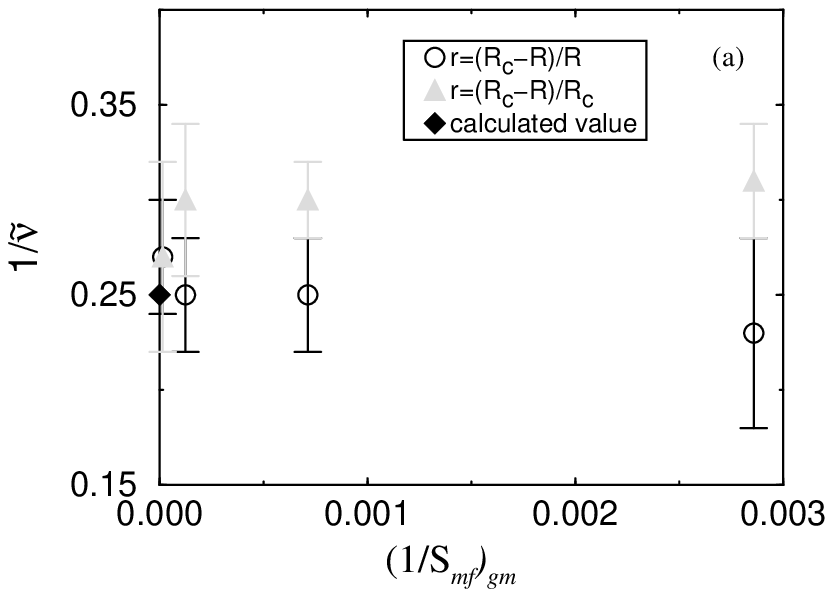,width=3truein} 
}
\nobreak
\centerline{
\psfig{figure=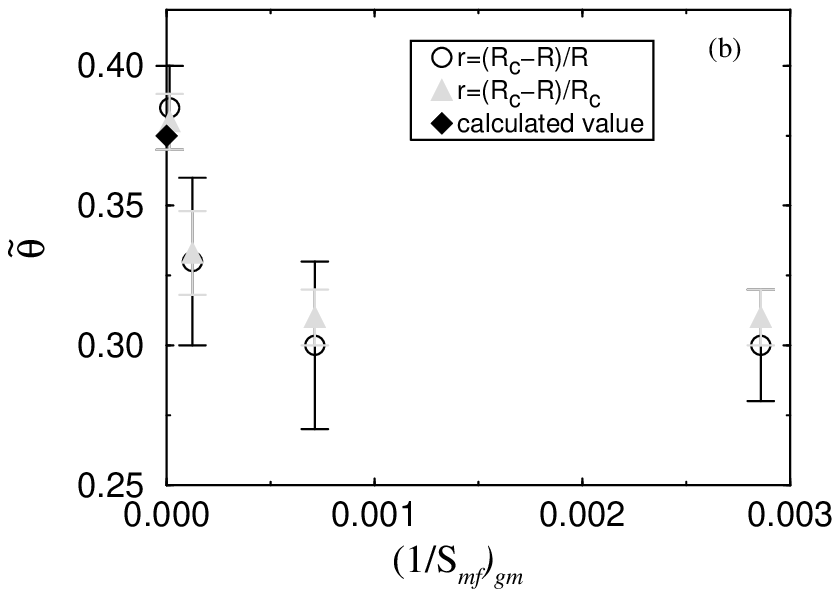,width=3truein} 
}
\caption[Mean field exponents $1/{\tilde \nu}$
        and ${\tilde \theta}$ from the spanning avalanches]
        {(a) and (b) {\bf $1/{\tilde \nu}$ and $\tilde \theta$ respectively,
        extracted from the mean field spanning avalanches collapses},
        as a function of the geometric average of $1/S_{mf}$ for three curves
        collapsed together (see text).
        The extrapolation (non-linear for $\tilde \theta$)
        to $1/S_{mf} \rightarrow 0$ agrees with the calculated
        values for the two exponents. \label{span_aval_mf}}
\end{figure}

As was mentioned earlier, in mean field there is no meaning to distance
or lattice, and thus there are no ``sides''. Purely for the purpose of
testing our extrapolation method for finite size scaling collapses in
the mean field simulation, we have defined a mean field ``spanning
avalanche'' to be one with more than $\sqrt {S_{mf}}$ spins flipping at
a field $H$, where $S_{mf}$ is the total number of spins in the system.
(Note that the mean field exponents are valid for dimensions $6$ and
above, but that in those dimensions distances do have a meaning.) Using
the above definition of a mean field spanning avalanche, it can be shown
(see appendix C) that the scaling form for their number is:
\begin{equation}
N_{mf}(S_{mf},R) \sim S_{mf}^{\tilde
\theta}\ {\cal N}_{\pm}^{mf} (S_{mf}^{1/{\tilde \nu}} |r|)
\label{span_aval_mf_eqn1}
\end{equation}
and that the values of the exponents ${\tilde \theta}$ and $1/{\tilde
\nu}$ are $3/8$ and $1/4$ respectively. $N_{mf}$ is the number of mean
field spanning avalanches, while ${\cal N}_{\pm}^{mf}$ is a universal
scaling function. The exponents ${\tilde \theta}$ and $1/{\tilde \nu}$
are defined by the arbitrary definition for a spanning avalanche.
Because of how they are defined, their values are different from the
mean field values of $1/\nu = 2$ and $\theta=1$, obtained from the
renormalization group\cite{Dahmen1,Dahmen2} and the exponent scaling
relation $1/\sigma=(d-\theta)\nu-\beta$\cite{Dahmen1,Dahmen3}.

Figure\ \ref{span_aval_mfa} shows the number of mean field spanning
avalanches as a function of disorder, for several sizes, as well as the
scaling collapse of the data. Note that the number of spanning
avalanches close to the critical disorder $R_c=\sqrt {2/\pi}$ increases
with the size $S_{mf}$ of the system, and that the peaks are getting
narrower. The scaling collapse in the inset, shows only the three
largest curves. For smaller sizes, the peaks do not collapse well with
the larger size systems presumably due to finite size effects. The
extrapolation plots for $\tilde \theta$ and $1/{\tilde \nu}$ are shown
in figures\ \ref{span_aval_mf}a and \ref{span_aval_mf}b. On the
horizontal axis of these two plots is the geometric mean of $1/S_{mf}$
for the three curves that are collapsed together, analogous to the
extrapolation method used for the integrated avalanche size
distribution. Note that the extrapolation to $1/S_{mf} \rightarrow 0$
for $\tilde \theta$ does not seem to be linear, and that the value of
$1/{\tilde \nu}$ from the linear extrapolation of the $r=(R_c-R)/R$ data
agrees better with the mean field value than the value obtained from the
linear extrapolation of the $r=(R_c-R)/R_c$ data.

Note that we measure the avalanche size distribution only for disorders
at which there are no ``mean field spanning avalanches'' (for a $10^6$
system, that is for $R \ge 0.912$), since that's what we do in
dimensions $2$ through $5$ (finite dimensions) to avoid large finite
size effects. For the second moments of the avalanche size distribution
measurements (see below), the spanning avalanches were removed (same as
in finite dimensions).

\begin{figure}
\centerline{
\psfig{figure=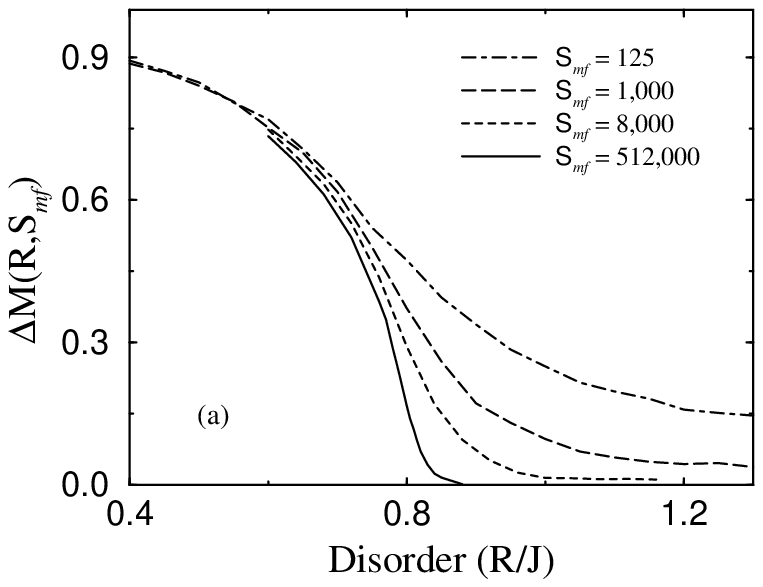,width=3truein} 
}
\nobreak
\centerline{
\psfig{figure=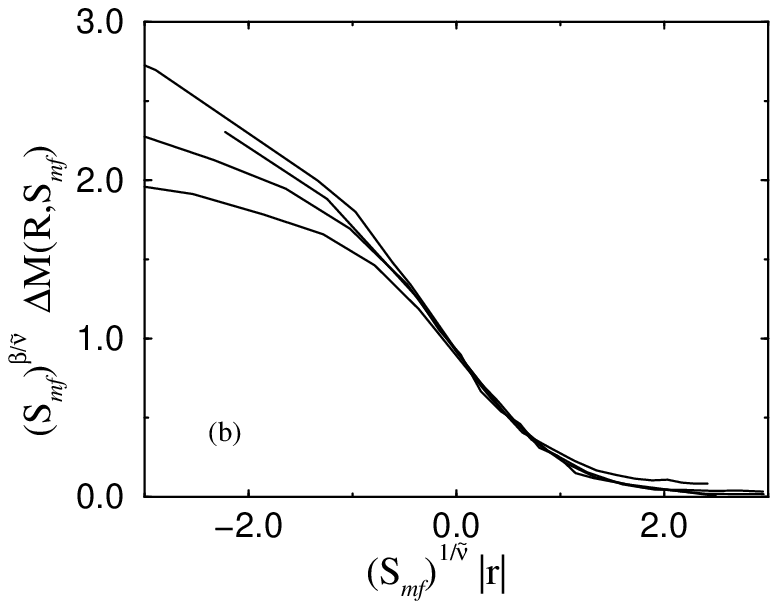,width=3truein} 
}
\caption[Magnetization change due to spanning avalanches in mean field]
        {(a) {\bf Change in the magnetization}
        due to spanning avalanches as a
        function of disorder $R$. The data is for several mean field
        system sizes.
        The critical disorder is
        $R_c =0.79788456$. The statistical errors are not larger than
        $0.005$ (in units of $\Delta M$).
        (b) Mean field scaling collapse of the change in the magnetization
        curves for sizes $S_{mf}=1000, 8000, 64000, 512000$. The
        exponents are $1/{\tilde \nu} = 0.25$ and $\beta/{\tilde \nu}= 0.125$
        and $r=(R_c-R)/R$. The part of the curve that is collapsed is for
        $R>R_c$. \label{deltaM_mofhfiga}}
\end{figure}

We have also measured the change in the magnetization $\Delta M$ due to
all the spanning avalanches, as a function of the disorder $R$ (figure\
\ref{deltaM_mofhfiga}a). This gives us an independent measurement of the
exponent $\beta$. In the thermodynamic limit above the critical
disorder, there are no spanning avalanches so the change in the
magnetization $\Delta M$ will be zero, while for small disorders the
change in the magnetization will converge to one. Close and below the
critical disorder $R_c$, at the critical field, the scaling form for the
change in the magnetization due to the spanning avalanches will be (from
equation\ (\ref{mf_mofh_equ1})):
\begin{equation}
\Delta M (H=H_c,R) \sim |r|^\beta.
\label{deltaM_mofheq1}
\end{equation}
For finite size systems, as shown in the figure, the change
in the magnetization is not zero above the critical disorder: the data
has to be analyzed using finite size scaling.

{\noindent The dependence on the
system size $S_{mf}$ can be brought in through a scaling function (see
references\ \cite{Goldenfeld,Barber}) that we call $\Delta {\cal
M}_{\pm}$:}
\begin{equation}
\Delta M(S_{mf},R) \sim |r|^\beta\
\Delta {\cal M}_{\pm}(S_{mf}^{1/{\tilde \nu}} |r|)
\label{deltaM_mofheq2}
\end{equation}
where $\tilde \nu$ is defined above, and $\pm$ refers to the sign of
$r$. We are free to define the scaling function $\Delta {\cal M}_{\pm}$
as:
\begin{equation}
\Delta {\cal M}_{\pm}(S_{mf}^{1/{\tilde \nu}} |r|)
\equiv\ \Bigl(S_{mf}^{1/{\tilde \nu}} |r|\Bigr)^{-\beta}\
{\Delta \widetilde {\cal M}_{\pm}} (S_{mf}^{1/{\tilde \nu}} |r|),
\label{deltaM_mofheq3}
\end{equation}
where ${\Delta \widetilde {\cal M}_{\pm}}$ is now a different scaling
function. The scaling form for the change of the magnetization $\Delta
M$ then becomes:
\begin{equation}
\Delta M(S_{mf},R) \sim S_{mf}^{-\beta/{\tilde \nu}}\
{\Delta \widetilde {\cal M}_{\pm}} (S_{mf}^{1/{\tilde \nu}} |r|).
\label{deltaM_mofheq4}
\end{equation}
Figure\ \ref{deltaM_mofhfiga}b shows a collapse of the data using this
scaling form. The collapse is done for disorders close to and above the
critical disorder, that is, for $r<0$. The scaling function in figure\
\ref{deltaM_mofhfiga}b, in the range of the collapse, is therefore
${\Delta \widetilde {\cal M}_{-}}$.

\begin{figure}
\centerline{
\psfig{figure=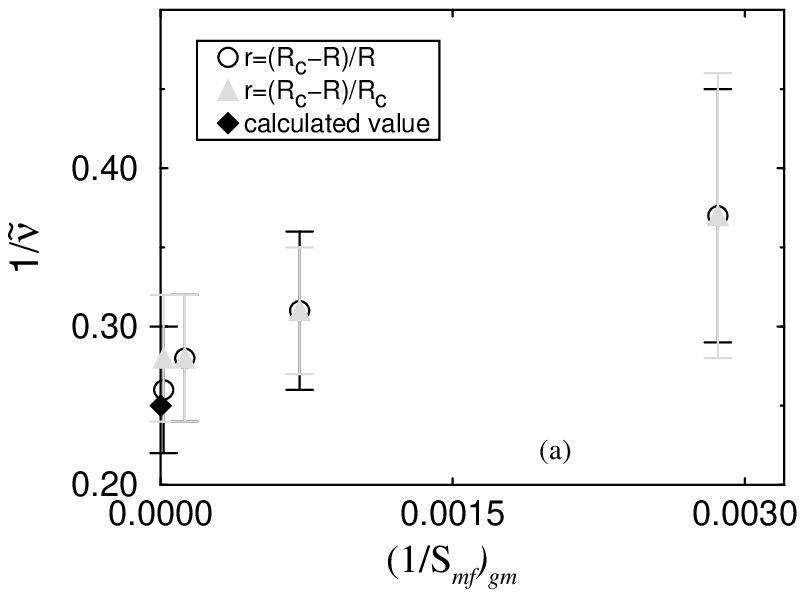,width=3truein} 
}
\nobreak
\centerline{
\psfig{figure=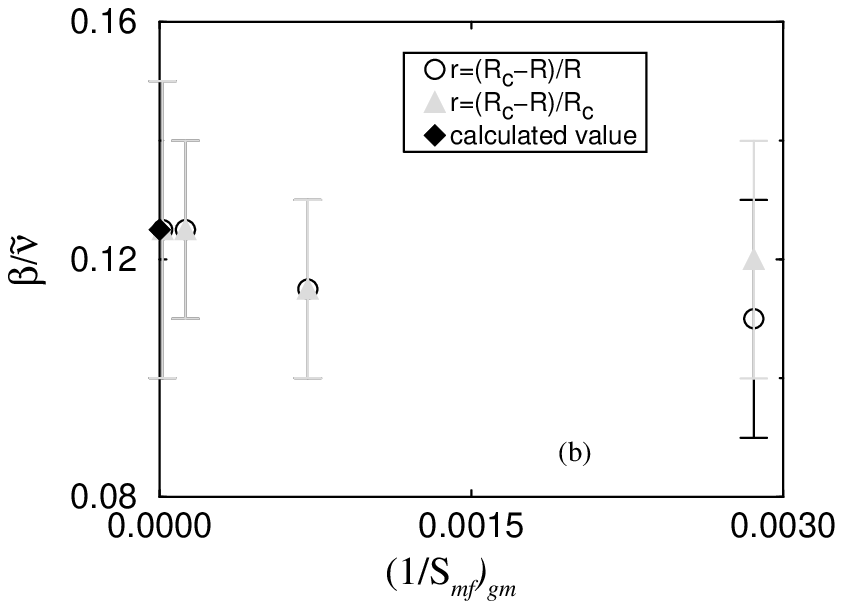,width=3truein} 
}
\caption[Mean field exponents $1/{\tilde \nu}$ and $\beta/{\tilde \nu}$
        from the magnetization change due to spanning avalanches]
        {(a) and (b) {\bf Mean field exponents $1/{\tilde \nu}$ and
        $\beta/{\tilde \nu}$ respectively,} from
        collapses of the magnetization change due to spanning avalanches
        (see text). The extrapolation to
        $(1/S_{mf})_{gm} = 0$ agrees with the calculated values.
        \label{deltaM_mofhfigb}}
\end{figure}

Values for the exponents $1/{\tilde \nu}$ and $\beta/{\tilde \nu}$
extracted from such collapses at several {\it geometric average}
reciprocal sizes are shown in figures\ \ref{deltaM_mofhfigb}a and
\ref{deltaM_mofhfigb}b. (These plots are done the same way as for the
spanning avalanches exponents.) The linear extrapolation to $1/S_{mf} =
0$ is in very good agreement with the calculated values. Note that the
extrapolation for $1/\tilde \nu$ of the $r=(R_c-R)/R$ data gives again a
better agreement with the calculated value than the extrapolation using
the $r=(R_c-R)/R_c$ data. The exponent $\beta$ in $3$, $4$, and $5$
dimensions is calculated from $\beta/\nu$, which is extracted from the
above kind of collapse. The obtained value is used to check the collapse
of the $M(H)$ and $dM/dH$ data curves.

\begin{figure}
\centerline{
\psfig{figure=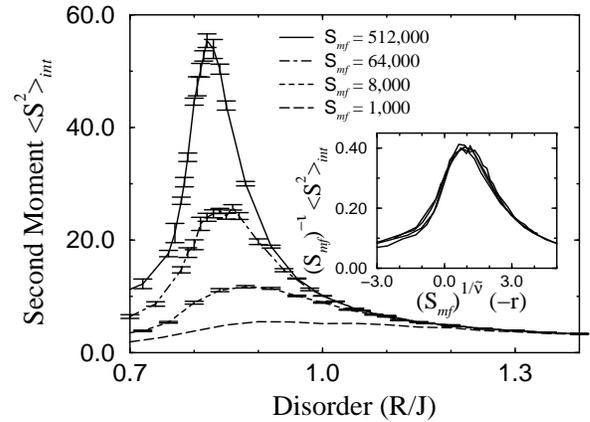,width=3truein} 
}
\caption[Second moments of the avalanche size distribution in
        mean field]
        {{\bf  Mean field second moments} of the avalanche size distribution
        integrated over the field $H$, for several different sizes.
        More than $20$ points are used for each curve; each point being an
        average of a few to several hundred random field configurations.
        The error bars for the $S_{mf}=1,000$ curve are too small to be
        shown. Curves at $S_{mf}=125$ and $343$ are not shown.
        The inset shows the collapse of these
        four curves at
        $\iota = -(\tau +\sigma\beta\delta - 3)/\sigma{\tilde \nu}
        = 3/8$ and $1/{\tilde \nu} = 1/4$,
        which are the mean field calculated values.
        \label{s2_mf_figa}}
\end{figure}

Another quantity that is related to the avalanches is the moment of the
size distribution. We have measured the second, third, and fourth moment,
and we will show how the second moment scales and collapses in mean
field. The second moment is defined as:
\begin{equation}
\langle S^2 \rangle = \int S^2\ D(S,R,H,S_{mf})\ dS
\label{s2_mf1}
\end{equation}
where $D(S,R,H,S_{mf})$ is the avalanche size distribution mentioned
above, but with the system size $S_{mf}$ included as a variable since we
are looking for the finite size scaling form, as is clear from the data
in figure\ \ref{s2_mf_figa}. Recall that only non-spanning avalanches
are included in the distribution function $D(S,R,H,S_{mf})$. Equation\
\ref{s2_mf1} can be written in terms of the scaling form for large sizes
$S$ of the avalanche size distribution $D$:
\begin{equation}
\langle S^2 \rangle \sim \int S^2\ S^{-\tau}\ {\bar {\cal D}_\pm}
(S^\sigma |r|, h/|r|^{\beta\delta}, S_{mf}^{1/\tilde \nu} |r|)\ dS
\label{s2_mf2}
\end{equation}
As we have seen before, the dependence on the system size in the scaling
function ${\bar {\cal D}_{\pm}}$ is given by $S_{mf}^{1/\tilde \nu} |r|$
where $\tilde \nu$ is defined above through the definition of a mean
field spanning avalanche. If we define
\begin{eqnarray}
{\bar {\cal D}_\pm}
(S^\sigma |r|,h/|r|^{\beta\delta}, S_{mf}^{1/\tilde \nu} |r|)\ =\ \nonumber \\
(S^\sigma |r|)^{-(2-\tau) \over \sigma}\ {\widetilde {\cal D}}_\pm
(S^\sigma |r|, h/|r|^{\beta\delta}, S_{mf}^{1/\tilde \nu} |r|)
\label{s2_mf2a}
\end{eqnarray}
where ${\widetilde {\cal D}}_\pm$ is a different scaling function, and
let $u = S|r|^{1/\sigma}$, we obtain:
\begin{equation}
\langle S^2 \rangle\ \sim\ |r|^{(\tau-3)/\sigma} \int {\widetilde {\cal D}}_\pm
(u^{\sigma}, h/|r|^{\beta\delta}, S_{mf}^{1/\tilde \nu} |r|)\ du\
\label{s2_mf3}
\end{equation}
The integral in equation (\ref{s2_mf3}) is a function of
$h/|r|^{\beta\delta}$ and $S_{mf}^{1/\tilde \nu} |r|$ only, so we can
write:
\begin{equation}
\langle S^2 \rangle\ \sim
|r|^{(\tau-3)/\sigma}\ {\cal S}_{\pm}^{(2)}( h/|r|^{\beta\delta},
S_{mf}^{1/\tilde \nu} |r|)
\label{s2_mf3a}
\end{equation}
which is the second moment scaling form, and ${\cal S}_{\pm}^{(2)}$ is a
universal scaling function.

\begin{figure}
\centerline{
\psfig{figure=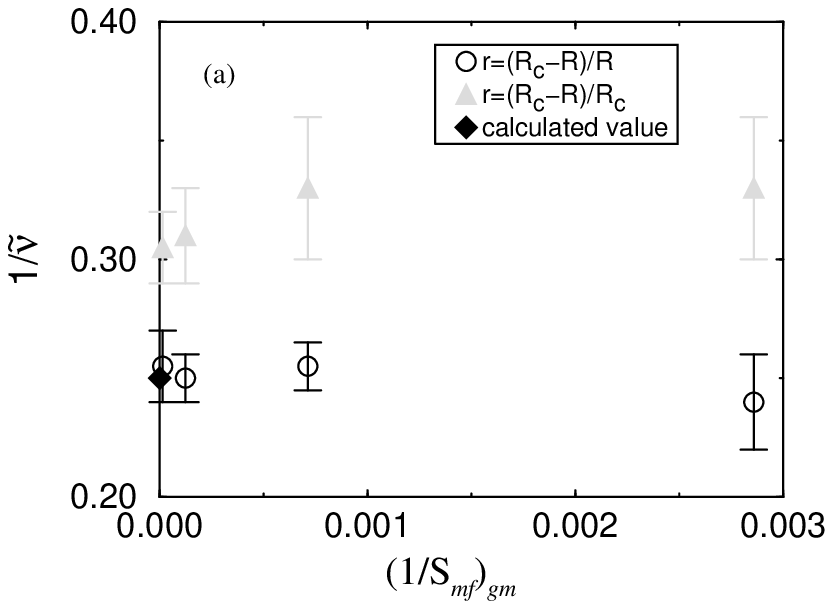,width=3truein} 
}
\nobreak
\centerline{
\psfig{figure=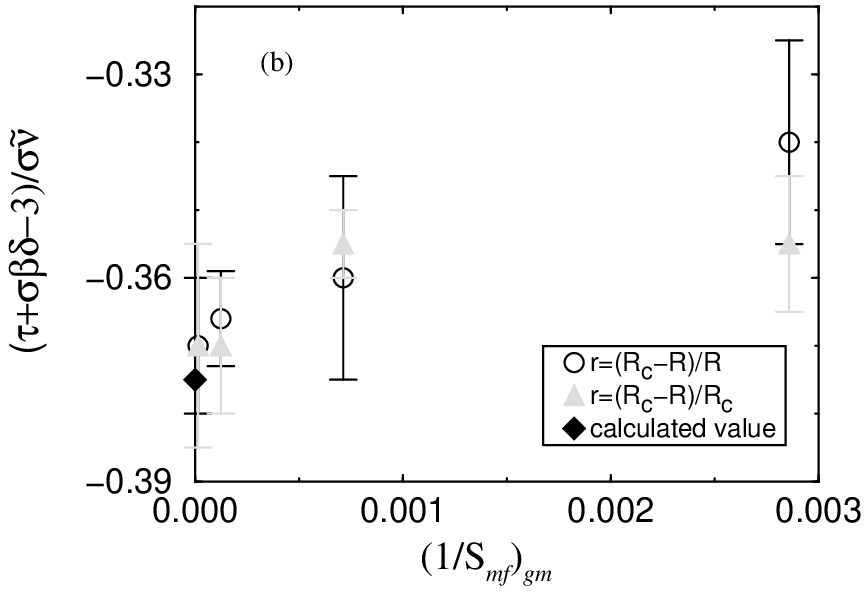,width=3truein} 
}
\caption[Mean field exponents $1/{\tilde \nu}$ and
        $(\tau +\sigma\beta\delta - 3)/\sigma{\tilde \nu}$ from the
        second moments of the avalanche size distribution]
        {(a) and (b) {\bf Values for $1/{\tilde \nu}$ and
        $(\tau +\sigma\beta\delta - 3)/\sigma{\tilde \nu}$
        respectively,} extracted from the collapses of the
        second moments of the avalanche
        size distribution. The exponents are plotted as a function
        of the geometric average of $1/S_{mf}$ for
        three curves collapsed at a time (see text).
        The extrapolation to large sizes
        agrees with the calculated values for these exponents.
        \label{s2_mf_fig}}
\end{figure}

In the simulation, we have measured the second moment of the
distribution integrated over the field $H$, whose scaling form can be
obtained by integrating the result of equation\ (\ref{s2_mf3a}):
\begin{equation}
\langle S^2 {\rangle}_{\it int} \sim |r|^{(\tau-3)/\sigma}
\int {\cal S}_{\pm}^{(2)}(h/|r|^{\beta\delta},
S_{mf}^{1/\tilde \nu} |r|)\ dh
\label{s2_mf4}
\end{equation}
As was done previously, we define $u=h/|r|^{\beta\delta}$, and call the
remaining integral:
\begin{eqnarray}
\int {\cal S}_{\pm}^{(2)}(h/|r|^{\beta\delta}, S_{mf}^{1/\tilde \nu} |r|)\ dh\
=\ \nonumber \\
(S_{mf}^{1/\tilde \nu} |r|)^{-(\tau + \sigma\beta\delta-3)/\sigma}\
{\widetilde {\cal S}}_{\pm}^{(2)}(S_{mf}^{1/\tilde \nu} |r|)
\label{s2_mf4a}
\end{eqnarray}
to obtain the second moment of the avalanche size distribution
integrated over the magnetic field $H$:
\begin{equation}
\langle S^2 {\rangle}_{\it int} \sim S_{mf}^{-(\tau +
\sigma\beta\delta-3)/{\sigma{\tilde \nu}}}\
{\widetilde {\cal S}}_{\pm}^{(2)}(S_{mf}^{1/\tilde \nu} |r|)
\label{s2_mf5}
\end{equation}
where ${\widetilde {\cal S}}_{\pm}^{(2)}$ is a universal scaling
function ($\pm$ indicates the sign of $r$). The mean field value for
$-(\tau + \sigma\beta\delta-3)/{\sigma{\tilde \nu}}$ is $3/8$.

Figure\ \ref{s2_mf_figa} shows the integrated second moments of
non-spanning mean field avalanches for several system sizes, and a
collapse using the scaling form in equation (\ref{s2_mf5}). Figures
\ref{s2_mf_fig}a and \ref{s2_mf_fig}b show the values for $1/{\tilde
\nu}$ and $-(\tau + \sigma\beta\delta-3)/{\sigma{\tilde \nu}}$
respectively, for several {\it geometric average} reciprocal sizes, and
show how well they linearly extrapolate to $1/S_{mf} \rightarrow 0$.
These plots are done the same way as for the mean field spanning
avalanches. Notice that for $1/\tilde \nu$, the linear extrapolation of
the data using $r=(R_c-R)/R$ gives a much better agreement with the
calculated value than the linear extrapolation of the data obtained
using $r=(R_c-R)/R_c$.

To summarize this section, we have shown that the values of the critical
exponents extracted from our mean field simulation by scaling collapses,
extrapolate to the expected (calculated) values for $R \rightarrow R_c$
and $1/S_{mf} \rightarrow 0$. Thus corrections to scaling due to finite
sizes as well as finite size effects near the critical point seem to be
avoided by extrapolation. The same extrapolation method is therefore
used for extracting exponents in $3$, $4$, and $5$ dimensions, which we
will see next. The results in $2$ dimensions will be shown last.



\narrowtext

\subsection{Simulation Results in $3$, $4$, and $5$ Dimensions}
\subsubsection{Magnetization Curves}

The magnetization as a function of the external field $H$ is measured
for different values of the disorder $R$. Initially all the spins are
pointing down ($s_i = -1$ for all $i$). The field is then slowly raised
from a large negative value, until a spin flips. When the first spin has
flipped, the external field is held constant while all the spins in the
avalanche are flipping. The change in the magnetization due to this
avalanche is just twice the size of the avalanche.

\begin{figure}
\centerline{
\psfig{figure=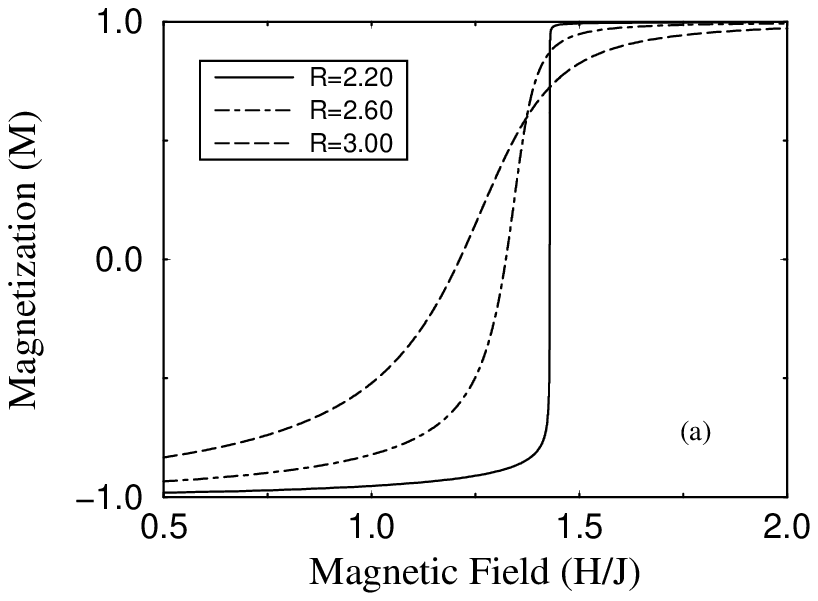,width=3truein} 
}
\nobreak
\centerline{
\psfig{figure=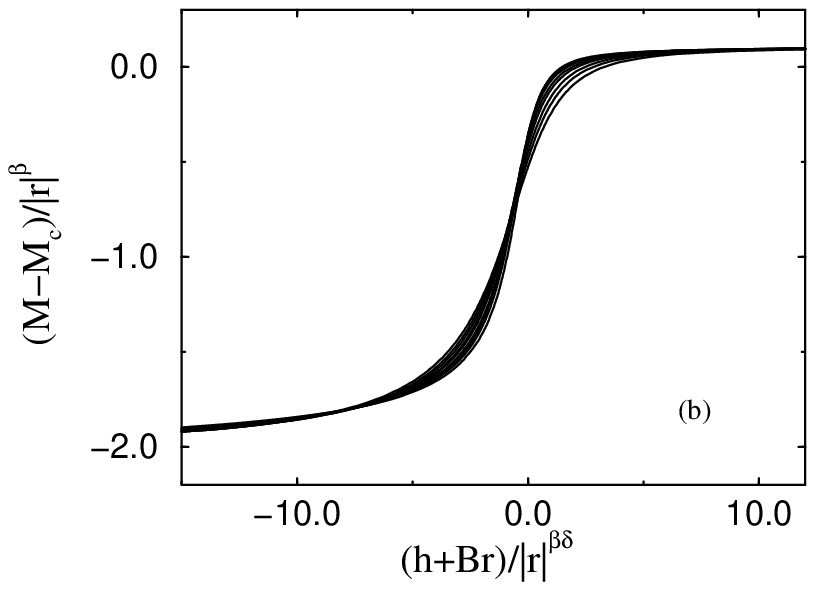,width=3truein} 
}
\caption[Magnetization curves in $3$ dimensions]
        {(a) {\bf Magnetization curves in $3$ dimensions}
        for size $L=320$, and three
        values of disorder. The curves are averages
        of up to $48$ different random field configurations. Note the
        discontinuity in the magnetization for $R=2.20$. In finite size
        systems, the discontinuity in the magnetization curve occurs even for
        $R>R_c$ ($R_c=2.16$ in $3$ dimensions).
        (b) ``Tilted'' scaling collapse (see text) of the magnetization curves
        in $3$ dimensions for size $L=320$. The disorders range from $R=2.35$ to
        $R=3.20$ ($R>R_c$).
        The critical magnetization is chosen as $M_c=0.9$ from an analysis
        of the magnetization curves and is kept fixed during the collapse. The
        exponents are $\beta=0.036$, $\beta\delta=1.81$, and the critical
        field and disorder are $1.435$ and $2.16$ respectively. The ``tilting"
        parameter $b$ is $0.39$. \label{3d_MofH_fig}}
\end{figure}

Figure \ref{3d_MofH_fig}a shows the magnetization curves obtained from
our simulation in $3$ dimensions for several values of the disorder $R$.
Similar plots can be obtained in $4$ and $5$ dimensions\cite{mosaic}. As
the disorder $R$ is decreased, a discontinuity (``jump'') in the
magnetization curve appears. The critical disorder $R_c$ is the value of
the disorder at which this discontinuity appears for the first time as
the amount of disorder is decreased, for a system in the thermodynamic
limit. For finite size systems, like the ones we use in our simulation,
the ``jump'' will occur earlier. The effective critical disorder for a
system of size $L$ is larger than the critical disorder $R_c$ ($1/L
=0$). The critical disorder $R_c$ is found from finite size scaling
collapses of the spanning avalanches and second moments of the avalanche
size distribution which will be covered later. The values are
listed in Table\ \ref{RH_table}.

We have seen in mean field that the magnetization curves near the
transition scale as
\begin{equation}
m(H,R) \sim |r|^\beta\ {\cal M}_{\pm}(h/|r|^{\beta\delta})
\label{mofh_3d_eq1}
\end{equation}
where $m=M(H,R)-M_c(H_c,R_c)$, $h=H-H_c$, and ${\cal M}_{\pm}$ is the
corresponding scaling function. The critical magnetization $M_c$ and
critical field $H_c$ are not universal quantities: in our mean field
simulation and the hard--spin mean field model for our system\
\cite{Dahmen1}, both are zero; however they are non--zero quantities in
a soft--spin model\ \cite{Dahmen1}.

In general, the scaling variables in\ (\ref{mofh_3d_eq1}) need not be
$r$ and $h$, but can instead be some ``rotated'' variables $r^\prime$
and $h^\prime$\ \cite{SouthAfrica} which to first approximation can be
written as:
\begin{equation}
r^\prime = r +a h
\label{mofh_3d_eq2}
\end{equation}
and:
\begin{equation}
h^\prime = h + br
\label{mofh_3d_eq3}
\end{equation}
(See appendix A for these and other corrections.) The constants $a$ and
$b$ are not universal and the critical exponents do not depend on them
(for the mean field data $a=b=0$).
In equation\ (\ref{mofh_3d_eq1}), the scaling variables $r$ and $h$
should be replaced by the ``rotated'' variables $r^\prime$ and
$h^\prime$, but since the measurements in our simulation are in terms of $r$
and $h$, we rewrite the scaling form in terms of those.
We find that in the leading order of scaling behavior, the
magnetization scales like:
\begin{equation}
M(H,R)-M_c\ \sim\ |r|^{\beta}\ 
{\widetilde {\cal M}}_{\pm}\Bigl((h+br)/|r|^{\beta\delta}\Bigr).
\label{mofh_3d_eq4}
\end{equation}
The correction $b\,r$ is dominant for $R \rightarrow R_c$, and can not
be ignored. The opposite is true for $a\,h$ (see appendix A).

From the previous equation, the parameters that need to be fitted are
$M_c$, $H_c$, $\beta$, $\beta\delta$, and the ``tilting'' constant $b$.
These should be found by collapsing the magnetization curves onto each
other. As in mean field, we find that collapses of magnetization curves
in $3$, $4$, and $5$ dimensions do not define well the value of the
critical magnetization $M_c$. Furthermore, we observe strong
correlations between the parameters, which lead to weak constraints on
their values.

To remove the dependence on the critical magnetization $M_c$, we can
look at the collapse of $dM/dH$ which scales like:
\begin{equation}
{dM \over dH}(H,R)\ \sim\ |r|^{\beta-\beta\delta}\ {\widetilde {\cal M}}_{\pm}
\Bigl((h+br)/|r|^{\beta\delta}\Bigr)
\label{mofh_3d_eq5}
\end{equation}
Although $M_c$ does not appear in the above form, the other parameters
are still not uniquely defined by the collapse. We find that we need to
extract $\beta$ from the magnetization discontinuity ($\Delta M$)
collapses, and $\beta\delta$ and $H_c$ from the binned avalanche size
distribution collapses rather than from the magnetization curves
themselves. Using the values obtained from these collapses, and the
value of $R_c$, the ``tilting'' constant $b$ is then found from
magnetization curve collapses (figure\ \ref{3d_MofH_fig}b).

\begin{figure}
\centerline{
\psfig{figure=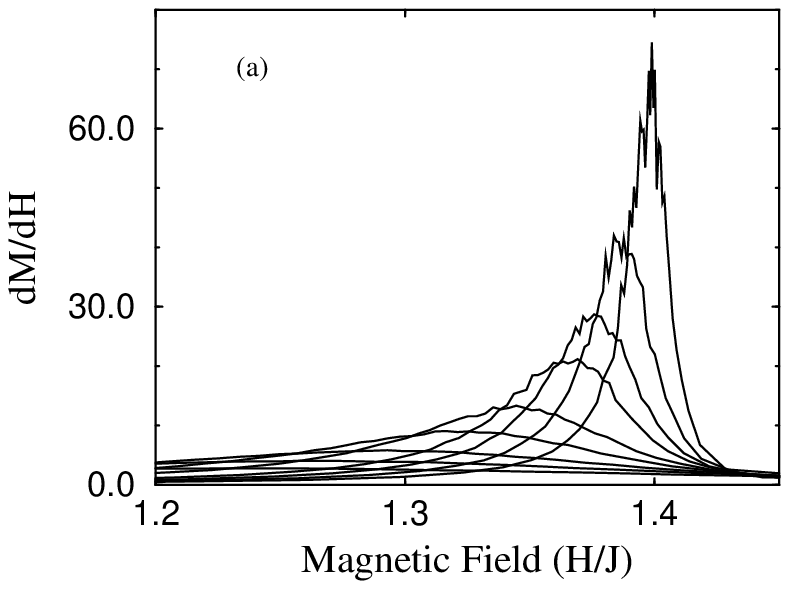,width=3truein} 
}
\nobreak
\centerline{
\psfig{figure=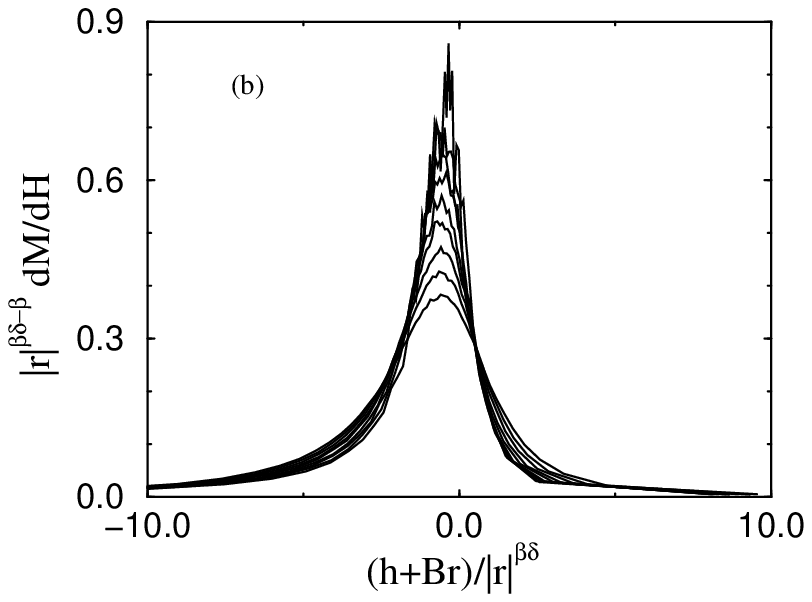,width=3truein} 
}
\caption[$dM/dH$ curves in $3$ dimensions]
        {(a) {\bf Derivative with respect to the field $H$ of the magnetization}
        $M$, for disorders $R=$ $2.35$, $2.4$, $2.45$, $2.5$, $2.6$, $2.7$,
	$2.85$, $3.0$,
        and $3.2$ (highest to lowest peak), {\bf in $3$ dimensions}.
        The curves are smoothed by $10$ data points before they are
        collapsed.
        (b) Scaling collapse of the data in (a) with $\beta =0.036$,
        $\beta\delta = 1.81$, $b=0.39$, $H_c= 1.435$, and $R_c=2.16$.
        While the curves are not collapsing onto a single curve, neither
        did they for the mean field theory curves
        (figure\ \protect\ref{mf_dmdh_fig}a). This is because the curves
        are still far from the critical disorder $R_c$. \label{dMdH_3d_fig}}
\end{figure}

Figure\ \ref{dMdH_3d_fig}a shows the curves for the derivative of the
magnetization with respect to the field $H$, and figure\
\ref{dMdH_3d_fig}b shows the scaling collapse using the same exponent
and parameter values as in figure\ \ref{3d_MofH_fig}b. The collapsed
curves have disorders larger than the critical disorder: below $R_c$,
the fluctuations are larger and the collapses are less reliable.

Since we found that $b \neq 0$ ($b=0.39$ in $3$d), the scaling variables
are indeed some $r^\prime$ and $h^\prime$, and not the variables we
measure: $r$ and $h$. Therefore, the scaling functions will in general
be functions of a different combination of scaling variables from the
ones we used in mean field, where the scaling variables are $r$ and $h$.
However, we find in appendix A that the measurements that are integrated
over the external field $H$ remove the ``tilt'' parameter $b$ (other
analytic corrections might still be important though). This is true for
the integrated avalanche size distribution, the avalanche correlation
(integrated over the field), the number of spanning avalanches, the
moments of the avalanche size distribution, and the time distribution of
avalanche sizes. In the sections that treat these measurements, we will
ignore the ``rotation'' of axis to simplify the presentation. Note that
the change in the magnetization $\Delta M$ due to the spanning
avalanches is integrated over only a small range of external fields
(wherever there are spanning avalanches). On the other hand, the
binned avalanche size distribution is not integrated over the field $H$,
and we therefore examine this measurement more carefully.




\narrowtext

\subsubsection{Avalanche Size Distribution}

\paragraph{Integrated Avalanche Size Distribution}

\begin{figure}
\centerline{
\psfig{figure=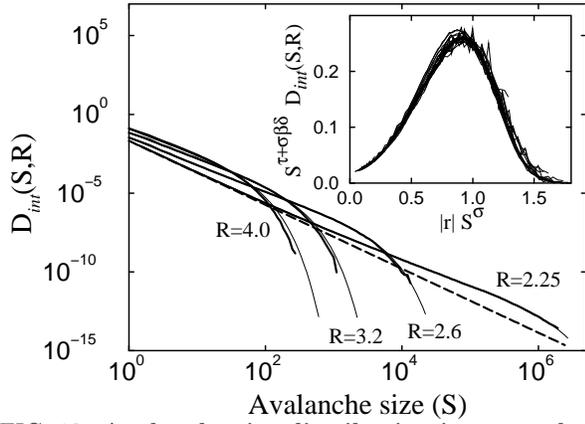,width=3truein}
}
\caption[Integrated avalanche size distribution curves in $3$ dimensions]
        {{\bf Avalanche size distribution integrated over the field $H$
        in $3$ dimensions}, for $320^3$
        spins and disorders $R=4.0, 3.2$, and $2.6$.
        The last curve is at $R=2.25$, for
        a $1000^3$ spin system. The $320^3$ curves are averages
        over up to $16$
        initial random field configurations. All curves are
        smoothed by $10$ data points
        before they are collapsed. The inset shows the scaling collapse
        of the integrated avalanche size distribution curves
        in $3$ dimensions, using $r=(R_c-R)/R$, $\tau+\sigma\beta\delta=2.03$,
        and $\sigma=0.24$,
        for sizes $160^3$, $320^3$, $800^3$, and $1000^3$, and
        disorders ranging from $R=2.25$ to $R=3.2$ ($R_c = 2.16$).
        The two top curves in the collapse, at
        $R=3.2$, show noticeable corrections to scaling. The
        thick dark curve through the collapse is the fit to the data (see
        text). In the
        main figure, the distribution curves obtained from the fit to the
        collapsed data are plotted (thin lines) alongside the raw data
        (thick lines). The straight dashed line is the expected asymptotic
        power law behavior: $S^{-2.03}$, which does not agree with the
        measured slope of the raw data due to the shape of the scaling
        function (see text). \label{aval_3fig}}
\end{figure}

In our model the spins often flip in avalanches, which are collective
flips of neighboring spins at a constant external field $H$. These
avalanches come in different sizes. The integrated avalanche size
distribution is the size distribution of all the avalanches that occur
in one branch of the hysteresis loop (for $H$ from $-\infty$ to
$\infty$). Figure\ \ref{aval_3fig}\cite{Perkovic} shows some of the raw
data (thick lines) in $3$ dimensions. Note that the curves follow a
power law behavior over several decades. Even $50\%$ away from
criticality (at $R=3.2$), there are still two decades of scaling, which
implies that the critical region is large. In experiments, a few decades
of scaling could be interpreted in terms of self-organized criticality
(SOC). However, our model and simulation suggest that several decades of
power law scaling can still be present rather ${\it far}$ from the
critical point (note that the size of the critical region is
non--universal). In the figure, the cutoff in the power law which
diverges as the critical disorder $R_c$ is approached ($R_c=2.16$ in $3$
dimensions), is a signature that the system is away from criticality,
and that a parameter can be tuned (here $R$) to bring it to the
transition. This cutoff scales as $S \sim |r|^{-1/\sigma}$, where $S$ is
the avalanche size and $r=(R_c-R)/R$ is the reduced disorder.

\begin{figure}
\centerline{
\psfig{figure=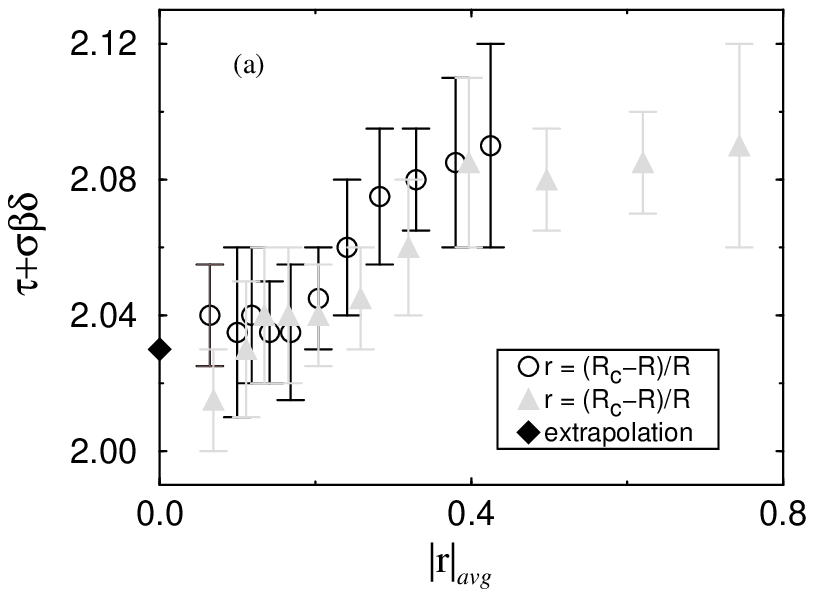,width=3truein}
}
\nobreak
\centerline{
\psfig{figure=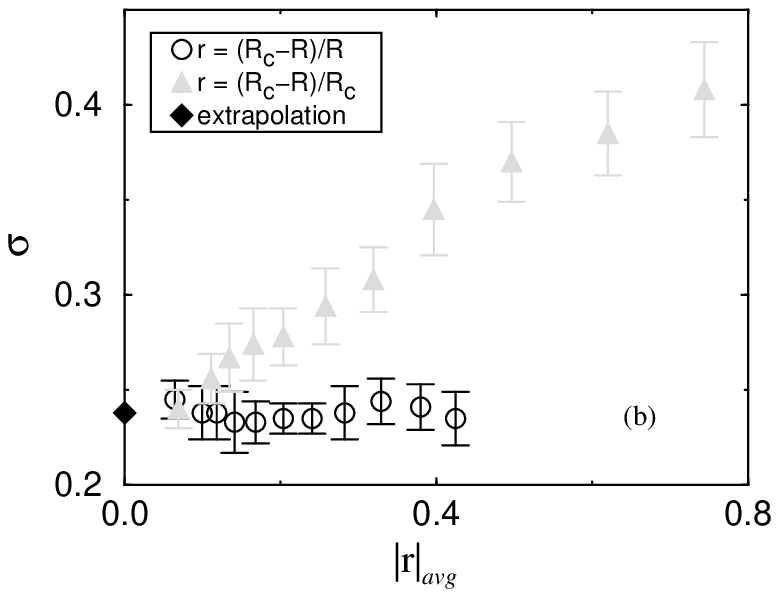,width=3truein}
}
\caption[Integrated avalanche size distribution exponents
        $\tau+\sigma\beta\delta$ and $\sigma$ in $3$ dimensions]
        {(a) and (b) $\tau+\sigma\beta\delta$ and $\sigma$ respectively,
        from collapses of the {\bf integrated
        avalanche size distribution curves} for a $320^3$ spin system.
        The data is plotted
        as in mean field. The two closest points to $|r|_{\it avg}=0$ are for
        a $800^3$ system, for a collapse using curves with disorder:
        $2.26$, $2.28$, $2.30$, $2.32$, $2.34$, and $2.36$.
        The extrapolation to $|r|_{\it avg}=0$ gives:
        $\tau + \sigma \beta \delta = 2.03$ and $\sigma = 0.24$.
        \label{3d_aval_expfig}}
\end{figure}

The power law for the curves of figure\ \ref{aval_3fig} can be obtained
through scaling collapses. A plot is shown in the inset of figure\
\ref{aval_3fig}. The scaling form is (see mean field section)
\begin{equation}
D_{int}(S,R) \sim S^{-(\tau+\sigma\beta\delta)}\
\bar{{\cal D}}^{(int)}_{-} (S^{\sigma}|r|)
\label{aval_3d_eq}
\end{equation}
where $\bar{{\cal D}}_{-}^{(int)}$ is the scaling function (the $-$ sign
indicates that the collapsed curves are for $R > R_c$). The critical
exponents $\tau+\sigma\beta\delta=2.03$ and $\sigma=0.24$ are obtained
from collapses and linear extrapolation of the extracted values to
$R=R_c$ (figures\ \ref{3d_aval_expfig}a and \ref{3d_aval_expfig}b), as
was done in mean field. (Although the ``real'' scaling variables are
$r^{\prime}$ and $h^{\prime}$, when integrating over the field $H$ we
recover the same form as in mean field; see appendix A.) Table\
\ref{measured_exp_table} lists all the exponents extracted from scaling
collapses, and extrapolated to $R \rightarrow R_c$ and $1/L \rightarrow
0$.

We have mentioned earlier that the mean field scaling function
$\bar{\cal D}_{-}^{(int)}(X)$, with $X=S^{\sigma}|r|$ and $r<0$, is a
polynomial for small $X$ and gives an exponential in $X^{1/\sigma}$
($1/\sigma=2$ in mean field) multiplied by $X^\beta$ ($\beta=1/2$ in
mean field) for large $X$ (see mean field section and appendix B). As we
have done in mean field, we can try to fit the scaling function
$\bar{\cal D}_{-}^{(int)}$ in dimensions $5$ and below with a product of
a polynomial and an exponential function. This is done in $3$ dimensions
in the inset of figure\ \ref{aval_3fig} (thick black line through the
data). The {\it phenomenological} fit is:
\begin{eqnarray}
\bar{{\cal D}}_{-}^{\it (int)}(X)\ =\ e^{-0.789X^{1/\sigma}}\ \times
\nonumber \\
(0.021+0.002X+0.531X^2-0.266X^3+0.261X^4)
\label{aval_fit_3d}
\end{eqnarray}
with $1/\sigma=4.20$ which is obtained from scaling collapses. The
distribution curves obtained using the above fit are plotted (thin lines
in figure\ \ref{aval_3fig}) alongside the raw data (thick lines). They
agree remarkably well even far above $R_c$. We should recall though,
from the mean field discussion (see figure\ \ref{mf_aval_fit}), that the
fitted curve to the collapsed data can differ from the ``real'' scaling
function even for large sizes and close to the critical disorder (in
mean field the error was about $10\%$). We expect a similar behavior in
finite dimensions.

\begin{figure}
\centerline{
\psfig{figure=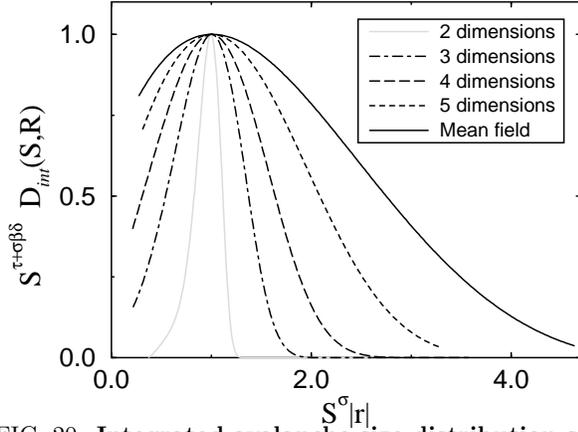,width=3truein}
}
\caption[Integrated avalanche size distribution scaling functions in $2$,
        $3$, $4$, and $5$ dimensions, and mean field]
        {{\bf Integrated avalanche size distribution scaling functions in
        $2$, $3$, $4$, and $5$ dimensions, and mean field.} The curves are
        fits (see text) to the scaling collapses done with exponents
        from Table\ \protect\ref{measured_exp_table} and\
        \protect\ref{conj_meas_2d_table}. The peaks are
        aligned to fall on (1,1). Due to the ``bump'' in the scaling function
        the power law exponent can not be extracted from a linear fit to
        the raw data. \label{bump_345fig}}
\end{figure}

The scaling function in the inset of figure\ \ref{aval_3fig} has a
peculiar shape: it grows by a factor of ten before cutting off. The
consequence of this shape is that in the simulations, it takes many
decades in the size distribution for the slope to converge to the
asymptotic power law. This can be seen from the comparison between a
straight line fit through the $R=2.25$ ($1000^3$!) curve in figure\
\ref{aval_3fig} and the asymptotic power law $S^{-2.03}$ obtained from
scaling collapses and the extrapolation (thick dashed straight line in
the same figure). A similar ``bump'' exists in other dimensions and mean
field as well. Figure\ \ref{bump_345fig} shows the scaling functions in
different dimensions and in mean field. In this graph, the scaling
functions are normalized to one and the peaks are aligned (the scaling
forms allow this). The curves plotted in figure\ \ref{bump_345fig} are
not raw data but fits to the scaling collapse in each dimensions, as was
done in the inset of figure\ \ref{aval_3fig}. The mean field and $3$
dimensions curves are given by equations\ (\ref{int_aval3a}) and
(\ref{aval_fit_3d}) respectively. For $5$, $4$, and $2$ dimensions, we
have respectively:

\begin{eqnarray}
\bar{{\cal D}}_{-}^{\it (int)}(X)\ =\ e^{-0.518 X^{1/\sigma}}\ \times
\nonumber \\
(0.112 + 0.459 X - 0.260 X^2 + 0.201 X^3 - 0.050 X^4)
\label{aval_fit_5d}
\end{eqnarray}
\begin{eqnarray}
\bar{{\cal D}}_{-}^{\it (int)}(X)\ =\ e^{-0.954 X^{1/\sigma}}\ \times
\nonumber \\
(0.058 + 0.396 X + 0.248 X^2 - 0.140 X^3 + 0.026 X^4)
\label{aval_fit_4d}
\end{eqnarray}
\begin{eqnarray}
\bar{{\cal D}}_{-}^{\it (int)}(X)\ =\ e^{-1.076 X^{1/\sigma}}\ \times
\nonumber \\
(0.492 - 4.472 X + 14.702 X^2 - \nonumber \\
20.936 X^3 + 11.303 X^4)
\label{aval_fit_2d}
\end{eqnarray}
with $1/\sigma=2.35, 3.20$, and $10.0$. The errors in the fits are in
the same range as for the mean field simulation data (see figure\
\ref{mf_aval_fit}). The $2$ dimensional fit plotted in grey will be
covered further in the next section.

From figure\ \ref{bump_345fig} we can conclude that in each dimension
(and in mean field!), a straight line fit to the integrated avalanche size
distribution data is going to give the {\it wrong} critical exponent,
and that only by doing scaling collapses and an extrapolation the
asymptotic value can be found. This is shown for $3$ dimensions in
figure\ \ref{aval_3fig}, and was found to be true in other dimensions as
well. We will next see that this is different for the {\it binned}
avalanche size distribution. The value for the slope obtained from a
linear fit to the data agrees very well with the value obtained from the
scaling collapses.

\begin{figure}
\centerline{
\psfig{figure=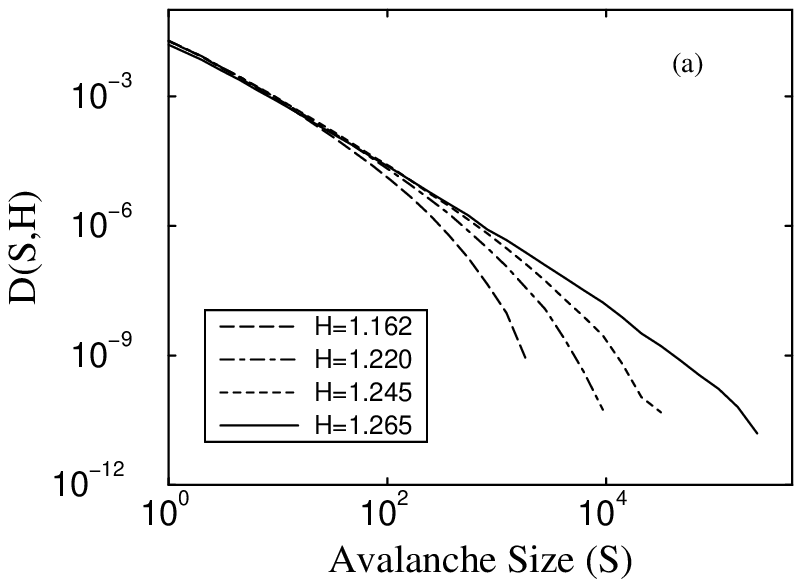,width=3truein}
}
\nobreak
\centerline{
\psfig{figure=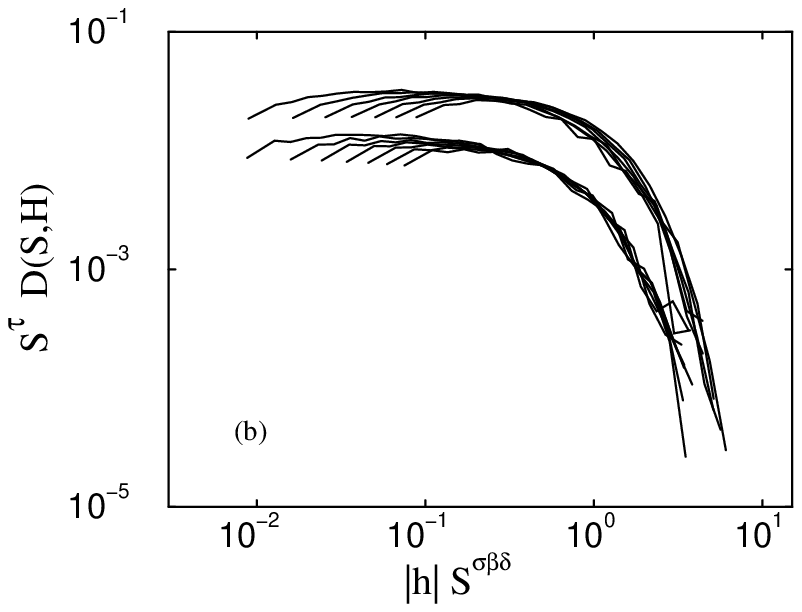,width=3truein}
}
\caption[Binned in $H$ avalanche size distribution curves in $4$ dimensions]
        {(a) {\bf Binned in $H$ avalanche size distribution in $4$ dimensions}
        for a system of $80^4$ spins at $R=4.09$ ($R_c=4.10$). The
        critical field is $H_c=1.265$. The curves are averages over
        close to $60$ random field configuration. Only a few curves are shown.
        (b) Scaling collapse of the binned avalanche size distribution
        for $H<H_c$ (upper collapse) and $H>H_c$ (lower collapse). The
        critical exponents are
        $\tau=1.53$ and $\sigma\beta\delta=0.54$, and the critical field is
        $H_c=1.265$. The bins are at fields: $1.162$, $1.185$, $1.204$,
        $1.220$, $1.234$, $1.245$, $1.254$, $1.276$, $1.285$, $1.296$,
        $1.310$, $1.326$, $1.345$, and
        $1.368$. Notice that the two scaling functions do not have a ``bump''
        (see text). \label{bin_aval_4dfig}}
\end{figure}

\paragraph{Binned in $H$ Avalanche Size Distribution}

The avalanche size distribution can also be measured at a field $H$ or
in a small range of fields centered around $H$. We have measured this
${\it binned}$ in $H$ avalanche size distribution for systems at the
critical disorder $R_c$ ($r=0$). To obtain the scaling form, we start
from the distribution of avalanches at field $H$ and disorder $R$
(eqn. \ref{int_aval0}):
\begin{equation}
D(S,R,H) \sim S^{-\tau}\ {\cal D}_{\pm}(S^\sigma |r|, |h|/|r|^{\beta\delta})
\label{aval_distr1}
\end{equation}
where as before ${\cal D}_{\pm}$ is the scaling function and $\pm$
indicates the sign of $r$. (For most of our data, we can ignore the
corrections due to the ``rotation'' of axis as explained in appendix A.)
The scaling function can be rewritten as ${\hat {\cal
D}}_{\pm}\Bigl(S^\sigma |r|, (S^\sigma |r|)^{\beta\delta}
|h|/|r|^{\beta\delta}\Bigr)$, where ${\hat D}_{\pm}$ is a new scaling
function. Letting $R \rightarrow R_c$, the scaling for the avalanche
size distribution at the field $H$, measured at the critical disorder
$R_c$ is:
\begin{equation}
D(S,H) \sim S^{-\tau}\ {\hat {\cal D}}_{\pm}(|h| S^{\sigma\beta\delta})
\label{aval_distr2}
\end{equation}

\begin{figure}
\centerline{
\psfig{figure=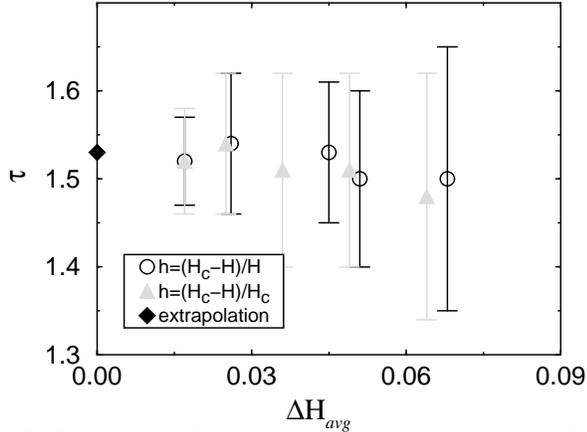,width=3truein}
}
\caption[Exponent $\tau$ from the binned in $H$ avalanche size distribution
        curves in $4$ dimensions]
        {{\bf Values for the exponent
        $\tau$ extracted from the  binned in $H$ avalanche size
        distribution curves in $4$ dimensions}, for a $80^4$ spin system
        at $R=4.09$ ($R_c=4.10$). The critical field is $H_c=1.265$.
        The exponent $\tau$ is found from this linear extrapolation to
        $\Delta H_{avg} = 0$. The exponent $\sigma\beta\delta$ is
        calculated from the value of $\tau+\sigma\beta\delta$,
        extracted from the
        integrated avalanche size distribution, and the value of $\tau$
        from this plot. \label{bin_aval_4dfigc}}
\end{figure}

Figure\ \ref{bin_aval_4dfig}a shows the binned in $H$ avalanche size
distribution curves in $4$ dimensions, for values of $H$ below the
critical field $H_c$. (The curves and analysis are similar in $3$ and
$5$ dimensions; results in $4$ dimensions are used here for variety.)
The simulation was done at the best estimate of the critical disorder
$R_c$ ($4.1$ in $4$ dimensions). The binning in $H$ is logarithmic and
started from an approximate critical field $H_c$ obtained from the
magnetization curves; better estimates of $H_c$ are obtained from the
binned distribution data curves and their collapses. Our best estimate
for the critical field $H_c$ in $4$ dimensions is $1.265 \pm 0.007$. The
scaling form for the logarithmically binned data is the same as in
equation\ (\ref{aval_distr2}), if the log-binned data is normalized by
the size of the bin. Figure\ \ref{bin_aval_4dfig}b shows the scaling
collapse for our data, both below {\it and} above the critical field
$H_c$. The ``top'' collapse gives the shape of the ${\hat {\cal D}_{-}}$
($H<H_c$) function, while the ``bottom'' collapse gives the ${\hat {\cal
D}_{+}}$ ($H>H_c$) function. Above the critical field $H_c$, there are
spanning avalanches in the system\ \cite{note4}. These are not included
in the binned avalanche size distribution collapse shown in figure\
\ref{bin_aval_4dfig}b.

The exponent $\tau$ which gives the power law behavior of the binned
avalanche size distribution is obtained from an extrapolation similar to
previous ones (figure\ \ref{bin_aval_4dfigc}), but with the field $H$
($\Delta H_{avg}$ in figure\ \ref{bin_aval_4dfigc} is the algebraic
average of $H-H_c$ for three curves collapsed together) as the variable
instead of the disorder $R$. The exponent $\sigma\beta\delta$ is found
to be very sensitive to $H_c$, while $\tau$ is not. We have therefore
used the values of $\tau + \sigma\beta\delta$ and $\sigma$ from the
integrated avalanche size distribution collapses, and $\tau$ from the
binned avalanche size distribution collapses to further constrain $H_c$
(by constraining $\sigma\beta\delta$), and to calculate $\beta\delta$.
The latter is then used to obtain collapses of the magnetization curves.
We should mention here that $H_c$ in all the dimensions is difficult to
find and that it is influenced by finite sizes. The values listed in
Table\ \ref{RH_table} are the best estimates obtained from the largest
system sizes we have. Nevertheless, systematic errors for $H_c$ could be
larger than the errors given in Table\ \ref{RH_table}. This implies
possible systematic errors for $\sigma\beta\delta$ which depends on
$H_c$, and for $\beta\delta$ which is calculated from
$\sigma\beta\delta$. These could also be larger than the errors listed
in Table\ \ref{calculated_exp_table}.

\begin{figure}
\centerline{
\psfig{figure=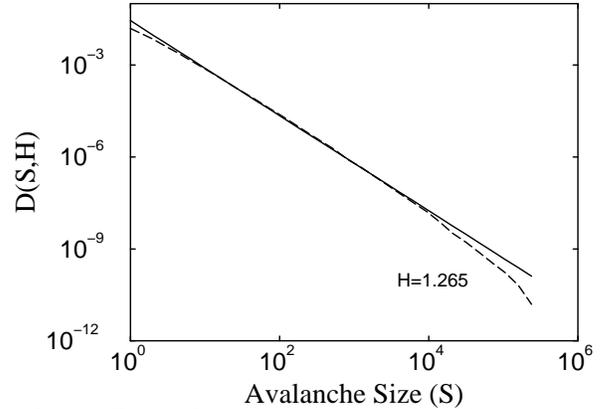,width=3truein}
}
\caption[Linear fit to a binned in $H$ avalanche size distribution curve in
        $4$ dimensions]
        {{\bf Binned avalanche size distribution curve (dashed line)
        in $4$ dimensions}, for a system of $80^4$ spins at $R_c=4.09$.
        The magnetic
        field is $H=1.265$. The straight solid line is a linear fit
        to the data for $S < 13,000$ spins. The slope from the fit
        is $1.55$ (this varies by not more than $3\%$ as the range
        over which the fit is done is changed),
        while the exponent $\tau$ obtained from the collapses
        and the extrapolation in figure\ \protect\ref{bin_aval_4dfigc}
        is $1.53 \pm 0.08$. \label{bin_aval_fit_4dfig}}
\end{figure}

From figure\ \ref{bin_aval_4dfig}b, we see that the two binned avalanche
size distribution scaling function do not have a ``bump'' as does the
scaling function for the integrated avalanche size distribution (inset
in figure \ref{aval_3fig}). Therefore, we expect that the exponent
$\tau$ which gives the slope of the distribution in figure\
\ref{bin_aval_4dfig}a can also be obtained by a linear fit through the
data curve closest to the critical field. Figure\
\ref{bin_aval_fit_4dfig} shows the curve for the $H=1.265$ bin (dashed
curve) as well as the linear fit. The slope from the linear fit is
$1.55$ while the value of $\tau$ obtained from the collapses and the
extrapolation in figure\ \ref{bin_aval_4dfigc} is $1.53 \pm 0.08$.
Fitting the binned distribution curves with a straight line to extract
the exponent $\tau$ is also possible in other dimensions and mean field
as well.



\narrowtext

\subsubsection{Avalanche Correlation}

The avalanche correlation function $G(x,R,H)$ measures the probability
that a flipping spin will trigger, through an avalanche of spins,
another spin a distance $x$ away. From the renormalization group
description\cite{Dahmen1,Dahmen2}, close to the critical point and for
large distances $x$, the correlation function is given by (corrections
are subdominant as explained in appendix A):
\begin{equation}
G(x,R,H) \sim {1 \over {x^{d-2+\eta}}}\ {\cal G}_{\pm}(x/{\xi(r,h)})
\label{correl_equ1}
\end{equation}
where $r$ and $h$ are respectively the reduced disorder and field,
${\cal G_{\pm}}$ ($\pm$ indicates the sign of $r$) is the scaling
function, $d$ is the dimension, $\xi$ is the correlation length, and
$\eta$ is called the ``anomalous dimension''. The correlation length
$\xi (r,h)$ is a macroscopic length scale in the system which is roughly
on the order of the mean linear extent of the avalanches for a system
away from the critical point.

\begin{figure}
\centerline{
\psfig{figure=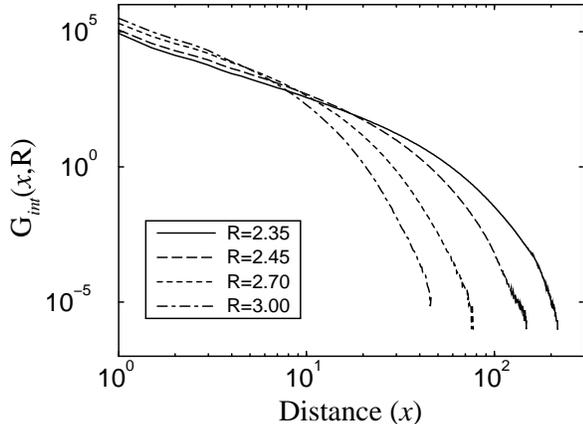,width=3truein}
}
\caption[Avalanche correlation curves in $3$ dimensions]
        {{\bf Avalanche correlation function integrated over the field $H$
        in $3$ dimensions}, for $L=320$. The curves are averages of up to $19$
        random field configurations. The critical disorder $R_c$ is $2.16$.
        \label{correl_3d_fig}}
\end{figure}

{\noindent At the critical field $H_c$ (h=0) and near
$R_c$, the correlation length scales like $\xi \sim |r|^{-\nu}$, while
for small field $h$ it is given by $\xi \sim |r|^{-\nu}\ {\cal
Y}_{\pm}(h/|r|^{\beta\delta})$ where ${\cal Y}_{\pm}$ is a universal
scaling function. The avalanche correlation function should not be
confused with the cluster or ``spin-spin'' correlation which measures
the probability that two spins a distance $x$ away have the same value.
(The algebraic decay for this other, spin-spin correlation function at
the critical point ($r=0$ and $h=0$), is $1/{x^{d-4+{\tilde
\eta}}}$\cite{Dahmen1}.)}

\begin{figure}
\centerline{
\psfig{figure=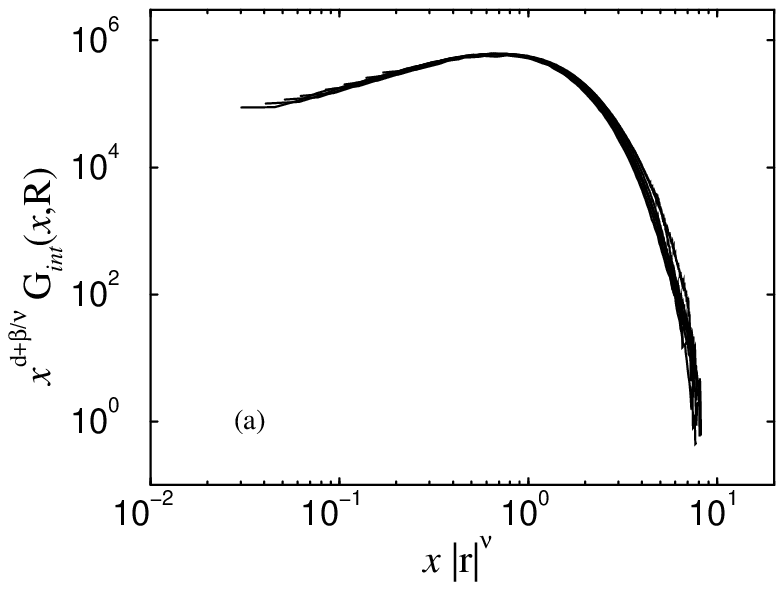,width=3truein}
}
\nobreak
\nobreak
\centerline{
\psfig{figure=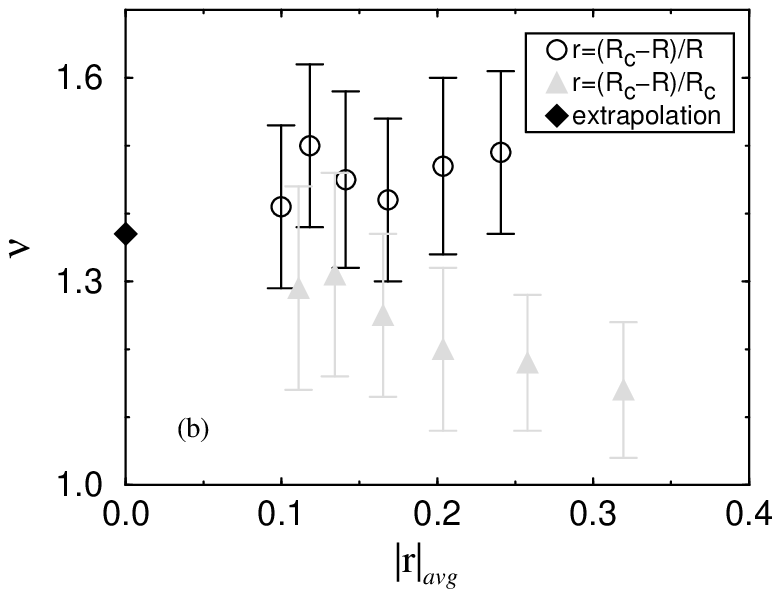,width=3truein}
}
\caption[Scaling collapse of the avalanche correlation curves in $3$
        dimensions, and exponent $\nu$]
        {(a) Scaling collapse of the {\bf avalanche correlation function
        integrated over the field $H$, in $3$ dimensions}
        for $L=320$. The values of the disorders
        range from $R=2.35$ to $R=3.0$, with $R_c=2.16$. The exponents are:
        $\nu=1.39 \pm 0.20$ and $d + \beta/\nu = 3.07 \pm 0.30$.
        (b) Exponent $\nu$ extracted from collapses of avalanche
        correlation curves
        (see (a)).
        The extrapolated value at $|r|_{avg}=0$ is
        $1.37 \pm 0.18$. \label{correl_collapse_3d_fig}}
\end{figure}

We have measured the avalanche correlation function integrated over the
field $H$, for $R>R_c$. For every avalanche that occurs between
$H=-\infty$ and $H =+\infty$, we keep a count on the number of times a
distance $x$ occurs in the avalanche. To decrease the computational time
not every pair of spins is selected; instead we do a statistical average
for $S$ pairs where $S$ is the size of the avalanche. Our simulation
seems to indicate that the difference between this statistical average
and the exact measurement is less than the fluctuations obtained from
measurements of the correlation function for different realizations of
the random field distribution. The data is saved in ``distance'' bins
separated by $0.5$ and  starting at a distance of $1.0$ (the self
correlation is not included), and is normalized by the number of
neighbors at each distance. The spanning avalanches are not included in
our correlation measurement. Figure\ \ref{correl_3d_fig} shows several
avalanche correlation curves in $3$ dimensions for $L=320$. The scaling
form for the avalanche correlation function integrated over the field
$H$, close to the critical point and for large distances $x$, is
obtained by integrating equation (\ref{correl_equ1}):
\begin{equation}
G_{\it int}(x,R) \sim \int {1 \over {x^{d-2+\eta}}}\
{\cal G}_{\pm}\Bigl(x/{\xi(r,h)}\Bigr)\ dh
\label{correl_equ2}
\end{equation}
Near the critical point $\xi(r,h) \sim |r|^{-\nu} {\cal Y}_{\pm}
(h/|r|^{\beta\delta})$.
Defining $u=h/|r|^{\beta\delta}$, equation (\ref{correl_equ2}) becomes:
\begin{equation}
G_{\it int}(x,R) \sim |r|^{\beta\delta}{x^{-(d-2+\eta)}} \int
{\cal G}_{\pm}\Bigl(x/|r|^{-\nu} {\cal Y}_{\pm} (u)\Bigr)\ du
\label{correl_equ3}
\end{equation}
The integral ($\cal I$) in equation (\ref{correl_equ3}) is a function of
$x|r|^{\nu}$ and can be written as:
\begin{equation}
{\cal I} = 
(x|r|^{\nu})^{-\beta\delta/\nu}\ {\widetilde {\cal G}}_{\pm}(x|r|^{\nu})
\label{correl_equ4}
\end{equation}
to obtain the scaling form:
\begin{equation}
G_{\it int}(x,R) \sim {1 \over x^{d+\beta/\nu}}\
{\widetilde {\cal G}}_{\pm}(x|r|^{\nu})
\label{correl_equ5}
\end{equation}
where we have used the scaling relation $(2-\eta)\nu=\beta\delta-\beta$
(see\ \cite{Dahmen1,Dahmen3} for the derivation).

\begin{figure}
\centerline{
\psfig{figure=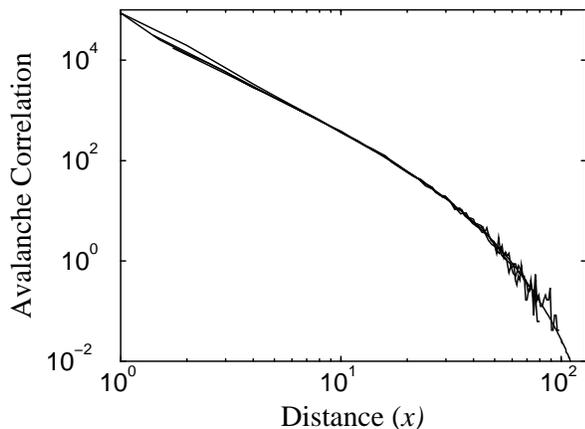,width=3truein}
}
\caption[Avalanche correlation function anisotropies in $3$ dimensions]
        {{\bf Anisotropies in the avalanche correlation function}. The curves
        are for a system of $320^3$ spins at $R=2.35$. Four curves are
        shown on the graph: one is the avalanche correlation function integrated
        over the field $H$ (as in figure\ \protect\ref{correl_3d_fig}),
        while the other three are measurements of the correlation along
        the three axis, the six face diagonals, and the four body diagonals.
        Avalanches involving more than four spins show no noticeable
        anisotropy: the critical point appears to have spherical symmetry.
        The same result is found in $2$ dimensions.
        \label{correl_aniso_3d_fig}}
\end{figure}

Figure\ \ref{correl_collapse_3d_fig}a shows the integrated avalanche
correlation curves collapse in $3$ dimensions for $L=320$ and $R>R_c$.
The exponent $\nu$ is obtained from such collapses by extrapolating to
$R = R_c$ (figure\ \ref{correl_collapse_3d_fig}b) as was done for other
collapses. The exponent $\beta/\nu$ can be obtained from these collapses
too, but it is much better estimated from the magnetization
discontinuity covered below. The value of $\beta/\nu$, listed in Table\
\ref{measured_exp_table} alongside all the other exponents, is derived
from the magnetization discontinuity collapses only.

We have also looked for possible anisotropies in the integrated
avalanche correlation function in $2$ and $3$ dimensions. The
anisotropic integrated avalanche correlation functions are measured
along ``generalized diagonals'': one along the three axis, the second
along the six face diagonals, and the third along the four body
diagonals. We compare the integrated avalanche correlation function and
the anisotropic integrated avalanche correlation functions to each
other, and find no anisotropies in the correlation, as can be seen from
figure\ \ref{correl_aniso_3d_fig}.



\narrowtext

\subsubsection{Spanning Avalanches}

The critical disorder $R_c$ was defined earlier as the disorder $R$ at
which an ${\it infinite}$ avalanche first appears in the system, in the
thermodynamic limit, as the disorder is lowered. At that point, the
magnetization curve will show a discontinuity at the magnetization
$M_c(R_c)$ and field $H_c(R_c)$. For each disorder $R$ below the
critical disorder, there is ${\it one}$ infinite avalanche that occurs
at a critical field $H_c(R)>H_c(R_c)$\ \cite{Dahmen1,Dahmen2}, while
above $R_c$ there are only finite avalanches. This is the behavior for
an infinite size system. In a finite size system far below and above
$R_c$ the above picture is still true, but close to the critical
disorder, as we approach the transition, the avalanches get larger and
larger, and we expect that one of them will be on the order of the
system size and span the system from one ``side'' to another in at least
one direction. This avalanche is not the infinite avalanche; it is only
the largest avalanche that occurs close to the critical point. If the
system was larger, this avalanche would be non--system spanning. Such an
avalanche (which spans the system) we call a spanning avalanche.

\begin{figure}
\centerline{
\psfig{figure=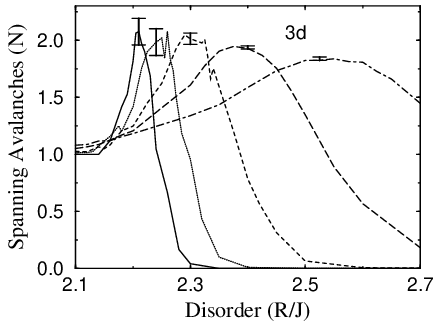,width=3truein}
}
\nobreak
\centerline{
\psfig{figure=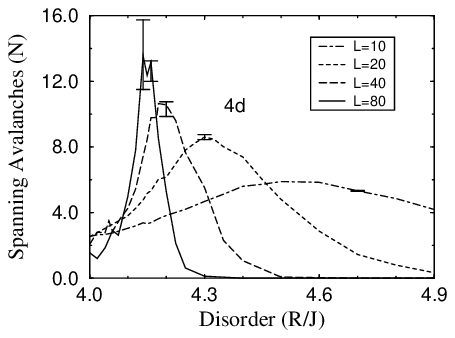,width=3truein}
}
\nobreak
\centerline{
\psfig{figure=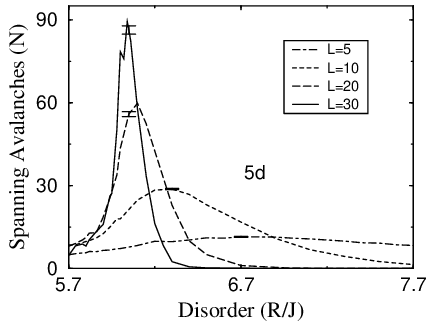,width=3truein}
}
\nobreak
\centerline{
\psfig{figure=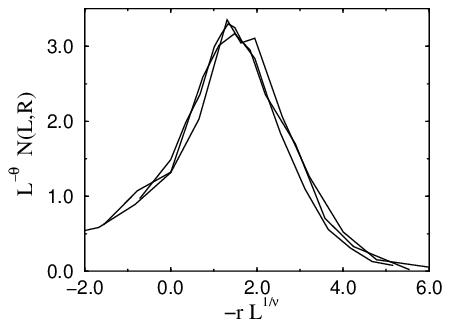,width=3truein}
}
\caption[Spanning avalanches in $3$, $4$, and $5$ dimensions]
        {(a) {\bf Number of spanning avalanches $N$ in $3$ dimensions,}
        occurring in the system between
        $H= -\infty$ to $H=\infty$, as a function of the disorder $R$,
        for linear sizes $L$: $20$ (dot-dashed), $40$ (long
        dashed), $80$ (dashed), $160$ (dotted), and $320$ (solid). The critical
        disorder $R_c$ is at $2.16$. The error bars for each curve tend to be
        smaller than the peak error bar for disorders above the peak and larger
        for disorders below the peak. They are not given here for clarity.
        Note that the number of avalanches
        increases only slightly as the size is increased.
        (b) {\bf Number of spanning avalanches in $4$
        dimensions.} The critical disorder is $4.1$.
        (c) {\bf Number of spanning avalanches in $5$ dimensions.}
        The critical disorder is $5.96$. Both in $4$ and $5$ dimensions,
        the peaks grow and shift towards $R_c$ as the size of the
        system is increased.
        (d) Collapse of the spanning avalanche curves in $4$ dimensions
        for linear sizes $L=20,40$, and $80$.
        The exponents are $\theta = 0.32$ and
        $\nu = 0.89$, and the critical disorder is
        $R_c = 4.10$. The collapse is done using $r = (R_c-R)/R$.
        \label{span_aval_345fig}}
\end{figure}

In our numerical simulation, we find that for finite sizes $L$, there
are not one but ${\it many}$ such avalanches in $4$ and $5$ dimensions
(and maybe $3$), and that their number increases as the system size
increases. Figures\ \ref{span_aval_345fig}(a-c) show the number of
spanning avalanches as a function of disorder $R$, for different sizes
and dimensions. In $4$ and $5$ dimensions, the spanning avalanche curves
become more narrow as the system size is increased. Also, the peaks
shift toward the critical value of the disorder ($4.1$ and $5.96$
respectively), and the number of spanning avalanches at $R_c$ increases.
This suggests that in $4$ and $5$ dimensions, for $L \rightarrow
\infty$, there will be one infinite avalanche below $R_c$, none above,
and an infinite number of spanning avalanches at the critical disorder
$R_c$. (These spanning avalanches are infinite avalanches for $L
\rightarrow \infty$.) In $3$ dimensions, the results are not conclusive,
which can be noticed from figure\ \ref{span_aval_345fig}a, but also from
the value of the spanning avalanche exponent $\theta = 0.15 \pm 0.15$
defined below (a value of $0$ implies only one infinite or spanning
avalanche at $R_c$ as $L \rightarrow \infty$).

In percolation, a similar multiplicity of infinite clusters\
\cite{Arcangelis,Stauffer} (as the system size is increased) is found
for dimensions above $6$ which is the upper critical dimension (UCD).
The UCD is the dimension at and above which the mean field exponents are
valid. Below $6$ dimensions, there is only one such infinite cluster.
The existence of a diverging number of infinite clusters in percolation
is associated with the breakdown of the hyperscaling relation above $6$
dimensions. Since a hyperscaling relation is a relation between critical
exponents that includes the dimension $d$ of the system, it is always
only satisfied up to and including the upper critical dimension. In our
system, the upper critical dimension is also $6$, but we find spanning
avalanches in dimensions even below that. In a comment by Maritan {\it
{et al.}}\cite{Maritan}, it was suggested that our system should satisfy
the hyperscaling relation: $d\nu-\beta = 1/\sigma$ which is also the one
found in percolation\ \cite{Stauffer}. But since our system has spanning
avalanches below the upper critical dimension, this hyperscaling
relation breaks down below $6$ dimensions. Due to the existence of many
spanning avalanches near $R_c$, the new ``violation of hyperscaling''
relation for dimensions $3$ and above becomes\ \cite{Dahmen1,Dahmen3}:
\begin{equation}
(d-\theta)\nu - \beta = 1/\sigma
\label{span_aval_eqn1}
\end{equation}
where $\theta$ is the ``breakdown of hyperscaling'' or spanning
avalanches exponent defined below. One can check that our exponents in
$3$, $4$, and $5$ dimensions and mean field satisfy this equation (see
Tables~\ref{measured_exp_table} and~ \ref{calculated_exp_table}).

For the simulation, we define a spanning avalanche to be an avalanche
that spans the system in one direction. We average over all the
directions to obtain better statistics. Depending on the size and
dimension of the system and the distance from the critical disorder, the
number of spanning avalanches for a particular value of disorder $R$ is
obtained by averaging over as few as $5$ to as many as $2000$ different
random field configurations. We define the exponent $\theta$ such that
the number $N$ of spanning avalanches, at the critical disorder $R_c$,
increases with the linear system size as: $N \sim L^{\theta}$ ($\theta >
0$). The finite size scaling form\cite{Goldenfeld,Barber} for the number of
spanning avalanches close to the critical disorder is:
\begin{equation}
N(L,R) \sim L^{\theta}\ {\cal N}_{\pm}(L^{1/\nu}|r|)
\label{span_aval_eqn2}
\end{equation}
where $\nu$ is the correlation length exponent and ${\cal N}_{\pm}$ is
the corresponding scaling function ($\pm$ indicates the sign of $r$).
The corrections to scaling are subdominant as explained in appendix A.
The collapse is shown in figure\ \ref{span_aval_345fig}d. The values for
$\theta$ and $\nu$ from collapses of curves of sizes $L=20, 30, 40,$ and
$80$ in $4$ dimensions, are shown in Table\ \ref{span_exp_4d_table}. (We
show the results and collapses in $4$ dimensions here since the
existence of spanning avalanches in $3$ dimensions is not conclusive.)
These values are used along with the results from other collapses to
obtain Table\ \ref{measured_exp_table}. In the analysis of the avalanche
size distribution, magnetization, and correlation functions for $R>R_c$,
how close we chose to come to the critical disorder $R_c$ was determined
by the spanning avalanches: we include no values $R$ below the first
value which exhibited a spanning avalanche.



\subsubsection{Magnetization Discontinuity}

We have mentioned earlier that in the thermodynamic limit, at and below
the critical disorder $R_c$, there is a critical field $H_c(R)>H_c(R_c)$
at which the infinite avalanche occurs. Close to the critical
transition, for $r$ small and $H=H_c(R)$, the change in the
magnetization due to the infinite avalanche scales as (equation
(\ref{mofh_3d_eq1})):
\begin{equation}
\Delta M(R) \sim\ r^{\beta}
\label{deltaM_eq1}
\end{equation}
where $r=(R_c-R)/R$, while above the transition, there is no infinite
avalanche.

\begin{figure}
\centerline{
\psfig{figure=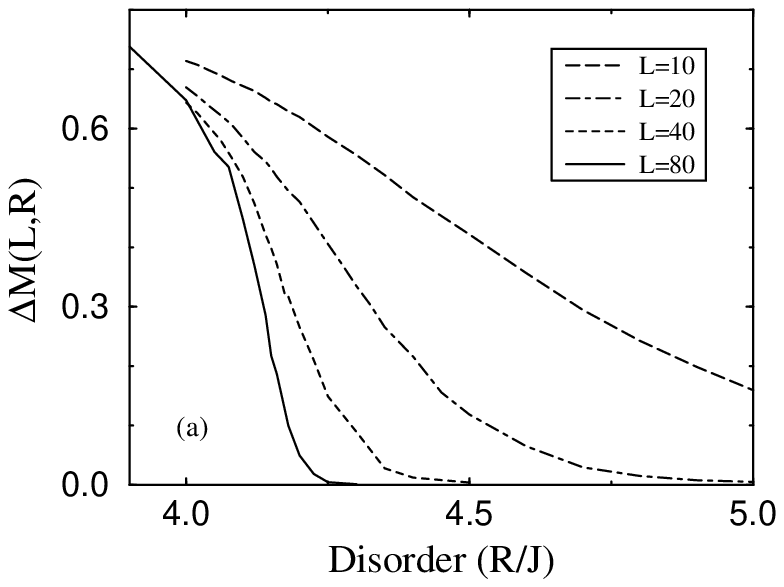,width=3truein}
}
\nobreak
\centerline{
\psfig{figure=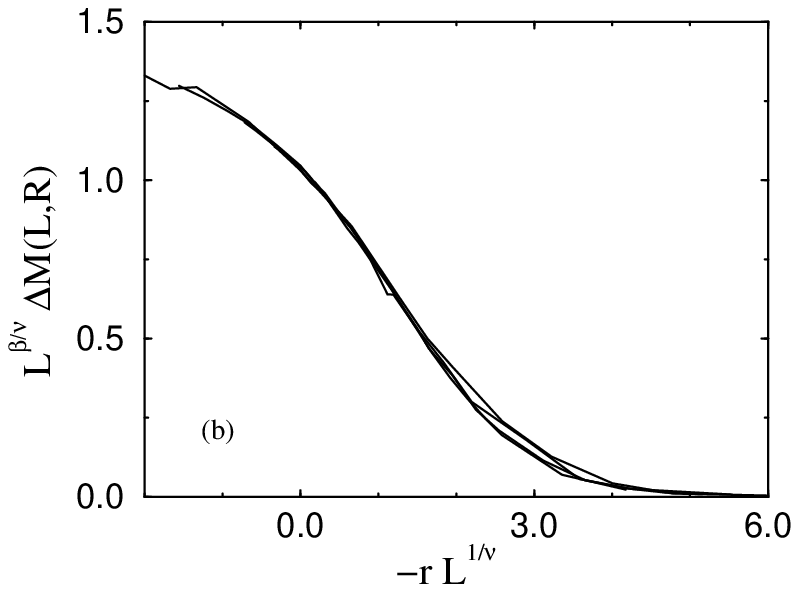,width=3truein}
}
\caption[Magnetization change due to spanning avalanches, in $4$ dimensions]
        {(a) {\bf Change in the} {\bf magnetization} {\bf due to the}
        {\bf spanning} {\bf avalanches in} {\bf $4$ dimensions,}
        for several linear sizes $L$,
        as a function of the disorder $R$.
        (b) Scaling collapse of the curves in (a) using $r=(R_c-R)/R$.
        The exponents are
        $1/\nu = 1.12$ and $\beta/\nu = 0.19$, and the critical disorder is
        $R_c = 4.1$. \label{deltaM_4d_fig}}
\end{figure}

{\noindent In finite size systems, the transition is not as sharp: we
have spanning avalanches above the critical disorder. If we measure the
change in the magnetization due to all the spanning avalanches (the
infinite avalanche is included too), the scaling form for that quantity
is going to depend on the system size $L$ analogous to the scaling of
the number of spanning avalanches:}
\begin{equation}
\Delta M(L,R) \sim\ |r|^{\beta}\ \Delta {\cal M}_{\pm}(L^{1/\nu}|r|)
\label{deltaM_eq2}
\end{equation}
where $\Delta {\cal M}_{\pm}$ is a universal scaling function. (Since
$\Delta M(L,R)$ is measured at $h^{\prime}=0$, corrections to scaling
are subdominant; see also appendix A.) Defining a new universal scaling
function $\Delta \widetilde {\cal M}_{\pm}$:
\begin{equation}
\Delta {\cal M}_{\pm}(L^{1/\nu}|r|) \equiv\ (L^{1/\nu}|r|)^{-\beta}\
\Delta \widetilde {\cal M}_{\pm}(L^{1/\nu}|r|)
\label{deltaM_eq3}
\end{equation}
we obtain the scaling form:
\begin{equation}
\Delta M(L,R) \sim\ L^{-\beta/\nu}\ \Delta {\widetilde {\cal M}_{\pm}}
(L^{1/\nu}|r|)
\label{deltaM_eq4}
\end{equation}

Figures\ \ref{deltaM_4d_fig}a and \ref{deltaM_4d_fig}b show the change
in the magnetization due to the spanning avalanches in $4$ dimensions,
and a scaling collapse of that data (similar results exist in $3$ and
$5$ dimensions). Notice that as the system size increases, the curves
approach the $|r|^{\beta}$ behavior. The exponents $1/\nu$ and
$\beta/\nu$ are extracted from scaling collapses (figure\
\ref{deltaM_4d_fig}b) and are listed in table\ \ref{deltaM_4d_table}.
The value of $\beta$ is calculated from $\beta/\nu$ and the knowledge of
$\nu$, and is the value used for collapses of the magnetization curves
(see earlier).

\begin{figure}
\centerline{
\psfig{figure=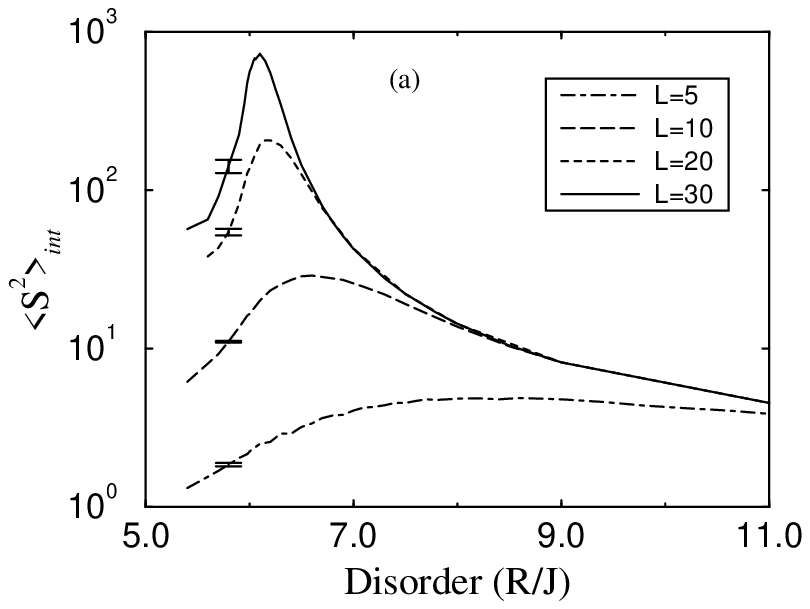,width=3truein}
}
\nobreak
\centerline{
\psfig{figure=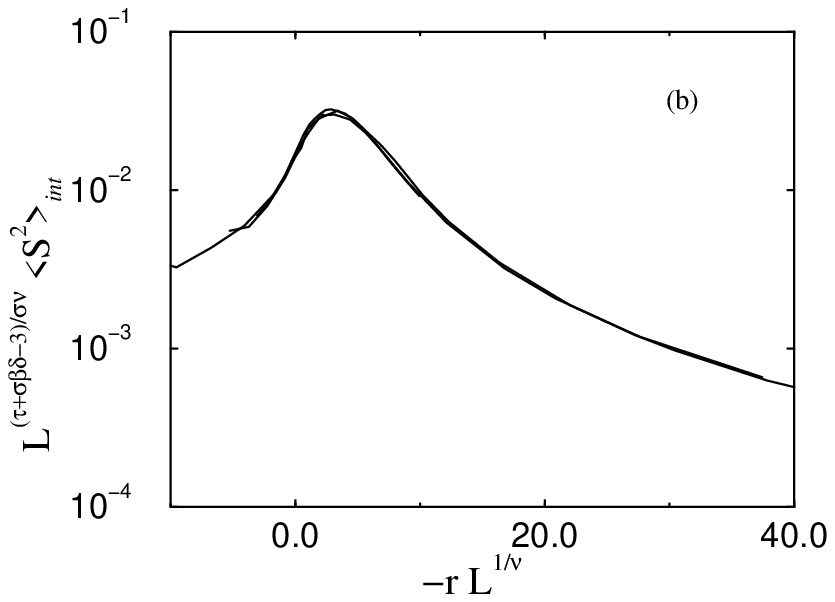,width=3truein}
}
\caption[Second moments of the avalanche size distribution
        in $5$ dimensions]
        {(a) {\bf Second moments}
        of the avalanche size distribution integrated over
        the field $H$, {\bf in $5$ dimensions.}
        Error bars are largest for smaller
        disorders (shown on the curves).
        The curves have between $24$ and $50$ points, and the value of the
        second moment for each
        disorder is averaged over $3$ to $100$ different random field
        configurations.
        (b) Scaling collapse of the $L=10, 20$, and $30$ curves from (a) using
        $r=(R_c-R)/R$.
        The exponents are $1/\nu = 1.47$ and $\rho = -(\tau+\sigma\beta\delta
        - 3)/\sigma\nu = 2.95$,
        and the critical disorder is $R_c = 5.96$. \label{s2_5d_fig}}
\end{figure}



\narrowtext

\subsubsection{Moments of the Avalanche Size Distribution}

The second moment of the avalanche size distribution was defined earlier
(see the mean field simulation section). We found that the scaling form
of the integrated over $H$ second moment is (equation\ \ref{s2_mf5}):
\begin{equation}
\langle S^2 {\rangle}_{\it int} \sim L^{-(\tau+\sigma\beta\delta
- 3)/\sigma\nu}\
{\widetilde {\cal S}}_{\pm}^{(2)}(L^{1/\nu}|r|)
\label{s2_5d}
\end{equation}
where $L$ is the linear size of the system, $r$ is the reduced disorder,
$\widetilde {\cal S}_{\pm}^{(2)}$ is the scaling function, and $\nu$ is
the correlation length exponent. The corrections are subdominant
(appendix A). We can similarly define the third and fourth moment, with
the exponent $-(\tau+\sigma\beta\delta - 3)/\sigma\nu$ replaced by
$-(\tau+\sigma\beta\delta-4)/\sigma\nu$ and $-(\tau+\sigma\beta\delta
-5)/\sigma\nu$ respectively. Figures\ \ref{s2_5d_fig}a and
\ref{s2_5d_fig}b show the second moments data in $5$ dimensions for
sizes $L=5, 10, 20,$ and $30$, and a collapse (again, results in $3$ and
$4$ dimensions are similar and we have chosen to show the curves in $5$
dimensions for variety). The curves are normalized by the average
avalanche size integrated over all fields $H$: $\int_{-\infty}^{+\infty}
\int_{1}^{\infty} S\ D(S,R,H,L)\ dS\ dH$. The spanning avalanches and
the infinite avalanche are not included in the calculation of the
moments. The collapse does not include the $L=5$ curve because, due to
finite size effects, this curve does not collapse well with the larger
size curves. Table\ \ref{s2_5d_exp_table} shows the values of the
exponents and $R_c$ from the collapses. The exponents for the third and
fourth moment can be calculated from this table, and we find that they
agree with the values obtained from their respective collapses.



\narrowtext

\subsubsection{Avalanche Time Measurement}

The exponents we have measured so far are static scaling exponents: they
do not depend on the dynamics of the model. If we measure the time an
avalanche takes to occur, we are making a dynamical measurement. The
time measurement in the numerical simulation is done by increasing the
time ``meter'' by one for each shell of spins in the avalanche; it
corresponds to a synchronous dynamics, where, when all unstable spins
are flipped, time is incremented by one, and the new list of unstable
spins is generated. The scaling relation between the time $t$ it takes
an avalanche to occur and the size $S$ of that avalanche for small
disorder $r$ can be found by noting that the characteristic duration of
an avalanche is proportional to the correlation length $\xi$ to the
power $z$\ \cite{dynamicz,MaBinney}:
\begin{equation}
t \sim \xi^z
\label{time_equ1}
\end{equation}
The exponent $z$ is known as the dynamical critical exponent. Equation\
(\ref{time_equ1}) gives the scaling for the time it takes for a spin to
``feel'' the effect of another a distance $\xi$ away. Since the
correlation length $\xi$ scales like $r^{-\nu}$ close to the critical
disorder, and the characteristic size $S$ as $r^{-1/\sigma}$, the time
$t$ then scales with large sizes as:
\begin{equation}
t \sim S^{\sigma \nu z}
\label{time_equ2}
\end{equation}

\begin{figure}
\centerline{
\psfig{figure=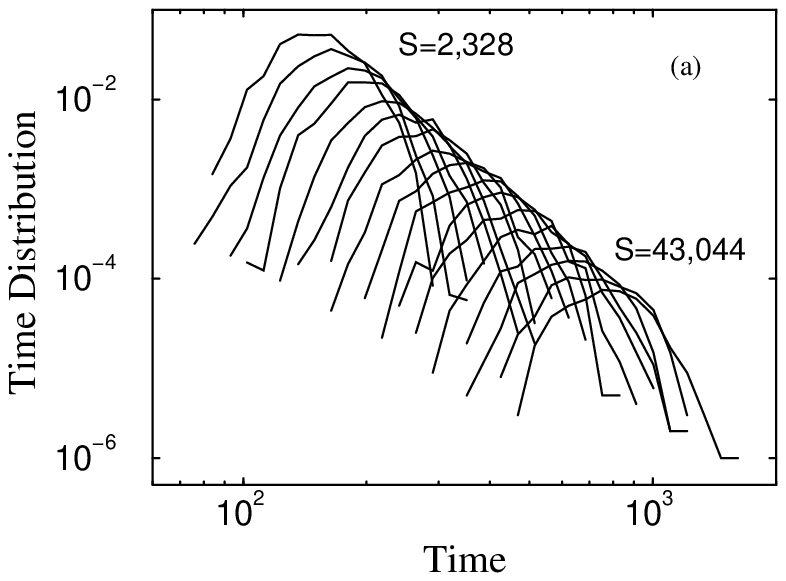,width=3truein}
}
\nobreak
\centerline{
\psfig{figure=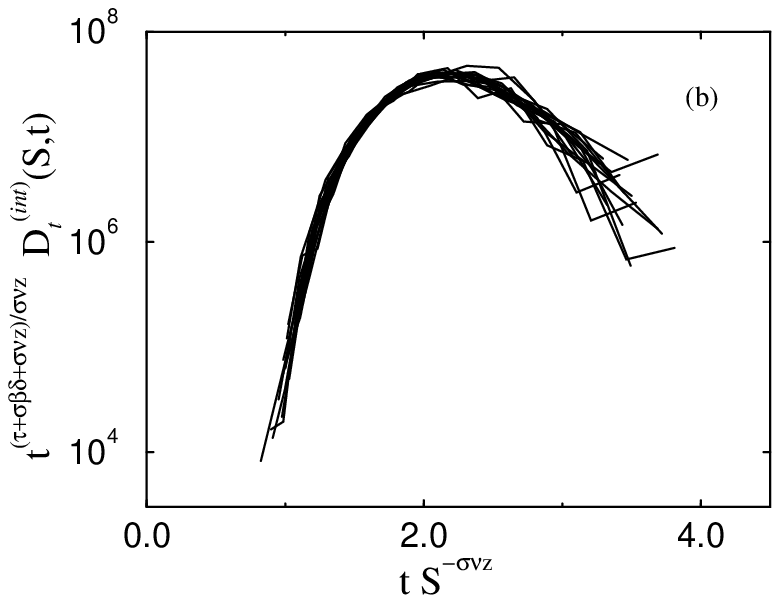,width=3truein}
}
\caption[Avalanche time distribution curves in $3$ dimensions]
        {(a) {\bf Avalanche time distribution curves in $3$ dimensions,}
        for avalanche size bins
        from about $2000$ to $40000$ spins (from upper left to lower right
        corner). The system size is $800^3$ at $R=2.26$. The curves
        are from only one random field configuration.
        (b) Scaling collapse of curves in (a). The values of the exponents
        are $\sigma\nu z = 0.57$ and $(\tau+\sigma\beta\delta+
        \sigma\nu z)/\sigma\nu z = 4.0$. \label{time_3d_fig}}
\end{figure}

In our simulation, we measure the distribution of times for each
avalanche size $S$. The distribution of times $D_t(S,R,H,t)$ for an
avalanche of size $S$ close to the critical field $H_c$ and critical
disorder $R_c$ is
\begin{equation}
D_t(S,R,H,t) \sim S^{-q}\ {\bar {\cal D}}_{\pm}^{(t)} (S^{\sigma}|r|,
h/|r|^{\beta\delta}, t/S^{\sigma\nu z})
\label{time_equ3}
\end{equation}
where $q=\tau +\sigma\nu z$, and is defined such that
\begin{eqnarray}
\int_{-\infty}^{+\infty} \!\! \int_{1}^{\infty} D_t(S,R,H,t)\ dH\ dt\ =\
\nonumber \\
S^{-(\tau + \sigma\beta\delta)}\ {\bar {\cal D}}_{\pm}^{(int)}(S^{\sigma}|r|)
\label{time_equ4}
\end{eqnarray}
where ${\bar {\cal D}}_{\pm}^{(int)}$ was defined in the integrated
avalanche size distribution section. The avalanche time distribution
integrated over the field $H$, at the critical disorder ($r=0$) is:
\begin{equation}
D_t^{(int)}(S,t)\ \sim\
t^{-{(\tau + \sigma\beta\delta + \sigma\nu z) /\sigma\nu z}}\ 
{\cal D}_t^{(int)}(t/S^{\sigma\nu z})
\label{time_equ5}
\end{equation}
which is obtained from equation (\ref{time_equ3}) in a derivation
analogous to the one for the integrated avalanche size distribution
scaling form.

Figures\ \ref{time_3d_fig}a and \ref{time_3d_fig}b show the avalanche
time distribution integrated over the field $H$ for different avalanche
sizes, and a collapse of these curves using the above scaling form, for
a $800^3$ system at $R=2.260$ (just above the range where spanning
avalanches occur). The data is saved in logarithmic size bins, each
about $1.2$ times larger than the previous one. The time is also
measured logarithmically (next bin is $1.1$ times larger than the
previous one). The extracted value for $z$ in $3$ dimensions is $1.68
\pm 0.07$. The results for other dimensions are listed in Table\
\ref{measured_exp_table}.



\narrowtext

\subsection{Simulation Results in $2$ Dimensions}

The critical transition in the shape of the hysteresis loop is observed
in the simulation, and expected from the renormalization group
\cite{Dahmen1,Dahmen2}, in $3$, $4$, and $5$ dimensions. We also found
that the upper critical dimension, at and above which the mean field
exponents become correct, is six. Furthermore, in one dimension, we
expect that in a thermodynamic system, with an unbounded distribution of
random fields, there will be no infinite avalanche for $R > 0$. That
will be so because if there is any randomness, there will be a spin in
the linear chain that will have the ``right'' value for its random field
to stop the first avalanche. For a bounded distribution of random
fields, the scaling behavior near the transition will not be universal\
\cite{Dahmen1}; instead, it will depend on the exact shape of the tails
of the distribution of random fields. Then, the question that remains
is: what happens in two dimensions?

From the simulation and a few arguments that we are about to show, we
conjecture that the two dimensional exponents will have the values:
$\tau + \sigma\beta\delta =2$, $\tau=3/2$, $1/\nu=0$, and $\sigma\nu
=1/2$. (The other exponents (except $z$) can be found from exponent
relations\ \cite{Dahmen1,Dahmen3} using these values.) The ``arguments''
are as follows.

\begin{figure}
\centerline{
\psfig{figure=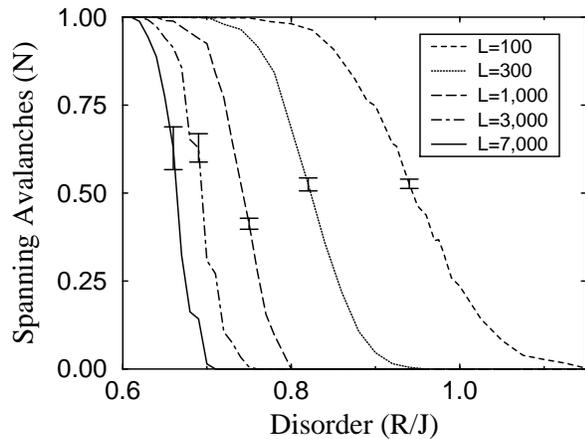,width=3truein}
}
\caption[Spanning avalanches in $2$ dimensions]
        {{\bf Number of spanning avalanches in $2$ dimensions}
        as a function of disorder
        $R$, for several system sizes. The data points are averages between
        as little as $10$ to as many as $2200$ random field configurations.
        Some typical error bars near the center of the curves are shown;
        error bars are smaller toward the ends. Note that there is no
        more than one spanning avalanche. \label{span_2d_fig}}
\end{figure}

It is quite possible that two is the {\it lower critical dimension}
(LCD) for our system. At the lower critical dimension, the critical
exponents are often ratios of small integers, and it is often possible
to derive exact solutions. Since the geometry in $2$ dimensions allows
for at most one system spanning avalanche, the ``breakdown of
hyperscaling'' exponent $\theta$ (see section IV B) must be zero, and
the hyperscaling relation\ \cite{Dahmen1,Dahmen3} is restored:
\begin{equation}
{1 \over \sigma\nu} = d - {\beta \over \nu}
\label{hyper_2d_equ1}
\end{equation}
We know that this relation is violated in $4$ and $5$ dimensions, and is
probably violated in $3$ dimensions. In dimensions above two, the
hyperscaling relation is modified by the exponent $\theta$ which gives a
measure of the number of spanning avalanches near the critical
transition, as a function of the system size. Figure\ \ref{span_2d_fig}
shows the number of spanning avalanches in $2$ dimensions for several
system sizes. Notice that, as assumed, there is not more than one
spanning avalanche in the system.

\begin{figure}
\centerline{
\psfig{figure=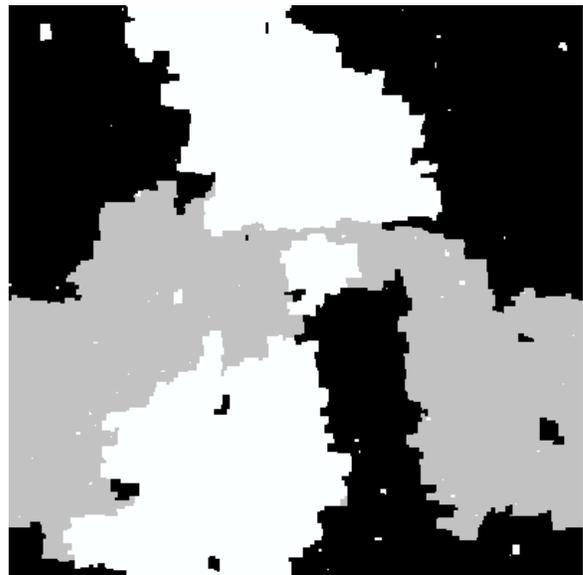,width=3truein}
}
\caption[Simulation of a $400^2$ spin system at $R=0.8$]
        {Simulation in $2$ dimensions of a $400^2$ spin system at $R=0.8$.
        The figure shows the configuration of the system after a spanning
        avalanche has just occurred (grey region). The dark area corresponds
        to spins
        that have not yet flipped, while the white area are spins that have
        flipped earlier. Notice that the spanning avalanche (grey area)
        seems compact. \label{compact_2d_fig}}
\end{figure}

\begin{figure}
\centerline{
\psfig{figure=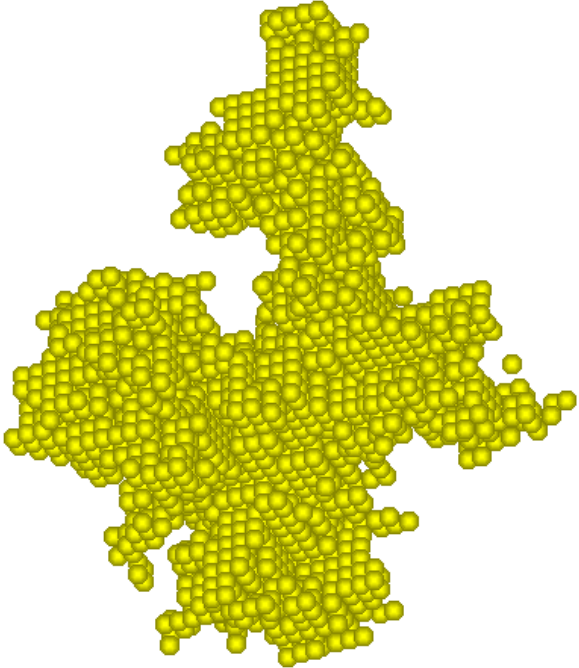,width=3truein}
}
\caption[Largest avalanche in the hysteresis loop in a $40^3$ system, near
        the critical point]
        {Largest avalanche occurring in the hysteresis loop
        in a $40^3$ spins system near the critical point.
        The avalanche is not compact. \label{notcompact_3d_fig}}
\end{figure}

We use two more arguments to derive the critical exponents. In $2$
dimensions, we find that the avalanches ``look'' compact (figure\
\ref{compact_2d_fig}). (The avalanches in $3$ dimensions are not compact
(figure\ \ref{notcompact_3d_fig}).) This implies that $1/\sigma\nu = d =
2$, which leads to $\beta/\nu=0$ from equation\ (\ref{hyper_2d_equ1}).
Furthermore, it is often the case that in the lower critical dimension,
the Harris criterion\cite{Dahmen1}
\begin{equation}
{\nu \over \beta\delta} \geq {2 \over d}
\label{Harris_eqn}
\end{equation}
becomes saturated (an equality); so in $2$ dimensions we expect
$\beta\delta/\nu=1$. From this and the previous result, the exponent
which gives the decay in space of the avalanche correlation function
\begin{equation}
\eta = 2 + {\beta \over \nu} - {\beta\delta \over \nu}
\label{eta_eqn}
\end{equation}
(see references\ \cite{Dahmen1,Dahmen3} for the derivation of all the
exponent relations) becomes equal to $\eta =1$.

Since at the LCD the correlation length typically diverges exponentially
as the critical point is approached, we expect $\nu \rightarrow
\infty$, and $\beta$ can be finite. Using the exponent relation\
\cite{Dahmen1,Dahmen3}:
\begin{equation}
\tau-2=\sigma\beta(1-\delta),
\label{hyper_2d_equ3a}
\end{equation}
we further find that $\tau=3/2$ and $\tau+\sigma\beta\delta=2$.

We must mention that our firm conjectures about the exponents in two
dimensions must be contrasted with our lack of knowledge about the
proper scaling forms. As mentioned above, at the LCD the correlation
exponent $\nu$ typically diverges, although some combinations of
critical exponents stay finite (hence $\sigma\nu = 1/2$). Those which
diverge and those which go to zero usually must be replaced by exponents
and logs, respectively. We have used three different RG-scaling {\it
ans\"atze} to model the data in two dimensions.  (1)~We used the
traditional scaling form $\xi \sim |R_c-R|^{-\nu}$, deriving $\nu =
5.3\pm 1.4$ and $R_c = 0.54\pm 0.04$. These collapses worked as well as
any, but the large value for $\nu$ (and larger value still for $1/\sigma
= 10\pm 2$) makes one suspicious. (2)~We used a scaling form suggested
by Bray and Moore\cite{BrayMoore} in the context of the equilibrium
thermal random field Ising model at the LCD, where $R_c=0$: if they
assume that $R$ is a marginal direction, then by symmetry the flows must
start with $R^3$, leading to $\xi \sim e^{(\tilde{a}/|R_c-R|^2)} \equiv
e^{(\tilde{a}/R^2)}$.  This form has the fewest free parameters, and
most of the collapses were about as good as the others (except notably
for the finite-size scaling of the moments of the avalanche size
distribution, which did not collapse well once spanning avalanches
became common). (3)~We developed another possible scaling form, based on
a finite $R_c$ and $R$ marginal, which generically has a quadratic flow
under coarse-graining: here $\xi \sim e^{(\tilde{b}/|R_c-R|)}$. We find
$R_c=0.42\pm0.04$. The rational behind these three forms is shown in
appendix A.

The results from data collapses in two dimensions were obtained from
measurements of the spanning avalanches, the second moments of the
avalanche size distribution, the integrated avalanche size
distributions, and the avalanche correlations. The magnetization curves
are also obtained from the simulation, but as in the higher dimensions,
the scaling region is small (around $H_c$ and $M_c$), and the collapses
do not define the exponents well.

Measurements that require the knowledge of the critical randomness are
the binned avalanche size distribution from which we extract the
exponents $\tau$ and $\beta\delta$, the critical magnetic field $H_c$,
and the avalanche time measurement which gives the exponent $z$. These
measurements were not obtained {\it at} the critical disorder because
$R_c$ is not well defined as was mentioned above, and because for low
disorders (less than $0.71$ for a $7000^2$ system), the system flips in
one infinite avalanche, and such measurements are therefore not
possible. We have nevertheless estimated the values of some of these
exponents and of $H_c$, from data obtained at a larger disorder (where
there is no spanning avalanche). From the avalanche size distribution
binned in $H$ at $R=0.71$ and $L=7000$, and the magnetization curves, we
find that the critical field $H_c$ is around $1.32$. A straight line fit
through the data agrees with a possible value of $\tau=3/2$ (the
conjectured value). From the time distribution of avalanche sizes for a
system of $30000^2$ spins, at $R=0.65$, we measured (from a straight
linear fit) the exponent $\sigma\nu z$ to be $0.64$. The other exponents
were obtained from scaling collapses as follows.

\begin{figure}
\centerline{
\psfig{figure=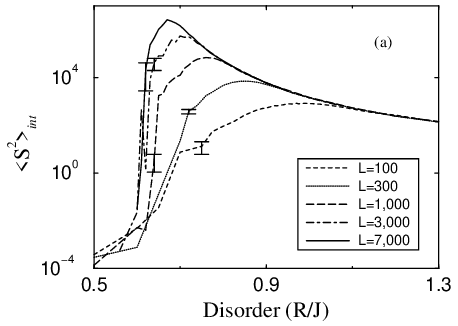,width=3truein}
}
\nobreak
\centerline{
\psfig{figure=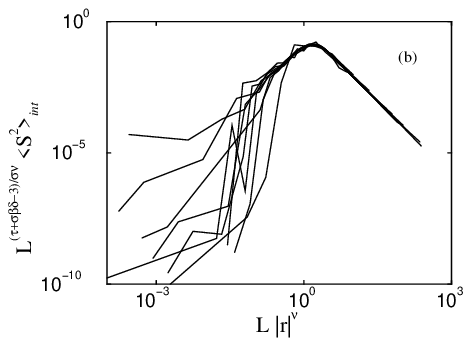,width=3truein}
}
\nobreak
\centerline{
\psfig{figure=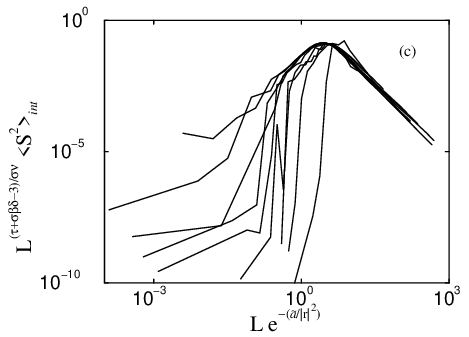,width=3truein}
}
\nobreak
\centerline{
\psfig{figure=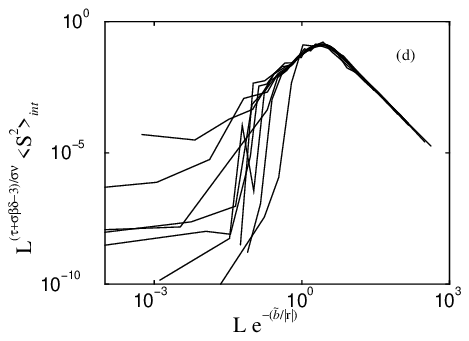,width=3truein}
}
\caption[Second moments of the avalanche size distribution in $2$ dimensions]
        {(a) {\bf Second moments of the avalanche size distribution in $2$
        dimensions,} integrated
        over the external field $H$, for several system sizes. The data
        points are averages over up to $2200$ random field
        configurations. Error bars are
        smaller than shown for larger disorders.
        (b), (c), and (d) Scaling collapses of the second moments of the
        avalanche size distribution in $2$ dimensions, integrated over the
        field $H$. The curves that are collapsed are of size:
        $50^2$, $100^2$, $300^2$, $500^2$, $1000^2$, $3000^2$, $5000^2$,
        $7000^2$, and $30000^2$.
        See text for the scaling forms, and the values of the exponents
        and parameters. \label{s2_2d_fig}}
\end{figure}

Figure\ \ref{s2_2d_fig}a shows the second moments of the avalanche size
distribution for several system sizes. The collapses using the three
different scaling forms are shown in figures\ \ref{s2_2d_fig}(b-d). The
first one (figure\ \ref{s2_2d_fig}b) is:
\begin{equation}
{\langle S^2 \rangle}_{int} \sim L^{-(\tau+\sigma\beta\delta-3)/\sigma\nu}\
{\check {\cal S}}_{int}^{(2)}(L\ |r|^{\nu})
\label{2d_equ01}
\end{equation}
which is the kind of scaling form used in $3$, $4$, and $5$ dimensions.
This form assumes $\xi \sim |r|^{-\nu}$. The exponents are $(\tau +
\sigma\beta\delta-3)/\sigma\nu = -1.9$ and $\nu=5.25$, and $r=(R_c -
R)/R$ with $R_c=0.54$. The second scaling form (figure\
\ref{s2_2d_fig}c) is:
\begin{equation}
{\langle S^2 \rangle}_{int} \sim L^{-(\tau+\sigma\beta\delta-3)/\sigma\nu}\
{\bar {\cal S}}_{int}^{(2)}(L\ e^{-\tilde{a}/|R_c-R|^2})
\label{2d_equ02}
\end{equation}
which is obtained from $\xi \sim e^{(\tilde{a}/|R_c-R|^2)}$. The values
of the exponents and parameters are: $(\tau +
\sigma\beta\delta-3)/\sigma\nu = -1.9$, $\tilde{a} = 3.4$ ($\tilde{a}$
is not universal), and $R_c \equiv 0$ (by assumption; see previous
paragraph). Notice that this collapse is not as good as the other two; a
better collapse is obtained with $R=0.15$ and $\tilde{a}=2.0$. If this
is the correct scaling form and $R_c=0$, this discrepancy can be due to
finite size effects. The third scaling form is (figure\
\ref{s2_2d_fig}d):
\begin{equation}
{\langle S^2 \rangle}_{int} \sim L^{-(\tau+\sigma\beta\delta-3)/\sigma\nu}\
{\hat {\cal S}}_{int}^{(2)}(L\ e^{-\tilde{b}/|R_c-R|})
\label{2d_equ03}
\end{equation}
which is obtained from $\xi \sim e^{(\tilde{b}/|R_c-R|)}$. The values of
the exponents and parameters are: $(\tau +
\sigma\beta\delta-3)/\sigma\nu = -1.9$, $\tilde{b} = 2.05$ ($\tilde{b}$
is also non-universal), and $R_c= 0.42$. As it is clear from the last
three figures, collapses with these different scaling forms are
comparable. Notice that the exponent $(\tau+\sigma\beta\delta
-3)/\sigma\nu$ is the same for the three collapses, but that $1/\nu$ is
zero for the last two (by assumption) while it is $0.19$ for the first
collapse. Let's now look at the collapses of the integrated avalanche
size distribution curves, which are not finite size scaling
measurements.

\begin{figure}
\centerline{
\psfig{figure=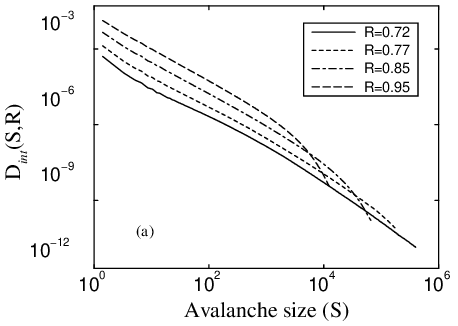,width=3truein}
}
\nobreak
\centerline{
\psfig{figure=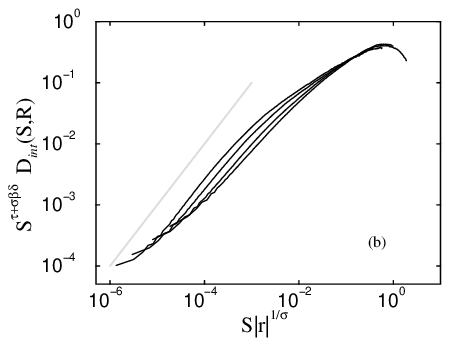,width=3truein}
}
\nobreak
\centerline{
\psfig{figure=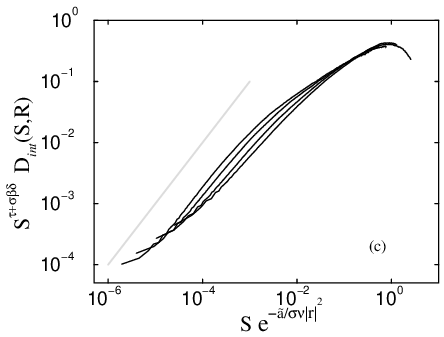,width=3truein}
}
\nobreak
\centerline{
\psfig{figure=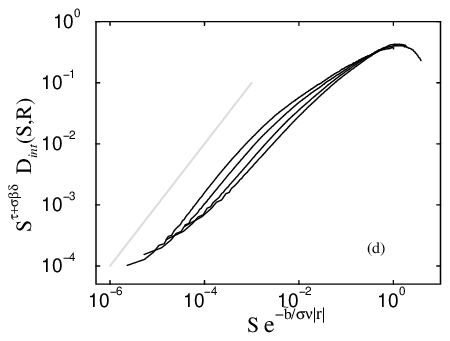,width=3truein}
}
\caption[Integrated avalanche size distribution curves in $2$ dimensions]
        {(a) {\bf Integrated avalanche size distribution} curves for several
        disorders {\bf in $2$ dimensions,}
        at the system size $L=7000$. The curves are averages over
        $10$ to $20$ random field configurations, and have been smoothed.
        (b), (c), and (d) Scaling collapses of the data from (a) using the
        three scaling forms and the exponents from the text.
        The collapsed curves have disorders:
        $0.72$, $0.74$, $0.77$, and $0.80$. The straight
        grey line in each of the plots has a slope of one. \label{aval_2d_fig}}
\end{figure}

Figure \ref{aval_2d_fig}a shows the integrated avalanche size
distribution curves for a $7000^2$ spin system, at several values of the
disorder $R$. Earlier, in figure\ \ref{bump_345fig}, we saw the fit to
the scaling collapse of such curves, done using the same scaling form as
in $3$, $4$, and $5$ dimensions:
\begin{equation}
D_{int}(S,R)\ \sim\ S^{-(\tau + \sigma\beta\delta)}\
{\bar {\cal D}}^{(int)}_{-}
(S^{\sigma} |r|)
\label{2d_equ1}
\end{equation}
(The $-$ sign indicates that the collapsed curves are for $r<0$, ie.
$R>R_c$.) However, $S^\sigma |r|$ might not be the appropriate scaling
argument in $2$ dimensions. First, from figure\ \ref{bump_345fig}, the
scaling curve in $2$ dimensions differs dramatically from the scaling
curves in higher dimensions for small arguments $X=S^\sigma |r|$. The
mean field scaling function $\bar{\cal D}_{-}^{(int)}(X)$ is a
polynomial for small $X$, and we expected (and found) a similar behavior
in $5$, $4$ and $3$ dimensions (but notice that the scaling function in
$3$ dimensions is starting to look like the curve in $2$ dimensions for
small $X$). In $2$ dimensions, if we collapse our data (figure\
\ref{aval_2d_fig}b) using the scaling form:
\begin{equation}
D_{int}(S,R)\ \sim\ S^{-(\tau + \sigma\beta\delta)}\
{\cal D}^{(int)}_{-}
(S|r|^{1/\sigma})
\label{2d_equ2}
\end{equation}
with $\tau + \sigma\beta\delta=2.04$, $1/\sigma=10$, and $r=(R_c -
R)/R$, we find that the scaling function for small $\tilde
X=S|r|^{1/\sigma}$ looks linear with power one! This might imply that
the scaling function ${\cal D}^{(int)}_{-}(S|r|^{1/\sigma})$ (eqn.\
(\ref{2d_equ2})) is the one that is analytic for small arguments in $2$
dimensions.

Second, we conjectured above that the values for $\sigma$ and $1/\nu$
are probably zero in $2$ dimensions, and that only the combination
$\sigma\nu$ is finite ($\sigma\nu$ $=$ $1/2$). It follows that, for the other
two scaling forms we use, the arguments of the scaling function should
be $S e^{-\tilde{a}/\sigma\nu|R-R_c|^2}$ and $S
e^{-\tilde{b}/\sigma\nu|R-R_c|}$, and {\it not} $S^\sigma
e^{-\tilde{a}/\nu|R-R_c|^2}$ and $S^\sigma e^{-\tilde{b}/\nu|R-R_c|}$
respectively. This is analogous to using $S|r|^{1/\sigma}$ in the
scaling form\ (\ref{2d_equ2}). We should mention here that both equation
(\ref{2d_equ1}) and equation (\ref{2d_equ2}) give the same scaling
exponents $\tau+\sigma\beta\delta$ and $\sigma$, and that in all our
scaling collapses we have assumed that the same scaling argument is
valid for small and large $\tilde X$ (and in between). This in general,
does not have to be true.

\begin{figure}
\centerline{
\psfig{figure=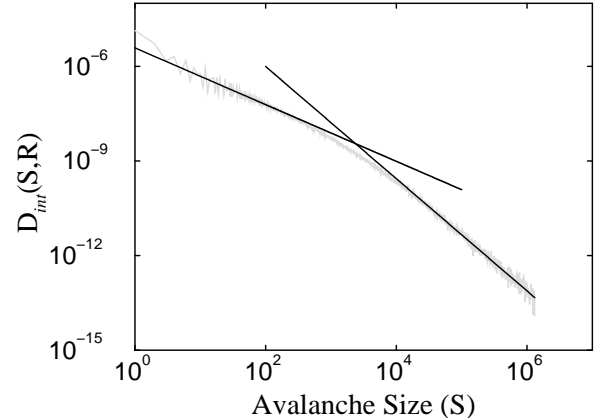,width=3truein}
}
\caption[Linear fits to an integrated avalanche size distribution curve in
        $2$ dimensions]
        { {\bf Integrated avalanche size distribution curve in $2$ dimensions},
        for a system of $30000^2$ spins, at $R=0.650$. Shown are two linear
        fits to the data: one for small sizes and the other for large sizes.
        The slope for the fit at small $S$ is $0.90$. The fit was done
        for sizes in the range $[10,250]$. The slope differs by less than
        $5\%$ when the range is changed ($S$ is never larger than $400$ though)
.
        The slope for the fit at large $S$ is $1.78$. The slope differs by
        less than $2\%$ when the range is changed ($S$ is never smaller
        than $10000$). The conjectured value for $\tau+\sigma\beta\delta$ is
        $2$ which is different from $1.78$. This is similar to
        the behavior we saw in $3$, $4$, and $5$ dimensions.
        On the other hand, for small sizes we expect the exponent
        $\tau+\sigma\beta\delta -1=1$ (see text). Again, the two measurements
        don't completely agree,
        but the slope from our data does seem to indicate
        such a behavior. \label{raw_aval_2d_fig}}
\end{figure}

Equation\ (\ref{2d_equ2}) is therefore one of the three scaling forms we
use. The second scaling form is:
\begin{equation}
D_{int}(S,R)\ \sim\ S^{-(\tau + \sigma\beta\delta)}\ {\cal D}^{(int)(2)}_{-}
\Bigl(S e^{-\tilde{a}/\sigma\nu|R_c-R|^2}\Bigr)
\label{2d_equ3}
\end{equation}
shown in figure\ \ref{aval_2d_fig}c, with $\tau +
\sigma\beta\delta=2.04$, $\tilde{a}/\sigma\nu=7.0$ (this implies that
$\sigma\nu=0.49$), and $r=R_c-R$ with $R_c \equiv 0$ by assumption. And
finally, the third scaling form we use is:
\begin{equation}
D_{int}(S,R)\ \sim\ S^{-(\tau + \sigma\beta\delta)}\ {\cal D}^{(int)(1)}_{-}
\Bigl(S e^{-\tilde{b}/\sigma\nu|R_c-R|}\Bigr)
\label{2d_equ6}
\end{equation}
shown in figure\ \ref{aval_2d_fig}d, with $\tau +
\sigma\beta\delta=2.04$, $\tilde{b}/\sigma\nu = 4.0$ (which makes
$\sigma\nu=0.51$), and $r=R_c-R$ with $R_c = 0.42$. Again, not only are
all three collapses comparable, but the exponents extracted from them
are as well. The exponent for the slope of the distribution is
$\tau+\sigma\beta\delta=2.04$ for the three collapses, and the exponent
combination $\sigma\nu$ is around $0.51$ (for the first collapse
$\sigma=0.10$, while $\nu=5.25$ from the equivalent second moment
collapse).

Figures\ \ref{aval_2d_fig}(b-d) show that the scaling function ${\cal
D}^{(int)}_{-}$ seems to be linear with slope one for small arguments
(the grey lines have slope one) and that the constant term in the
polynomial expansion is zero (or close to zero). This leads to a
singular scaling function correction to the avalanche size distribution
exponent $\tau+\sigma\beta\delta$ for small non--zero $\tilde X$:
\begin{equation}
D_{int}(S,R)\ \sim\ S^{-(\tau+\sigma\beta\delta)}\
{\cal D}_{-}^{(int)}(\tilde X)\ \sim\ S^{-(\tau+\sigma\beta\delta)+1}
\label{2d_equ7}
\end{equation}
(Note that we could have used ${\cal D}^{(int)(1)}_{-}$ or ${\cal
D}^{(int)(2)}_{-}$ as well.)

Recall that because of the ``bump'' in the avalanche size distribution
scaling function in $3$, $4$, and $5$ dimensions, and in mean field, the
slope of the raw data curves did not agree with the value of the
exponent $\tau+\sigma\beta\delta$. In $2$ dimensions, this is still
true, but we also find a singular behavior for ${\cal
D}_{-}^{(int)}(\tilde X)$, which changes the slope of the data curve for
small $\tilde X$. In figure\ \ref{raw_aval_2d_fig}, an integrated
avalanche size distribution curve for a system of $30000^2$ spins, at
$R=0.65$, is plotted along with the linear fits to the data for small
and large size $S$. For large $S$, the slope is close to but not equal
to $2$, while for small $S$, the slope is close to one!

\begin{figure}
\centerline{
\psfig{figure=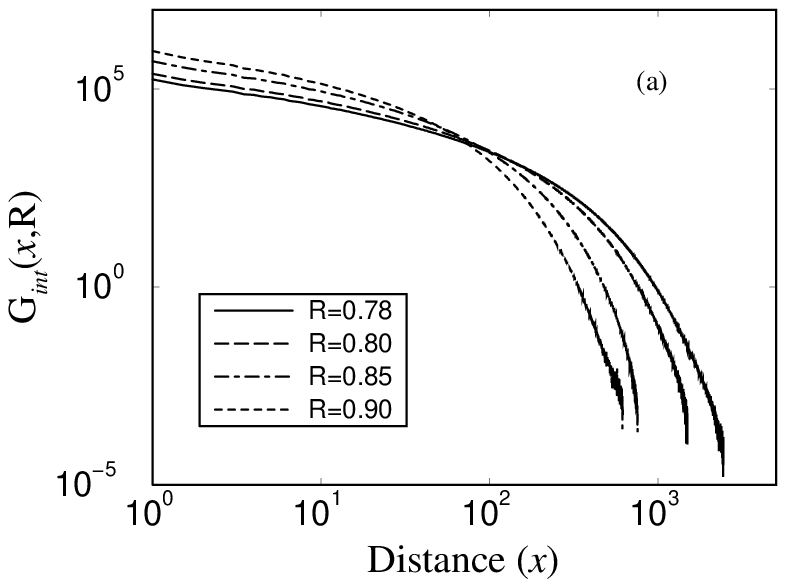,width=3truein}
}
\nobreak
\centerline{
\psfig{figure=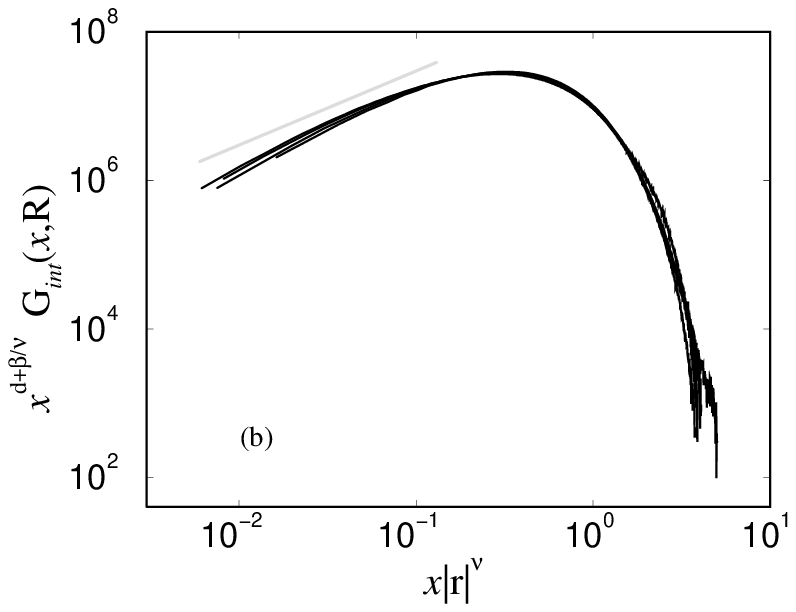,width=3truein}
}
\nobreak
\caption[Avalanche correlation curves in $2$ dimensions]
        {(a) {\bf Avalanche correlation function in $2$ dimensions,}
        integrated over the external field $H$, for several disorders $R$
        and the system size $L=30000$.
        Only the curve with the smallest disorder is an average over several
        random field configuration.
        (b) Scaling collapse of the avalanche correlation curves in $2$
        dimensions, for a system of $30000^2$ spins. The exponent values
        are: $\nu=5.25$ and $\beta/\nu=0$. The critical disorder is $R_c=0.54$,
        and $r=(R-R_c)/R$. Notice that for small $x|r|^\nu$, the scaling
        function looks singular with a power close to one (the straight line
        has a slope of one). \label{correl_2d_fig}}
\end{figure}

The avalanche correlation data (see figure\ \ref{correl_2d_fig}a) is
collapsed with three different scaling forms as well. These forms are
analogous to the ones used for the second moments collapses, but with
the distance $x$ taking the place of the system size $L$. The collapses
and the extracted exponents from these three forms are again very
similar, and only one of the collapses is shown in figure\
\ref{correl_2d_fig}b. The value of $\beta/\nu$ from these collapses is
$0.03 \pm 0.06$. If we compare figure\ \ref{correl_2d_fig}b with the
collapse of the avalanche correlation in $3$ dimensions (fig.\
\ref{correl_collapse_3d_fig}a), we find that the scaling function in $2$
dimensions seems to be singular with slope one for small distances, as
is the integrated avalanche size distribution for small sizes.

The spanning avalanches data are also analyzed using three scaling forms
similar to those used for the second moments of the avalanche size
distribution collapses. The exponent $\theta$ is poorly defined from
these collapses (and is therefore not listed in Table\
\ref{conj_meas_2d_table}), although the data does collapse for the
exact value: $\theta=0$.

The three collapses for all the measurements we have done are very
similar. This is not a surprise: it is always hard to distinguish large
power laws ($\nu$ and $1/\sigma$ are large in the ``linear argument''
scaling form (eqns.\ (\ref{2d_equ01}) and (\ref{2d_equ2}))) from
exponentials. Although some of the exponents have very different values
in the three collapses, the average of the exponents from the three
methods agree within the error bars with each method (see figure\
\ref{exp_compare}) and our conjectures. In conclusion, although we do
not know the correct scaling form for the data in $2$ dimensions, the
possible three scaling forms we mention give exponent values that are
compatible with each other and with our conjectures (see Table\
\ref{conj_meas_2d_table}). (Table\ \ref{conj_2d_table} gives the
conjectured values for the exponents that have not been measured in the
collapses.) Much larger system sizes might be necessary to obtain more
conclusive results.

\narrowtext

\section{Comparison with the analytical results}

We have compared the simulation results with the renormalization group
analysis of the same system\ \cite{Dahmen1,Dahmen2}. According to the
renormalization group the upper critical dimension (UCD), at and above
which the critical exponents are equal to the mean field values, is six.
Close to the UCD, it is possible to do a $6-\epsilon$ expansion
($\epsilon$ is small and greater than $0$), and obtain estimates for the
critical exponents and the magnetization scaling function, which can
then be compared with our numerical results. Furthermore, at dimension
eight there is a prediction for another transition. Below eight
dimensions, there is a discontinuity in the slope of the magnetization
curve as it approaches the ``jump'' in the magnetization ($R < R_c$),
while above eight dimensions the approach is smooth.

\begin{figure}
\centerline{
\psfig{figure=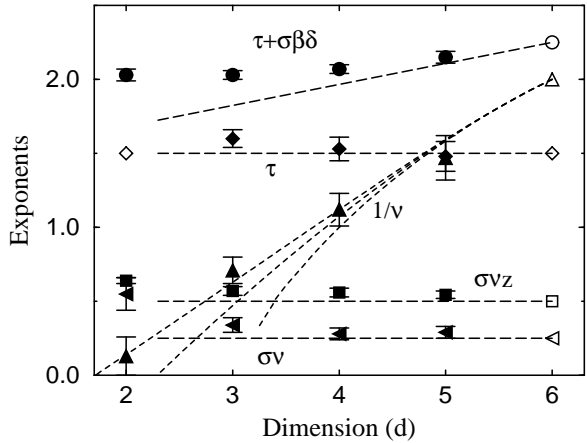,width=3truein}
}
\caption[Comparison between the critical exponents from the simulation
        and the $\epsilon$ expansion]
        {Numerical values (filled symbols) of the exponents $\tau +
\sigma\beta\delta$, $\tau$, $1/\nu$, $\sigma\nu z$, and $\sigma\nu$
(circles, diamond, triangles up, squares, and triangle left) in $2$, $3$, $4$,
and $5$ dimensions. The empty symbols are values for these exponents in
mean field (dimension 6). Note that the value of $\tau$ in $2$d is the
conjectured value: we have not extracted $\tau$ from scaling collapses
(see text).
We have simulated sizes up to $30000^2$, $1000^3$, $80^4$, and $50^5$,
where for $320^3$ for example, more than $700$ different random field
configurations were measured.
The long-dashed lines are the $\epsilon$
expansions to first order for the exponents $\tau + \sigma\beta\delta$,
$\tau$, $\sigma\nu z$, and $\sigma\nu$. The short-dashed lines are Borel
sums\protect\cite{LeGuillou-Kleinert} for $1/\nu$. The lowest is the
variable-pole Borel sum from LeGuillou {\it et
al.}\protect\cite{LeGuillou-Kleinert}, the middle uses the method of
Vladimirov {\it et al.} to fifth order, and the upper uses the method of
LeGuillou {\it et al.}, but without the pole and with the correct fifth
order term.
The error bars denote systematic errors in finding the
exponents from extrapolation of the values obtained from collapses of curves
at different disorders $R$.
Statistical errors are smaller. \label{exp_compare}}
\end{figure}

Figure\ \ref{exp_compare} shows the numerical and analytical results for
five of the critical exponents obtained in dimensions two to six (in six
dimensions, the values are the mean field ones). The other exponents can
be obtained from scaling relations\cite{Dahmen1,Dahmen3}. The exponent
values in figure\ \ref{exp_compare} are obtained by extrapolating the
results of scaling collapses to either $R \rightarrow R_c$ or $1/L
\rightarrow 0$ (see section on simulation results). In two dimensions,
which is possibly the lower critical dimension, the plotted values are
averages from the three different scaling forms used to collapse the
data and extract the exponents. The error bars shown span all three {\it
ans\"atze}, and are compatible with our conjectures from the previous
section. The long-dashed lines are the $\epsilon$ expansions to first
order for $\tau + \sigma\beta\delta$, $\tau$, $\sigma\nu z$, and
$\sigma\nu$. The three short-dashed lines\cite{Dahmen1} are Borel
sums\protect\cite{LeGuillou-Kleinert} for $1/\nu$. The lowest is the
variable-pole Borel sum from LeGuillou {\it et
al.}\protect\cite{LeGuillou-Kleinert}, the middle uses the method of
Vladimirov {\it et al.} to fifth order, and the upper uses the method of
LeGuillou {\it et al.}, but without the pole and with the correct fifth
order term\ \cite{Dahmen1}. Notice that the numerical values converge
nicely to the mean field predictions, as the dimension approaches six,
and that the agreement between the numerical values and the $\epsilon$
expansion is quite impressive.

\begin{figure}
\centerline{
\psfig{figure=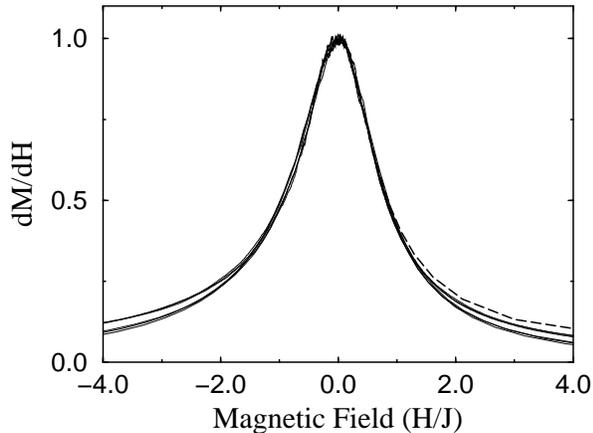,width=3truein}
}
\caption[Comparison between simulated $dM/dH$ curves in $5$ dimensions, and
        the $dM/dH$ curve obtained from the $\epsilon$ expansion]
        {Comparison between six simulation curves (thin lines)
        and the $dM/dH$ curve (thick dashed line)
        obtained from a parametric form\ \protect\cite{Zinn-Justin}
        to third order
        in $\epsilon$. The six curves are for a system of $30^5$ spins
        at disorders: $7.0, 7.3,$ and $7.5$ ($R_c=5.96$ in $5$ dimensions),
        and for a system of $50^5$ spins at disorders: $6.3, 6.4$, and $6.5$
        (for larger fields, these are closer to the dashed line in the
        figure). All the curves have been
        stretched/shrunk in the horizontal and vertical direction to lie
        on each other, and shifted horizontally. \label{5d_dmdh_fig}}
\end{figure}

The $\epsilon$ expansion can be an even more powerful tool if it can
predict the scaling functions. This has been done for the magnetization
scaling function of the pure Ising model in $4-\epsilon$ dimensions\
\cite{DombWallace,Zinn-Justin}. Since the $\epsilon$ expansion for our
model is the same as the one for the {\it equilibrium} RFIM\
\cite{Dahmen1}, and the latter has been mapped to {\it all} orders in
$\epsilon$ to the corresponding expansion of the regular Ising model in
two lower dimensions\ \cite{Dahmen1,Aharony,Parisi}, we can use the
results obtained in\ \cite{DombWallace,Zinn-Justin}. This is done in
figure\ \ref{5d_dmdh_fig}, which shows the comparison between the
$dM/dH$ curves obtained in $5$ dimensions at $R=6.3, 6.4, 6.5, 7.0, 7.3,
7.5$ ($R_c=5.96$) (the curves have been stretched/shrunk to lie on top
of each other, and shifted horizontally so that the peaks align), and
the parametric form (thick dashed line) for the scaling function of
$dM/dH$, to third order in $\epsilon$, where $\epsilon =1$ in $5$
dimensions (see\ \cite{Zinn-Justin}). As we see, the agreement is very
good in the scaling region (close to the peaks).This brings up the
possibility of using the $\epsilon$ expansion for the scaling function
to extract the critical exponents from simulation or experimental data.
So far though, only the scaling function for the magnetization has been
obtained.

\begin{figure}
\centerline{
\psfig{figure=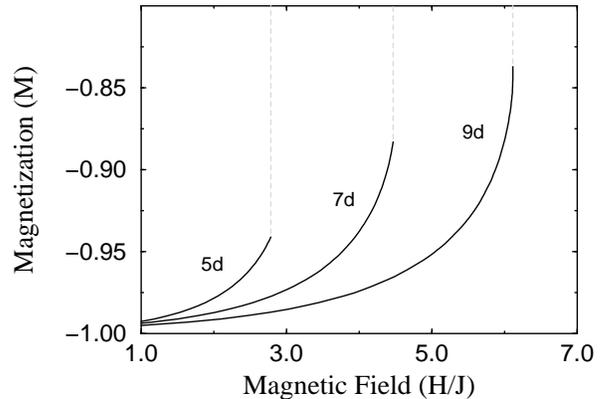,width=3truein}
}
\caption[Magnetization curves showing the approach to the ``infinite jump''
        in $5$, $7$, and $9$ dimensions]
        {{\bf Magnetization curves in $5$, $7$, and $9$ dimensions.}
        The disorders for these curves are $R=3.3$, $4.7$, and $6.0$ for
        $30^5$, $10^7$, and $5^9$ size systems respectively. The dashed
        lines represent the ``jump'' in the magnetization. Notice that in
        $9$ dimensions the approach to the ``jump'' seems to be continuous.
        \label{transition_8d}}
\end{figure}

\begin{figure}
\centerline{
\psfig{figure=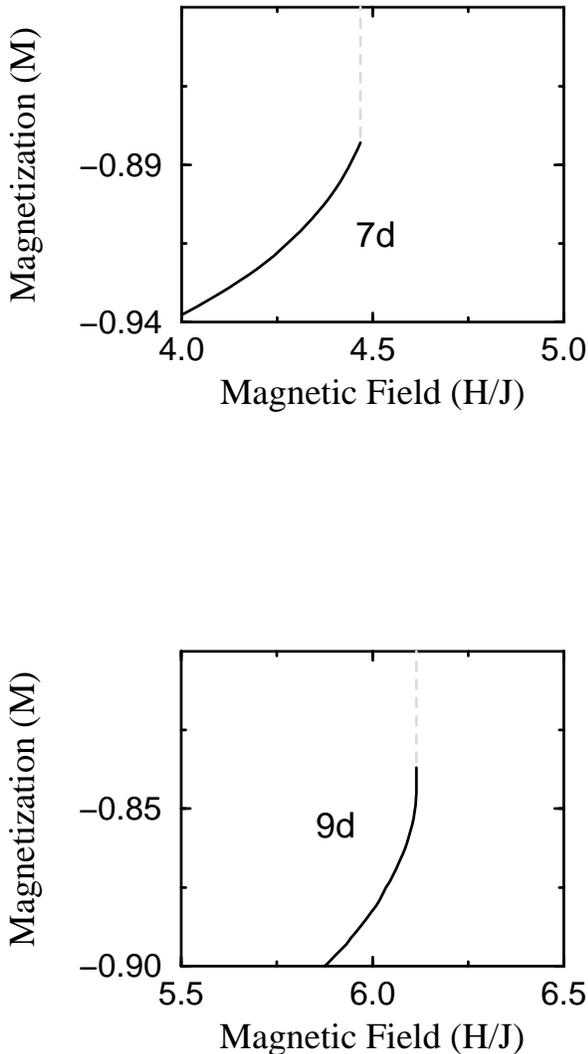,width=3truein}
}
\caption[Closeup of magnetization curves with the ``infinite jump''
        in $7$ and $9$ dimensions]
        {(a) and (b) Closeup of the {\bf magnetization curves in $7$ and $9$
        dimensions} respectively from figure\ \protect\ref{transition_8d}.
        In $8$ dimensions, there is a prediction from the renormalization
        group\ \protect\cite{Dahmen1} that there is a transition
        in the way the jump is approached (see text). \label{closeup_8d}}
\end{figure}

\begin{figure}
\centerline{
\psfig{figure=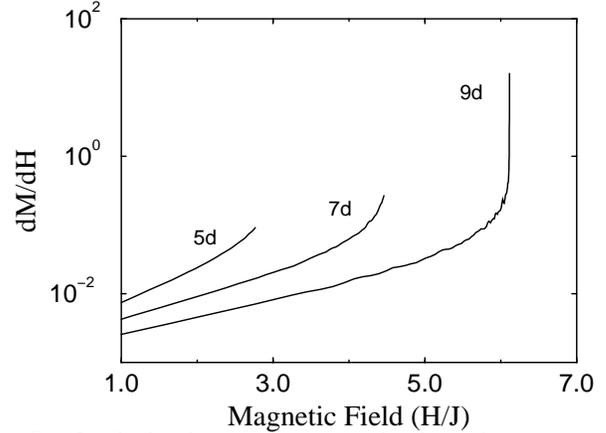,width=3truein}
}
\caption[$dM/dH$ for magnetization curves with the ``infinite jump'',
        in $7$ and $9$ dimensions]
        {Derivative of the magnetization with respect to the field $H$, for
        the curves in figure\ \protect\ref{transition_8d}.
        The approach to the ``infinite jump'' seems to be continuous
        in $9$ dimensions. Note that the
        vertical axis is logarithmic. \label{slope_8d}}
\end{figure}

As another check between the simulation and the renormalization group,
we have looked for the predicted transition in eight dimensions. Figure\
\ref{transition_8d} shows the magnetization curves in $5$, $7$, and $9$
dimensions (system sizes: $30^5$, $10^7$, and $5^9$) for values of the
disorder equal to ${2 \over 3}d$, where $d$ is the dimension. These
values of disorder are below the critical disorder in dimensions below
six, and are expected to be below for dimensions $7$ and $9$ as well.
For $5$d and $7$d, the approach to the ``jump'' in the magnetization is
discontinuous. Above the eight dimension, the approach is continuous
(see close ups in figure\ \ref{closeup_8d}). This is as expected from
the renormalization group\ \cite{Dahmen1}. We have also looked at
$dM/dH$, which appears clearly to diverge in $d=9$ and not in $d=7$
(figure\ \ref{slope_8d}).



\narrowtext

\section{Conclusion}

We have used the zero temperature random field Ising model, with a
Gaussian distribution of random fields, to model a random system that
exhibits hysteresis. We found that the model has a transition in the
shape of the hysteresis loop, and that the transition is critical. The
tunable parameters are the amount of disorder $R$ and the external
magnetic field $H$. The transition is marked by the appearance of an
infinite avalanche in the thermodynamic system. Near the critical point,
($R_C$, $H_C$), the scaling region is quite large: the system can
exhibit power law behavior for several decades, and still not be near
the critical transition. This is important to keep in mind whenever
experimental data are analyzed. If a tunable parameter can be found, a
system that appears to be SOC, might in reality have a disorder induced
critical point.

We have extacted critical exponents for the magnetization, the
avalanche size distribution (integrated over the field and binned in the
field), the moments of the avalanche size distribution, the avalanche
correlation, the number of spanning avalanches, and the distribution of
times for different avalanche sizes. These values are listed in Table\
\ref{measured_exp_table} and Table\ \ref{conj_meas_2d_table}, and were
obtained as an average of the extrapolation results (to $R \rightarrow
R_c$ or $L \rightarrow \infty$) from several measurements. For example,
the correlation length exponent $\nu$ is the average value from three
different collapses: the correlation function, the spanning avalanches,
and the second moments of the avalanche size distribution, while the
critical disorder $R_c$ is estimated from both the spanning avalanches
collapses and the collapses of the moments of the avalanche size
distribution. As shown earlier, the numerical results compare well with
the $\epsilon$ expansion\ \cite{Dahmen1,Dahmen2}. Furthermore, the
renormalization group work predicts another transition in eight
dimensions, which we find in the simulation as well. Comparisons to
experimental Barkhausen noise measurements\ \cite{Perkovic} are very
encouraging, and a more comprehensive review of possible experiments
that exhibit disorder--driven critical phenomena similar to our model
is under way\ \cite{Dahmen3}.

Finally, we should mention that there are other models for avalanches in
disordered magnets. There is a large body of work on depinning
transitions and the motion of the single
interface\cite{depinning,SameDepinning}. In these models, avalanches
occur only at the growing interface. Our model though, deals with many
interacting interfaces: avalanches can grow anywhere in the system.
Similar models exist with random bonds\cite{RandomBonds,Vives} and
random anisotropies. In the random bonds model, the interaction
$J_{ij}$ between neighboring spins $i$ and $j$ is random.

The zero temperature random bond Ising model\ \cite{RandomBonds,Vives}
also exhibits a critical transition in the shape of the hysteresis loop,
where the mean bond strength is analogous to our disorder $R$. It has
been argued numerically\ \cite{Vives} and analytically\ \cite{Dahmen1},
that as long as there are no long-range forces\cite{Urbach} and
correlated disorder, the random bond and the random field Ising model
are in the same universality class. A comparison between our simulation
and the results in reference\ \cite{Vives} show that the $3$ dimensional
results agree quite nicely. However, in $2$ dimensions, there are large
differences, which we believe occur because of the small system sizes
used by the authors for their simulation (only up to $L=100$). We have
seen that our results (see section on the $2$ dimensional simulation)
are very size dependent. Looking back for example at figure\
\ref{span_2d_fig}, we find that for a system of $L=100$ spins, a
``good'' estimate for the critical disorder $R_c$ would indeed be $0.75$
as was found in\ \cite{Vives}. However, we find after increasing the
system size that the critical disorder $R_c$ is $0.54$ or lower.



We acknowledge the support of DOE Grant \#DE-FG02-88-ER45364 and NSF
Grant \#DMR-9419506. We would like to thank Sivan Kartha and Bruce W.
Roberts for their initial ideas on the "probabilities'' algorithm.
Furthermore, we would like to thank M. E. J. Newman, J. A. Krumhansl, J.
Souletie, and M. O. Robbins for helpful conversations. This work was
conducted on the SP1 and SP2 at the Cornell National Supercomputing
Facility (CNSF), funded in part by the National Science Foundation, by
New York State, and by IBM, and on IBM 560 workstations and the IBM J30
SMP system (both donated by IBM). We would like to thank CNSF and IBM
for their support. Further pedagogical information using Mosaic is
available at http://www.lassp.cornell.edu/sethna/hysteresis.



\narrowtext

\appendix
\section{Derivation of the various scaling forms and corrections}

In this paper we make extensive use of scaling collapses.  Many
variations are important to us: Widom scaling, finite-size scaling,
singular corrections to scaling, analytic corrections to scaling,
rotating axes, and exponentially diverging correlation length scaling.
The underlying theoretical framework for scaling is given by the
renormalization group, developed by Wilson and Fisher\cite{WilsonFisher}
in the context of equilibrium critical phenomena and by now well
explicated in a variety of texts\cite{Goldenfeld,MaBinney,texts}.

We have discovered that we can derive all the scaling forms and
corrections that have been important to us from two simple hypotheses
(found in critical regions): universality and invariance under
reparameterizations. {\sl Universality} is the statement that two
completely different systems will behave the same near their critical
point\ \cite{note5} (for example, they can have exactly the same kinds
of correlations). {\sl Reparameterization invariance} is the statement
that smooth changes in the units or methods of measurement should not
affect the critical properties. We use these properties to develop the
scaling forms and corrections we use in this paper. Each example we
cover will build on the previous ones while developing a new idea.

For our first example, consider some property $F$ of a system at its
critical point, as a function of a scale $x$.  $F$ might be the
spin-spin correlation function as a function of distance $x$ (or it
might be the avalanche probability distribution as a function of size
$x$, etc.)  If two different experimental systems are at the same
critical point, their $F$'s must agree.  It would seem clear that they
cannot be expected to be equal to one another: the overall scale of $F$
and the scale of $x$ will depend on the microscopic structure of the
materials.  The best one could imagine would be that
\begin{equation}
F_1(x_1) = A F_2( B x_2)
\label{single}
\end{equation}
where $A$ would give the ratio of, say, the squared magnetic moment per
domain of the two materials, and $B$ gives the ratio of the domain sizes.

Now, consider comparing a system with itself, but with a different
measuring apparatus.  Universality in this self-referential sense would
imply $F(x) = A F(B x)$, for suitable $A$ and $B$.  If instead of using
finite constant $A$ and $B$, we arrange for an infinitesimal change in
the measurement of length scales, we find:
\begin{equation}
F(x) = (1-\alpha \epsilon)\ F\Bigl((1-\epsilon) x\Bigr)
\label{small_single}
\end{equation}
where $\epsilon$ is small and $\alpha$ is some constant. Taking the
derivative of both sides with respect to $\epsilon$ and evaluating it at
$\epsilon = 0$, we find $-\alpha F = x F'$, so
\begin{equation}
F(x) \sim x^{-\alpha}.
\label{power}
\end{equation}
The function $F$ is a power--law! The underlying reason why power--laws
are seen at critical points is that power laws look the same at
different scales.

Now consider a new measurement with a distorted measuring apparatus. Now
$F(x) \sim {\cal A}\Bigl[F\Bigl({\cal B}(x)\Bigr)\Bigr]$ where ${\cal
A}$ and ${\cal B}$ are some nonlinear functions.  For example, one might
measure the number of microscopic domains $x$ flipped in an avalanche,
or one might measure the total acoustic power ${\cal B}(x)$ emitted
during the avalanche; these two ``sizes'' should roughly scale with one
another, but nonlinear amplifications will occur while the spatial
extent of the avalanche is small compared to the wavelength of sound
emitted: we expand ${\cal B}(x) = B x + b_0 + b_1/x + \ldots$ Similarly,
our microphone may be nonlinear at large sound amplitudes, or the
absorption of sound in the medium may be nonlinear: ${\cal A}(F) = A F +
a_2 F^2 + \ldots$ So,
\begin{eqnarray}
{\cal A}\Bigl[F\Bigl({\cal B}(x)\Bigr)\Bigr]\ \approx\ \nonumber \\
	  A \Bigl(F(B x) + F'(B x)(b_0 + b_1/x + \ldots)\ +\ \nonumber \\
		F''(B x)(\ldots) + \ldots \Bigr)
		+\ a_2 F^2(B x)\ + \ldots 
\label{nonlinear_single}
\end{eqnarray}
We can certainly see that our assumption of universality cannot hold
everywhere: for large $F$ or small $x$ the assumption of
reparameterization invariance (\ref{nonlinear_single}) prevents any
simple universal form.  Where is universality possible?  We can take the
power-law form $F(x) \sim x^{-\alpha} = x^{\log A/\log B}$ which is the
only form allowed by linear reparameterizations and plug it into
(\ref{nonlinear_single}), and we see that all these nonlinear
corrections are subdominant ({\it i.e.}, small) for large $x$ and small
$F$ (presuming $\alpha>0$).  If $\alpha>1$, the leading correction is
due to $b_0$ and we expect $x^{-\alpha-1}$ corrections to the universal
power law at small distances; if $0<\alpha<1$ the dominant correction
is due to $a_2$, and we expect corrections of order $x^{-2\alpha}$.
Thus our assumptions of universality and reparameterization invariance
both lead us to the power-law scaling forms and inform us as to some
expected deviations from these forms.  Notice that the simple rescaling
led to the universal power-law predictions, and that the more complicated
nonlinear rescalings taught us about the dominant corrections: this will
keep happening with our other examples.

For our second example, let us consider a property $K$ of a system, as a
function of some external parameter $R$, as we vary $R$ through the
critical point $R_c$ for the material (so $r=R-R_c$ is small).  $K$
might represent the second moment of the avalanche size distribution,
where $R$ would represent the value of the randomness; alternatively $K$
might represent the fractional change in magnetization $\Delta M$ at the
infinite avalanche $\ldots$ If two different experimental systems are
both near their critical points ($r_1$ and $r_2$ both small), then
universality demands that the dependence of $K_1$ and $K_2$ on
``temperature'' $R$ must agree, up to overall changes in scale. Thus,
using a simple linear rescaling $K(r) = (1-\mu \epsilon)
K\Bigl((1-\epsilon) r\Bigr)$ leads as above to the prediction
\begin{equation}
K(r) = r^{-\mu}.
\label{linear_single}
\end{equation}

Now let us consider nonlinear rescalings, somewhat different than the
one discussed above. In particular, the nonlinearity of our measurement
of $K$ can be dependent on $r$. So,
\begin{equation}
{\cal A}_r\Bigl(K(r)\Bigr) = a_0 + a_1 r + a_2 r^2 + \ldots +
			 a_{01} K(r) + \ldots
\label{nonlinear_k}
\end{equation}
If $\mu>0$, these analytic corrections don't change the dominant power
law near $r=0$.  However, if $\mu<0$, all the terms $a_n$ for $n<-\mu$
will be more important than the singular term!  Only after fitting them
to the data and subtracting them will the residual singularity be
measurable. For the fractional change in magnetization: $\Delta M \sim
r^\beta$ has $0< \beta < 1$ (at least above three dimensions), so we
might think we need to subtract off a constant term $a_0$, but $\Delta M
= 0$ for $R \ge R_c$, so $a_0$ is zero. On the other hand, in a previous
paper\cite{Dahmen1}, we discussed the singularity in the area of the
hysteresis loop: $Area \sim r^{2-\alpha}$, where $2-\alpha = \beta +
\beta \delta$ is an analogue to the specific heat in thermal systems.
Since $\alpha$ is near zero (slightly positive from our estimates of
$\beta$ and $\delta$ in 3, 4, and 5 dimensions), measuring it would
necessitate our fitting and subtracting three terms (constant, linear,
and quadratic in $r$): we did not measure the area for that reason.

For our third example, let's consider a function $F(x,r)$, depending on
both a scale $x$ and an external parameter $r$.  For example, $F$ might
be the probability $D_{int}$ that an avalanche of size $x$ will occur
during a hysteresis loop at disorder $r=R-R_c$.  Universality implies
that two different systems must have the same $F$ up to changes in
scale, and therefore that $F$ measured at one $r$ must have the same
form as if measured at a different $r$. To start with, we consider a
simple linear rescaling:
\begin{equation}
F(x,r) = (1 - \alpha \epsilon)\
	F\Bigl( (1-\epsilon) x, (1+\zeta \epsilon) r \Bigr).
\label{linear_double}
\end{equation}
Taking the derivative of both sides with respect to $\epsilon$ gives a
partial differential equation that can be manipulated to show $F$ has a
scaling form.  Instead, we change variables to a new variable $y =
x^\zeta r$ (which satisfies $y'=y$ to order $\epsilon$).  If $\tilde
F(x,y) \equiv F(x,r)$ is our function measured in the new variables,
then
\begin{equation}
F(x,r) = \tilde F(x,y) = (1-\alpha \epsilon)\
		 \tilde F\Bigl((1-\epsilon)x,y\Bigr)
\label{linear_double_tilde}
\end{equation}
and $-\alpha \tilde F = x \, \partial \tilde F/\partial x$ shows that at
fixed $y$, $F\sim x^{-\alpha}$, with a coefficient ${\cal F}(y)$ which
can depend on $y$.  Hence we get the scaling form
\begin{equation}
F(x,r) \sim x^{-\alpha}\ {\cal F}(x^\zeta r).
\label{double_scaling}
\end{equation}
This is just Widom scaling. The critical exponents $\alpha$ and $\zeta$,
and the scaling function ${\cal F}(x^\zeta r)$ are universal (two
different systems near their critical point will have the {\sl same}
critical exponents and scaling functions). We don't need to discuss
corrections to scaling for this case, as they are similar to those
discussed above and below (and because none were dominant in our cases).

Notice that if we sit at the critical point $r=0$, the above result
reduces to equation (\ref{power}) so long as ${\cal F}(0)$ is not zero
or infinity. If, on the other hand, ${\cal F}(y) \sim y^n$ as $y \to 0$,
the two-variable scaling function gives a singular correction to the
power--law near the critical point: $F(x,r) \sim x^{-\alpha}\ {\cal
F}(x^\zeta r) \sim x^{-\alpha + n \zeta}\ $ for $x <\!< r^{-1/\zeta}$:
only when $x \sim r^{-1/\zeta}$ will the power-law $x^{-\alpha}$ be
observed.  This is what happened in two dimensions to the integrated
avalanche size distribution (figures\ \ref{aval_2d_fig} and
\ref{raw_aval_2d_fig}) and the avalanche correlation functions (figure\
\ref{correl_2d_fig}b).


For the fourth example, we address finite-size scaling of a property $K$
of the system, as we vary a parameter $r$.  If we measure $K(r,L)$ for
a variety of sizes $L$ (say, all with periodic boundary conditions), we 
expect (in complete analogy to (\ref{double_scaling}))
\begin{equation}
K(r,L) \sim r^{-\mu}\ {\cal K}(r L^{1/\nu}).
\label{finite_size_scaling}
\end{equation}
Now, suppose our ``thermometer'' measuring $r$ is weakly size-dependent,
so the measured variable is ${\cal C}(r) = r + c/L + c_2/L^2 + \ldots$\ 
The effects on the scaling function is
\begin{eqnarray}
K\Bigl({\cal C}(r),L\Bigr) \sim r^{-\mu}\ \times \nonumber \\
	 \Bigl({\cal K}(r L^{1/\nu})\ +\ \nonumber \\ 
	 (c L^{1/\nu-1} + c_2 L^{1/\nu-2})\ {\cal K}'(r L^{1/\nu}) +
			\ldots \Bigr).
\label{finite_size_corrections_to_scaling}
\end{eqnarray}

In two and three dimensions, $\nu>1$ and these correction terms are
subdominant. In four and five dimensions, we find $1/2 < \nu < 1$, so we
should include the term multiplied by $c$ in equation
(\ref{finite_size_corrections_to_scaling}). However, we believe this
first term is zero for our problem. For a fixed boundary problem (all
spins ``up'' at the boundary) with a first order transition, there is
indeed a term like $c/L$ in $r(L)$\ \cite{finite_size_first_order}. At a
critical transition, the leading correction to $r(L)$ can be $c/L$ or a
higher power of $L$ ($1/L^2$ and so on). This seems to depend on the
model studied, the geometry of the system, and the boundary conditions
(free, periodic, ferromagnetic, $\ldots$)\ \cite{finite_size_critical}.
Furthermore, for the same kind of model, the coefficient $c$ itself
depends on the geometry and boundary conditions, and it can even vanish,
which leaves only higher order corrections. In a periodic boundary
conditions problem like ours, we expect that the correction is smaller
than $c/L$. Our finite-size scaling collapses for spanning avalanches
$N$, the second moments $\langle S^2\rangle$, and the magnetization jump
$\Delta M$, were successfully done by letting $c=0$.

For the fifth example, consider a property $K$ depending on two external
parameters: $r$ (the disorder for example) and $h$ (could be the
external magnetic field $H-H_c$). Analogous to (\ref{double_scaling}),
$K$ should then scale as
\begin{equation} 
K(r,h) \sim r^{-\mu}\ {\cal K}(h/r^{\beta\delta}).
\label{double_scaling_2}
\end{equation}
Consider now the likely dependence of the field $h$ on the disorder $r$.
A typical system will have a measured field which depends on the
randomness: $\tilde{\cal C}(h) = h +  b\, r + b_2 r^2 + \ldots$
(Corresponding nonlinearities in the effective value of $r$ are
subdominant.) This system will have
\begin{eqnarray}
K\Bigl(r,\tilde{\cal C}(h)\Bigr)\ =\ r^{-\mu}\ \times \nonumber \\
	 \Bigl( {\cal K}(h/r^{\beta\delta}) 
	    + (b\, r + b_2 r^2)\ r^{-\beta\delta}\ 
		{\cal K}'(h/r^{\beta\delta}) \Bigr).
\label{rotated_double_scaling}
\end{eqnarray}
Now, for our system $1 < \beta \delta < 2$ for dimensions three and
above. This means that the term multiplied by $b$ is dominant over the
critical scaling singularity: unless one shifts the measured $h$ to the
appropriate $h'=h + b\,r$, the curves will not collapse ({\it e.g.}, the
peaks will not line up horizontally).  We measure this (non-universal)
constant for our system (Table\ \ref{RH_table}), using the derivative of
the magnetization with field $dM/dH(r,h)$.  The magnetization $M(r,h)$
and the correlation length $\xi(r,h)$ should also collapse according to
equation (\ref{double_scaling_2}) (but with $h + b\,r$ instead of $h$);
we don't directly measure the correlation length, and the collapse of
$M(r,h)$ in figure\ \ref{3d_MofH_fig}b includes the effects of the tilt
$b$.  In two dimensions, $\beta \delta$ is large (probably infinite), so
in principle we should need an infinite number of correction terms: in
practise, we tried lining up the peaks in the curves (with no correction
terms); because we did not know $\beta$ (which we usually obtained
from $\Delta M$, which gives $\beta/\nu=0$ in two dimensions), we failed
to extract reliable exponents in two dimensions from $dM/dH$.

For the sixth example, suppose $F$ depends on $r$, $h$, and a size $x$.
Then from the previous analysis, we expect
\begin{equation}
F(x,r,h) \sim x^{-\alpha}\ {\cal F}(x^\zeta r,\, h/r^{\beta \delta}).
\label{triple_scaling}
\end{equation}
Notice that universality only removes one variable from the scaling
form. One could in practice do two--variable scaling collapses (and we
believe someone has probably done it), but for our purposes these more
general scaling forms are used by fixing one of the variables.  For
example, we measure the avalanche size distribution at various values of
$h$ (binned in small ranges), at the critical disorder $r=0$.  We can
make sense of equation (\ref{triple_scaling}) by changing variables from
$h/r^{\beta \delta}$ to $x^{\zeta \beta \delta} h$:
\begin{equation}
F(x,r,h) \sim x^{-\alpha} \tilde{\cal F}(x^\zeta r,\, x^{\zeta \beta \delta} h).
\label{triple_scaling_nice}
\end{equation}
Before we can set $r=0$, we must see what are the possible corrections
to scaling in this case. If the disorder $r$ depends on the field, then
instead of the variable $r$, we must use $r + a h$ (the analysis is
analogous to the one in example five; other corrections are
subdominant). Setting $r=0$ now, leaves $F$ dependent on its first
variable, as well as the second:
\begin{eqnarray}
F(x,r,h) & \sim\  x^{-\alpha}\ \tilde{\cal F}(x^\zeta (a h),\,
			x^{\zeta \beta \delta} h)
          \approx\  x^{-\alpha}\ \times \nonumber \\
	 &  \Bigl(\tilde{\cal F}(0,\ x^{\zeta \beta \delta} h)
	+ \nonumber \\  &  a h x^\zeta\
	 \tilde{\cal F}^{(1,0)}(0,\, x^{\zeta \beta \delta} h) \Bigr),
\label{triple_scaling_reduced_corrections}
\end{eqnarray}
where $\tilde{\cal F}^{(1,0)}$ is the derivative of $\tilde{\cal F}$
with respect to the first variable (keeping the second fixed).

For the binned avalanche size distribution, $x^\zeta$ is $S^\sigma$,
where $0 \le \sigma < 1/2$ as we move from two dimensions to five and
above. Thus, the correction term will only be important for rather large
avalanches, $S > h^{-1/\sigma}$, so long as we are close to the critical
point.  Expressed in terms of the scaling variable, important
corrections to scaling occur if the scaling variable $X =
S^{\sigma\beta\delta} h >  h^{1-\beta\delta}$. For us, $\beta \delta >
3/2$, and we only use fields near the critical field ($h < 0.08$), so
the corrections will become of order one when $X=4$ for the largest $h$
we use. In $3$ and $4$ dimensions, this correction does not affect our
scaling collapses, while in $5$ dimensions some of the data needs this
correction. We have tried to avoid this problem (since we don't measure
our data such that it can be used in a two--variable scaling collapse)
by concentrating on collapsing the regions in our data curves where this
correction is negligible.

A similar analysis can be done for the avalanche time distribution,
which has two ``sizes'' $S$ and $t$ and one parameter $r$ which is set
to zero; because we integrate over the field $h$ the correction in
(\ref{triple_scaling_reduced_corrections}) does not occur, and other
scaling corrections are small.

Finally, we discuss the unusual exponential scaling forms we developed
to collapse our data in two dimensions. If we assume that the critical
disorder $R_c$ is zero {\it and} that the linear term in the rescaling
of $r$ vanishes ($\zeta \epsilon r$ in equation\ (\ref{linear_double})
vanishes), then from symmetry the correction has to be cubic, and
equation\ (\ref{linear_double}) becomes:
\begin{equation}
F(x,r) = (1 - \alpha \epsilon)\
        F\Bigl( (1-\epsilon) x,\, (1+ k \epsilon\, r^2) r \Bigr).
\label{exponential_scaling_cubic}
\end{equation}
with $k$ (which is not universal) and $\alpha$ constants, and $\epsilon$
small.

Taking the derivative of both sides with respect to $\epsilon$ and
setting it equal to zero gives a partial differential equation for the
function $F$. To solve for $F$, we do a change of variable: $(x,r)
\rightarrow (x,y)$ with $y=x\ e^{-a^*/r^2}$. The constant $a^*$ is
determined by requiring that $y$ rescales onto itself to order
$\epsilon$: we find $a^*=1/2\,k$. We then have:
\begin{equation}
0 = -\alpha\ \tilde{F}(x,y) - {\partial{\tilde{F}} \over \partial{x}}\ x
\label{exponential_scaling_cubic_2}
\end{equation}
which gives
\begin{equation}
F(x,r) = x^{-\alpha}\ \tilde{\cal{F}}\Bigl(xe^{-1/2\,k\,r^2}\Bigr).
\label{exponential_scaling_cubic_3}
\end{equation}

This is one of the forms we use in $2$ dimensions for the scaling
collapse of the second moments $\langle S^2 \rangle_{int}$, the
avalanche size distribution $D_{int}$ integrated over the field $H$, the
avalanche correlation $G_{int}$, and the spanning avalanches $N$. We use
another form too which is obtained by assuming that the critical
disorder $R_c$ is not zero but that the linear term in the rescaling of
$r$ still vanishes. Instead of equation\
(\ref{exponential_scaling_cubic}), we have:
\begin{equation}
F(x,r) = (1 - \alpha \epsilon)\
        F\Bigl( (1-\epsilon) x,\, (1+ l \epsilon\, r) r \Bigr).
\label{exponential_scaling_square}
\end{equation}
The function $F$ becomes:
\begin{equation}
F(x,r) = x^{-\alpha}\ \tilde{\cal{F}}\Bigl(xe^{-1/l\,r}\Bigr).
\label{exponential_scaling_square_2}
\end{equation}
The corrections to scaling for the last two forms (equations\
(\ref{exponential_scaling_cubic_3}) and
(\ref{exponential_scaling_square_2})) are similar to the ones discussed
above. They are all are subdominant.

\section{Full derivation of the mean field scaling form for the
integrated avalanche size distribution}

The mean field scaling form for the integrated avalanche size
distribution $D_{int}(S,R)$ was obtained in section IV A using the {\it
scaling form} of the avalanche size distribution $D(S,R,H)$. The scaling
form for $D_{int}(S,R)$ can also be obtained by integrating the
avalanche probability distribution $D(S,t)$ (derived originally
in\cite{Sethna}) directly:
\begin{equation}
D_{int} (S,R) = \int_{-\infty}^{+\infty}
\rho(-JM-H)\ D(S,t)\ dH
\label{apA1_eq1}
\end{equation}
where $\rho (-JM-H)$ is the probability distribution for the random
fields, and $\rho (-JM-H)\ dH$ is the probability for a spin to flip
between fields $-JM(H) - H$ and $-JM(H+dH) - (H+dH)$. $D(S,t)$ is the
probability of having an avalanche of size $S$, a small ``distance'' $t
\equiv 2J \rho (-JM-H) - 1$ from the infinite avalanche at $\rho (-JM-H)
= 1/2J$, given that a spin has flipped at $-JM -
H$\cite{Sethna,Dahmen1}. (The scaling form for the non-integrated
avalanche size distribution $D(S,R,H)$ (eqn.\ref{int_aval0}) is obtained
from $D(S,t)$ by expressing $t$ as a function of $R$ and $H$
\cite{Sethna,Dahmen1}). $J$ is the coupling of a spin to all others in
the system, $H$ is the external magnetic field, and $R$ is the disorder.
The advantage of this procedure is that we can find out something about
the scaling function $\bar {\cal D}_{-}^{(int)}$.

The average mean field magnetization $M$ and the avalanche probability
distribution $D(S,t)$ are given by \cite{Sethna,Dahmen1}:
\begin{equation}
M(H,R) = 1 - 2 \int_{-\infty}^{-JM(H)-H} \rho(f)\ df,
\label{apA1_eq2}
\end{equation}
and
\begin{equation}
D(S,t) = {S^{S-2} \over (S-1)!}\ (t+1)^{S-1}\ e^{-S(t+1)}
\label{apA1_eq3}
\end{equation}
To solve equation (\ref{apA1_eq1}), let's define the variable
$y=(-JM-H)/({\sqrt 2}\ R)$ and rewrite the integral as:
\begin{eqnarray}
D_{int} (S,R)\ =\ {\sqrt 2}\ R\ \times \nonumber \\
\Biggl[\int_{-\infty}^{+\infty}
\rho ({\sqrt 2}Ry)\ D\Bigl(S,\ 2J \rho ({\sqrt 2}Ry) - 1\Bigr) \times
\nonumber \\
\ \Bigl(1 - 2J \rho ({\sqrt 2}Ry)\Bigr)\ dy \Biggr],
\label{apA1_eq4}
\end{eqnarray}
where we have used:
\begin{equation}
{dy \over dH} = {1 \over {\sqrt 2}\ R}\ \biggl(-J\ {2\ \rho(-JM-H)
\over {1-2J \rho(-JM-H)}} -1\biggr)
\label{apA1_eq5}
\end{equation}
Since we are interested in the behavior of the integrated avalanche
distribution for large sizes, the factorial in equation (\ref{apA1_eq3})
can be expanded using Stirling's formula. To first order, we have:
\begin{equation}
(S-1)!\ \approx\ {S^S\ {\sqrt {2 \pi}} \over e^S\ {\sqrt S}}
\label{apA1_eq6}
\end{equation}
Substituting this and the random field distribution function $\rho$,
\begin{equation}
\rho ({\sqrt 2}Ry) = {1 \over {\sqrt {2 \pi}} R}\ e^{-y^2},
\label{apA1_eq7}
\end{equation}
in equation (\ref{apA1_eq4}), we obtain:
\begin{eqnarray}
D_{int}(S,R)\ \approx\ C\ \biggl({R_c \over R}\biggr)^S\ \times
\nonumber \\
\int_{-\infty}^{+\infty}
e^{-S\ \bigl(y^2 + {R_c \over R}\ e^{-y^2}\bigr)}
\ \ \biggl(1-{R_c \over R}\  e^{-y^2}\biggr)\ dy
\label{apA1_eq8}
\end{eqnarray}
where $C=S^{-{3 \over 2}}\ e^S\ R_c/(\pi R {\sqrt 2})$, and $S$ is large.

For disorders above but close to the critical disorder $R_c$, we have:
\begin{eqnarray}
\biggl({R_c \over R}\biggr)^S\ =\ e^{S\ log\bigl({R_c \over R}\bigr)}\ \approx\
\nonumber \\
e^{S\ \Bigl(-\ {\bigl(1-{R_c \over R}\bigr) \over 1} -
\ {\bigl(1-{R_c \over R}\bigr)^2 \over 2}
- {\bigl(1 - {R_c \over R}\bigr)^3 \over 3} - ... \Bigr)}
\label{apA1_eq9}
\end{eqnarray}
If we assume that only terms up to $S\ (1-R_c/R)^2$ are important (terms
like $S\ (1-R_c/R)^3$ and $S\ (1-R_c/R)^4$ go to zero as $R \rightarrow
R_c$), and we note that the integrand in equation (\ref{apA1_eq8}) is an
even function of $y$, equation (\ref{apA1_eq8}) becomes:
\begin{eqnarray}
D_{int}(S,R)\ \approx\ 2\ C\ \times \nonumber \\
\Biggl[ \int_{0}^{+\infty}
e^{-S\ \Bigl({\bigl(1-{R_c \over R}\bigr) \over 1} +
{\bigl(1-{R_c \over R}\bigr)^2 \over 2}
+y^2 + {R_c \over R}\ e^{-y^2}\Bigr)} \times \nonumber \\
\ \ \biggl(1-{R_c \over R}\  e^{-y^2}\biggr)\ dy \Biggr]
\label{apA1_eq10}
\end{eqnarray}

The asymptotic behavior of the above integral, as $S \rightarrow
\infty$, is obtained using Laplace's method\cite{BenderOrszag}. The idea
is as follows. The asymptotic behavior as $S \rightarrow \infty$ of
the integral:
\begin{equation}
I(S) = \int_{a}^{b} f(x)\ e^{S \phi (x)}\ dx
\label{apA1_eq11}
\end{equation}
can be found by integrating over a small region $[c-\epsilon,\
c+\epsilon]$ (instead of the interval $[a,b]$) around the maximum of the
function $\phi (x)$ at $x=c$, since in the asymptotic expansion, the
largest contribution to the integral will be from this region. The
corrections will be exponentially small. The maximum of $\phi$ must be
in the interval $[a,b]$, $f(x)$ and $\phi(x)$ are assumed to be real
continuous functions, and $f(c) \ne 0$. $f(x)$ and $\phi (x)$ can now
both be expanded around $x=c$, and the integral solved. Often the
integral is easier to handle if the limit of integration is extended to
infinity. This will add only exponentially small corrections in the
asymptotic limit of $S \rightarrow \infty$.

Let's apply this method to equation (\ref{apA1_eq10}). The function in
the exponential has a maximum at $y=0$. The function $\biggl(1-{R_c
\over R}\ e^{-y^2}\biggr)$ is not zero there if $R \ne R_c$. We can thus
expand both functions in the integral of equation (\ref{apA1_eq10})
around $y=0$. Defining $u=y^2{\sqrt S}$, we obtain:
\begin{eqnarray}
D_{int}(S,R)\ \approx\ C_{1}\ \times \nonumber \\ \Biggl[ \int_{0}^{\epsilon}
e^{-{\sqrt S}\ \Bigl(\bigl(1-{R_c \over R}\bigr) u\ +\
{u^2 \over {2 {\sqrt S}}}{R_c \over R} -\ {u^3 \over {6\ S^2}}{R_c \over R}
+ ...\Bigr)}\ \times \nonumber \\
\biggl(\Bigl(1-{R_c \over R}\Bigr) +
{R_c \over R} {u \over {\sqrt S}} -\ {R_c \over 2R} {u^2 \over S} + ...\biggr)
\ {du \over {\sqrt u}} \Biggr]
\label{apA1_eq12}
\end{eqnarray}
where
\begin{equation}
C_{1} = {1 \over \pi {\sqrt 2}}\ {R_c \over R}\ S^{-{9 \over 4}}
\ \ e^{-{S \over 2}\ \bigl(1-{R_c \over R}\bigr)^2},
\label{apA1_eq13}
\end{equation}
$S$ is large, $R$ is close to but not equal to $R_c$, and only terms up
to $S (1-R_c/R)^2$ are non-vanishing. In the asymptotic limit of $S
\rightarrow \infty$ we can ignore terms with powers of $S$ in the
denominator, and look at the distribution for $R$ close to $R_c$. To
first order in $r = (R_c-R)/R$, $R_c/R \approx 1$ and $1-R_c/R \approx
-r$, which gives:
\begin{eqnarray}
D_{int}(S,R)\ \approx\ {1 \over \pi {\sqrt 2}}\ S^{-{9 \over 4}}
\ e^{-{S \over 2}\ (-r)^2}\ \times \nonumber \\ \int_{0}^{\infty}
e^{\Bigl(-(-r){\sqrt S}\ u\ -\ {u^2 \over 2}\Bigr)}\ \
\Bigl(-r {\sqrt S} + u\Bigr)
\ {du \over {\sqrt u}}
\label{apA1_eq14}
\end{eqnarray}
where we have expanded the integration to infinity. As mentioned above,
this will only add exponentially small corrections in the asymptotic
limit of $S \rightarrow \infty$. Equation\ (\ref{apA1_eq14}) is the
integrated avalanche size distribution in mean field for large sizes
$S$, and finite $S r^2$. We see right away that it gives the correct
scaling form:
\begin{equation}
D_{int}(S,R)\ \sim\ S^{-{9 \over 4}}\
\bar{\cal D}_{\pm}^{(int)} \Bigl({\sqrt S}\ |r|\Bigr)
\label{apA1_eq15}
\end{equation}
where $\pm$ indicates the sign of $r$, the exponent
$\tau+\sigma\beta\delta$ and $\sigma$ are $9/4$ and $1/2$ respectively,
and the scaling function $\bar{\cal D}_{\pm}^{(int)}$ is:
\begin{equation}
\bar{\cal D}_{\pm}^{(int)} \Bigl({\sqrt S}\ |r|\Bigr) =
e^{-{({\sqrt S}\ |r|)^2 \over 2}}\
\bar{\cal F}_{\pm} \Bigl({\sqrt S}\ |r|\Bigr).
\label{apA1_eq16}
\end{equation}
The function $\bar{\cal F}_{\pm} \Bigl({\sqrt S}\ |r|\Bigr)$ is
proportional to the integral in equation\ (\ref{apA1_eq14}). Note that
the above result is equivalent to the one obtained (eqn.\
\ref{int_aval3}) by integrating the {\it scaling form} of $D(S,R,H)$
over the field $H$.

What is the behavior of the scaling function $\bar{\cal
D}_{-}^{(int)}(X)$ for small and large positive arguments $X={\sqrt S}\
(-r) > 0$ ($R > R_c$)? From equations\ (\ref{apA1_eq14}) and
(\ref{apA1_eq16}), for small arguments we have a polynomial in $X$:
\begin{equation}
\bar{\cal D}_{-}^{(int)}(X) \approx A+BX+CX^2+{\cal O}(X^3)
\label{apA1_eq17}
\end{equation}
These parameters can be calculated numerically. We obtain in mean field:
\begin{eqnarray}
{\bar{\cal D}_{-}^{(int)}}\ & \approx\ &
0.232+0.243 X-0.174 X^2- \nonumber \\ 
 & & 0.101 X^3+0.051 X^4
\label{dist_mf2}
\end{eqnarray}
On the other hand, for large arguments we find:
\begin{equation}
\bar{\cal D}_{-}^{(int)}(X) \approx
{\pi}^{1/2} e^{-{X^2 \over 2}}\  {\sqrt X} \Bigl( 1 + {\cal O}
(X^{-2})\Bigr)
\label{apA1_eq18}
\end{equation}
In general, for all dimensions, in equation\ (\ref{apA1_eq18}) the
exponential is of $X^{1/\sigma}$ ($1/\sigma = 2$ in mean field), since
the exponent $\sigma$ gives the exponential cutoff to the power law
distribution for large $X$, and the power of $X$ is $\beta$ ($\beta =
1/2$ in mean field). One can see the latter by expanding the
distribution function $D_{int}(S,R)$ in terms of $1/S$ ($S$ is large),
analogous to\ \cite{Griffiths}:
\begin{equation}
D_{int}(S,R) = \sum_{n=1}^{\infty} f_n (r)\ S^{-n}
\label{apA1_equ18a}
\end{equation}
Since $X=$ $S^\sigma (-r)$, then we can write $S=$ $X^{1/\sigma}\
(-r)^{-1/\sigma}$ and obtain:
\begin{equation}
D_{int}(S,R) = \sum_{n=1}^{\infty} f_n (r)\ X^{-n/\sigma} (-r)^{n/\sigma}
\label{apA1_equ18b}
\end{equation}
The scaling function $\bar{\cal D}_{-}^{(int)}(X)$ scales like
$S^{(\tau+\sigma\beta\delta)}\ \times D_{int}(S,R)$:
\begin{eqnarray}
\bar{\cal D}_{-}^{(int)}(X) \sim\ \Biggl[ \sum_{n=1}^{\infty} f_n (r)\
X^{-n/\sigma} X^{(\tau+\sigma\beta\delta)/\sigma}\ \times \nonumber \\
(-r)^{n/\sigma}
(-r)^{-(\tau+\sigma\beta\delta)/\sigma} \Biggr]
\label{apA1_equ18c}
\end{eqnarray}
and since it is only a function of $X$, it must satisfy:
\begin{equation}
\bar{\cal D}_{-}^{(int)}(X) \sim\ \sum_{n=1}^{\infty} g_n\
X^{-n/\sigma} X^{(\tau+\sigma\beta\delta)/\sigma}
\label{apA1_equ18d}
\end{equation}
where $g_n$ is independent of $r$.

The exponent combination $(\tau+\sigma\beta\delta)/\sigma$ can be
rewritten as:
\begin{equation}
{\tau+\sigma\beta\delta \over \sigma} = {2 \over \sigma} +
{\tau+\sigma\beta\delta - 2 \over \sigma} = {2 \over \sigma} +
\beta
\label{apA1_equ18e}
\end{equation}
where we have used the scaling relation\cite{Dahmen1,Dahmen3}:
$\beta - \beta\delta = (\tau - 2)/\sigma$. Thus we have for the scaling
function $\bar{\cal D}_{-}^{(int)}(X)$:
\begin{equation}
\bar{\cal D}_{-}^{(int)}(X) \sim\ X^\beta \sum_{n=-1}^{\infty} g_n\
X^{-n/\sigma} = X^\beta\ {\cal K}(X^{1/\sigma})
\label{apA1_equ18f}
\end{equation}
which shows (compare to equation\ (\ref{apA1_eq18})) that the power of $X$
is indeed the exponent $\beta$.

We have used the results of the expansion of the mean field scaling
function ${\bar {\cal D}}_{-}^{(int)} (X)$ for small and large
parameters (equations\ (\ref{apA1_eq17}) and (\ref{apA1_eq18})), to
build a fitting function to the integrated avalanche size distribution
scaling functions in $2$, $3$, $4$, and $5$ dimensions, described in
section IV~ B.

Finally, note from equation\ (\ref{apA1_eq17}) that the scaling function:
\begin{equation}
{\cal D}_{-}^{(int)} (S r^2) = e^{-{(S r^2) \over 2}}\
{\cal F}_{-}(S r^2)
\label{apA1_eq19}
\end{equation}
used earlier in reference\cite{Dahmen1}, is not analytic for small
arguments $S r^2$, from which we conclude that the appropriate scaling
variable should be ${\sqrt S}\ (-r)$ and not $S r^2$. (Notice that this
no longer seems true in two dimensions; see section on $2$ dimensional
results.)

\section{Derivation of the mean field scaling form for the spanning avalanches}

We have defined earlier a mean field spanning avalanche to be an
avalanche larger than $\sqrt {S_{mf}}$, where $S_{mf}$ is the {\it
total} size of the system. We want to derive the scaling form for the
number of such avalanches in half of the hysteresis loop (for $H$ from
$-\infty$ to $+\infty$) as a function of the system size $S_{mf}$ and
the disorder $R$. The number of mean field spanning avalanches is
proportional to the probability of having avalanches of size larger than
$\sqrt {S_{mf}}$. Since we want the number of spanning avalanches, we
need to multiply this probability by the total number of avalanches. For
large system sizes, this scales with the system size $S_{mf}$
(corrections are subdominant). We thus obtain by integrating over
equation\ (\ref{apA1_eq15}) (which gives the scaling form for the
probability distribution of avalanches of size $S$ in the hysteresis
loop):
\begin{eqnarray}
N_{mf}(S_{mf},R)\ \sim\
S_{mf}\ \times \nonumber \\
\int_{\sqrt {S_{mf}}}^{\infty}\ S^{-{9 \over 4}}\
e^{-{\bigl({\sqrt S} |r|\bigr)^2 \over 2}}\
\bar{\cal F}_{\pm}\Bigl({\sqrt S} |r|\Bigr)\ dS
\label{apB_eq1}
\end{eqnarray}
Let's define $u={\sqrt S} |r|$, then equation\ (\ref{apB_eq1}) can be
written as:
\begin{eqnarray}
N_{mf}(S_{mf},R)\ \sim\ 2\ S_{mf}\ |r|^{5 \over 2} \times \nonumber \\
\int_{|r| S_{mf}^{1 \over 4}}^{\infty}\ u^{-{7 \over 2}}\
e^{-{u^2 \over 2}}\ \bar{\cal F}_{\pm}(u)\ du
\label{apB_eq2}
\end{eqnarray}
The integral ${\cal I}$ is a function of $S_{mf}^{1 \over 4} |r|$ only,
and we can write it as:
\begin{equation}
{\cal I} = \Bigl(S_{mf}^{1 \over 4} |r|\Bigr)^{-{5 \over 2}}\
{\cal N}_{\pm}^{mf}\Bigl(S_{mf}^{1 \over 4} |r|\Bigr)
\label{apB_eq3}
\end{equation}
to obtain the scaling form for the number $N_{mf}$ of mean field
spanning avalanches:
\begin{equation}
N_{mf}(S_{mf},R)\ \sim\
S_{mf}^{3 \over 8}\
{\cal N}_{\pm}^{mf}\Bigl(S_{mf}^{1 \over 4} |r|\Bigr)
\label{apB_eq4}
\end{equation}
From this scaling form, we can extract the exponents $\tilde \theta
=3/8$, and $1/\tilde \nu=1/4$. This form is used for collapses of the
spanning avalanche curves in mean field (see mean field section).


%

%
\begin{table}
\begin{tabular}{cr@{$\,\pm\,$}lr@{$\,\pm\,$}lr@{$\,\pm\,$}lc}
 &
\multicolumn{2}{c}{$3$d} &
\multicolumn{2}{c}{$4$d} &
\multicolumn{2}{c}{$5$d} &
\multicolumn{1}{c}{mean field} \\ \hline
$R_c$ & 2.16 & 0.03 & 4.10 & 0.02 & 5.96 & 0.02 & 0.79788456 \\
$H_c$ & 1.435 & 0.004 & 1.265 & 0.007 & 1.175 & 0.004 & 0 \\
$b$ & 0.39 & 0.08 & 0.46 & 0.05 & 0.23 & 0.08 & 0 \\
\end{tabular}
\vspace{0.25cm}
\caption{Numerical values for the critical disorders and fields,
	and the ``tilt'' parameter $b$ (see section on magnetization
	curves collapses)
        in $3$, $4$, and $5$ dimensions extracted from
	scaling collapses. The critical disorder is obtained from
	collapses of the spanning avalanches and the second moments of
	the avalanche size distribution. The critical field is obtained
	from the binned avalanche size distribution and the magnetization
	curves. $H_c$ is affected by finite sizes, and systematic
	errors could be larger than the ones listed here.
        The mean field values are calculated
	analytically\protect\cite{Sethna,Dahmen1}.
	The ``tilt'' $b$ is obtained from the $dM/dH$ collapses
	using the values for the critical disorder and field from this
	table and the values for the exponents from Table\
	\ref{calculated_exp_table}. Only the parameter $b$ is allowed to
	vary. The values in $2$ dimensions (which are not listed)
	are less accurate. Depending on the scaling form
	we obtain a critical disorder of $0$, $0.45$, or $0.54$. The critical
	field is around $1.32$ and is estimated from the binned in $H$
	avalanche size distribution and magnetization curves (see text).
	The ``tilt'' $b$ was not measured.
	$R_c$, $H_c$, and $b$ are not universal
	characteristics of the system.}
\label{RH_table}
\end{table}
\twocolumn[\hsize\textwidth\columnwidth\hsize\csname
@twocolumnfalse\endcsname

\begin{table}
\begin{tabular}{cr@{$\,\pm\,$}lr@{$\,\pm\,$}lr@{$\,\pm\,$}lc}
measured exponents&
\multicolumn{2}{c}{$3$d} &
\multicolumn{2}{c}{$4$d} &
\multicolumn{2}{c}{$5$d} &
\multicolumn{1}{c}{mean field}  \\ \hline
$1/\nu$ & 0.71 & 0.09 & 1.12 & 0.11 & 1.47 & 0.15 & 2 \\
$\theta$ & 0.015 & 0.015 & 0.32 & 0.06 & 1.03 & 0.10 & 1 \\
$(\tau+\sigma\beta\delta -3)/\sigma \nu$ & -2.90 & 0.16
& -3.20 & 0.24 & -2.95 & 0.13 & -3 \\
$1/\sigma$ & 4.2 & 0.3 & 3.20 & 0.25 & 2.35 & 0.25 & 2 \\
$\tau+\sigma\beta\delta$ & 2.03 & 0.03 & 2.07 & 0.03 & 2.15 & 0.04 & 9/4 \\
$\tau$ & 1.60 & 0.06 & 1.53 & 0.08 & 1.48 & 0.10 & 3/2 \\
$d + \beta/\nu$ & 3.07 & 0.30 & 4.15 & 0.20 & 5.1 & 0.4 & 7 (at
$d_c=6$)\\
$\beta/\nu$ & 0.025 & 0.020 & 0.19 & 0.05 & 0.37 & 0.08 & 1 \\
$\sigma\nu z$ & 0.57 & 0.03 & 0.56 & 0.03 & 0.545 & 0.025  & 1/2 \\
\end{tabular}
\vspace{0.25cm}
\caption{ Values for the exponents extracted from scaling
	collapses
	in $3$, $4$, and $5$ dimensions.
	The mean field values are calculated
	analytically\protect\cite{Sethna,Dahmen1}.
	$\nu$ is the
	correlation length exponent and is found from collapses of
	avalanche correlations,
	number of spanning avalanches, and moments of the avalanche
	size distribution data. The exponent $\theta$ is a measure of
	the number of spanning avalanches and is obtained from collapses
	of that data.
	$(\tau+\sigma\beta\delta-3)/\sigma\nu$ is obtained
	from the second moments of the avalanche size distribution collapses.
 	$1/\sigma$ is associated with the cutoff in the
	power law distribution of avalanche sizes integrated over the
	field $H$,
	while $\tau+\sigma\beta\delta$ gives the slope of that distribution.
 	$\tau$ is obtained from the binned avalanche size distribution
	collapses. $d+\beta/\nu$ is obtained
	from avalanche correlation collapses and $\beta/\nu$ from magnetization
	discontinuity collapses.
 	$\sigma\nu z$ is the exponent
	combination for the time distribution of avalanche sizes and is
	extracted from that data.}
\label{measured_exp_table}
\end{table}
]
\twocolumn[\hsize\textwidth\columnwidth\hsize\csname
@twocolumnfalse\endcsname

\begin{table}
\begin{tabular}{cr@{$\,\pm\,$}lr@{$\,\pm\,$}lr@{$\,\pm\,$}lc}
calculated exponents &
\multicolumn{2}{c}{$3$d} &
\multicolumn{2}{c}{$4$d} &
\multicolumn{2}{c}{$5$d} &
\multicolumn{1}{c}{mean field}  \\ \hline
$\sigma \beta \delta$ & 0.43 & 0.07 & 0.54 & 0.08 & 0.67 & 0.11 &
3/4 \\
$\beta\delta$ & 1.81 & 0.32 & 1.73 & 0.29 & 1.57 & 0.31 & 3/2 \\
$\beta$ & 0.035 & 0.028 & 0.169 & 0.048 & 0.252 & 0.060 & 1/2 \\
$\sigma\nu$ & 0.34 & 0.05 & 0.28 & 0.04 & 0.29 & 0.04 & 1/4 \\
$\eta = 2 + (\beta-\beta\delta)/\nu$ & 0.73 & 0.28 & 0.25 & 0.38 &
0.06 & 0.51 & 0 \\
\end{tabular}
\vspace{0.25cm}
\caption{Values for exponents in $3$, $4$, and $5$ dimensions
	that are not extracted directly from scaling collapses,
	but instead are derived from
	Table\ \protect\ref{measured_exp_table} and the
	exponent relations (see \protect\cite{Dahmen1,Dahmen3}).
	The mean field values are obtained
	analytically\protect\cite{Sethna,Dahmen1}. Both $\sigma\beta\delta$
	and $\beta\delta$ could have larger systematic errors than the
	errors listed here. See the binned avalanche size distribution
	section for details.}
\label{calculated_exp_table}
\end{table}
]
\twocolumn[\hsize\textwidth\columnwidth\hsize\csname
@twocolumnfalse\endcsname

\begin{table}
\begin{tabular}{cr@{$\,\pm\,$}lr@{$\,\pm\,$}lr@{$\,\pm\,$}lr@{$\,\pm\,$}l}
\multicolumn{1}{c} {Exponents} &
\multicolumn{4}{c} {L=10,20,40} &
\multicolumn{4}{c} {L=20,40,80} \\
and $R_c$ &
\multicolumn{2}{c} {$r=(R_c-R)/R$} &
\multicolumn{2}{c} {$r=(R_c-R)/R_c$} &
\multicolumn{2}{c} {$r=(R_c-R)/R$} &
\multicolumn{2}{c} {$r=(R_c-R)/R_c$} \\ \hline
$1/\nu$ & 0.96 & 0.07 & 1.07 & 0.05 & 1.05 & 0.10 & 1.12 & 0.06 \\
$\theta$ & 0.35 & 0.10 & 0.34 & 0.06 & 0.32 & 0.04 & 0.32 & 0.04 \\
$R_c$ & 4.09 & 0.02 & 4.09 & 0.01 & 4.095 & 0.015 & 4.10 & 0.01
\end{tabular}
\vspace{0.25cm}
\caption{Exponent values and critical disorder $R_c$
	from collapses of spanning avalanche curves in $4$
	dimensions. Three curves (different linear size $L$) are collapsed
	together, with $r=(R_c-R)/R$ and $r=(R_c-R)/R_c$.
	Tables\ \protect\ref{span_exp_4d_table},
	\protect\ref{deltaM_4d_table}, and
	\protect\ref{s2_5d_exp_table} give information equivalent to that
	given in for example figures\ \protect\ref{mf_aval_expfig}a
	and \protect\ref{mf_aval_expfig}b.
	Graphs showing two points with an extrapolation to
	$1/L \rightarrow 0$ seemed unnecessary.}
\label{span_exp_4d_table}
\end{table}
]
\twocolumn[\hsize\textwidth\columnwidth\hsize\csname
@twocolumnfalse\endcsname

\begin{table}
\begin{tabular}{cr@{$\,\pm\,$}lr@{$\,\pm\,$}lr@{$\,\pm\,$}lr@{$\,\pm\,$}l}
\multicolumn{1}{c} {Exponents} &
\multicolumn{4}{c} {L=10,20,40} &
\multicolumn{4}{c} {L=20,40,80} \\
   &
\multicolumn{2}{c} {$r=(R_c-R)/R$} &
\multicolumn{2}{c} {$r=(R_c-R)/R_c$} &
\multicolumn{2}{c} {$r=(R_c-R)/R$} &
\multicolumn{2}{c} {$r=(R_c-R)/R_c$} \\ \hline
$1/\nu$ & 1.10 & 0.04 & 1.24 & 0.08 & 1.10 & 0.05 & 1.11 & 0.05 \\
$\beta/\nu$ & 0.195 & 0.035 & 0.19 & 0.05 & 0.18 & 0.05 & 0.20 & 0.06 \\
\end{tabular}
\vspace{0.25cm}
\caption{ Exponent values for $1/\nu$ and $\beta/\nu$, obtained from scaling
        collapses of the change of the magnetization $\Delta M$ due to
	the spanning
        avalanches. Three curves of different size $L$
	are collapsed together with
        $r=(R_c-R)/R$ and $r=(R_c-R)/R_c$, where $R_c=4.10 \pm 0.02$.
	See also comment in Table\
	\protect\ref{span_exp_4d_table}.
        }
\label{deltaM_4d_table}
\end{table}
]
\twocolumn[\hsize\textwidth\columnwidth\hsize\csname
@twocolumnfalse\endcsname

\begin{table}
\begin{tabular}{cr@{$\,\pm\,$}lr@{$\,\pm\,$}lr@{$\,\pm\,$}lr@{$\,\pm\,$}l}
\multicolumn{1}{c} {Exponents} &
\multicolumn{4}{c} {L=5,10,20} &
\multicolumn{4}{c} {L=10,20,30} \\
   &
\multicolumn{2}{c} {$r=(R_c-R)/R$} &
\multicolumn{2}{c} {$r=(R_c-R)/R_c$} &
\multicolumn{2}{c} {$r=(R_c-R)/R$} &
\multicolumn{2}{c} {$r=(R_c-R)/R_c$} \\ \hline
$1/\nu$ & 1.40 & 0.05 & 1.60 & 0.10 & 1.41 & 0.07 & 1.53 & 0.13 \\
$-(\tau+\sigma\beta\delta - 3)/\sigma\nu$ & 2.75 & 0.10 & 2.70 & 0.10 &
2.93 & 0.08 & 2.90 & 0.08 \\
\end{tabular}
\vspace{0.25cm}
\caption{Exponent values from the collapses of second moments of the avalanche
	size distribution curves in $5$ dimensions. Three curves of different
	size $L$
	are collapsed together with $r=(R_c-R)/R$ and $r=(R_c-R)/R_c$,
	where $R_c =5.96 \pm 0.02$.
	See also comment in Table\ \protect\ref{span_exp_4d_table}.
	}
\label{s2_5d_exp_table}
\end{table}
]
\twocolumn[\hsize\textwidth\columnwidth\hsize\csname
@twocolumnfalse\endcsname

\begin{table}
\begin{tabular}{ccccccccc}
$2$d & $1/\nu$ & $(\tau+\sigma\beta\delta-3)/\sigma\nu$ &
$\sigma$ & $\tau+\sigma\beta\delta^*$ & $\tau$ &
$\sigma\nu$ & $\beta/\nu^{*}$ \\ \hline
conj. & 0 & -2 & 0 & 2 & 3/2 & 1/2 & 0 \\
meas. & $0.13 \pm 0.13$ & $-1.9 \pm 0.1$ & $0.10 \pm 0.02$ &
$2.04 \pm 0.04$ & $0.0 \pm 0.0$ & $0.51 \pm 0.08$ & $0.03 \pm 0.06$ \\
\end{tabular}
\vspace{0.25cm}
\caption{Conjectured and measured values for some exponents in $2$ dimensions.
        We don't have a conjectured value for the exponent combination
        $\sigma\nu z$, but the measured value is $0.64 \pm 0.02$.
        (*) Note that the distribution of avalanche sizes at $R_c$ in two
        dimensions, integrated over the loop, will scale as $S^{-(\tau
        +\sigma\beta\delta)+\omega}$, where the correction $\omega \sim 1$ is
        due to the singularity in the scaling function
        ${\cal D}^{(int)}_{-}(X) \sim X^{\omega}$ as $X \rightarrow 0$. See
        text section IV, figure\ \protect\ref{aval_2d_fig},
        and appendix A for details. A similar
        argument can be made for the avalanche correlation measurement
        (integrated over the field $H$),
        where due to the singularity of the scaling function
        ${\widetilde {\cal G}}_{-}$, the scaling
        for small $x|r|^{\nu}$ is $x^{-(d+\beta/\nu)+\tilde{\omega}}$,
        with $\tilde\omega \sim 1$ (see
        text and figure\ \protect\ref{correl_2d_fig}b).}
\label{conj_meas_2d_table}
\end{table}
]
\begin{table}
\begin{tabular}{cccccccc}
$2$ dimensions & $\theta$ &
$\sigma\beta\delta$ & $\beta\delta/\nu$ & $\sigma\nu$ & $1/\delta$ &
$\eta$ & $\bar \eta$ \\ \hline
conjectured & 0 & 1/2 & 1 & 1/2 & 0 & 1 & 2 \\
\end{tabular}
\vspace{0.25cm}
\caption{Conjectured values for some exponents in $2$ dimensions. These
        exponents were not extracted from collapses (see text).}
\label{conj_2d_table}
\end{table}

\begin{references}
\bibitem{note1} In the bulk of our simulations, we inadvertently
	did something incorrect. We calculated the
	probability $P^{flip}(n_{\uparrow})$ as an integral
	over the range $\Delta H$, and {\it not}
	at the new field $H + \Delta H$ as we should have.
	This method is correct for $\Delta H$ small, and so is correct in the
	thermodynamic limit and near the critical point. The results
	can however differ if $\Delta H$ is large
	(which occurs for the first few
	spins that flip). We have checked and found that both methods
	lead to the same results within the statistical error bars;
	indeed, in 87\% of the runs at $R=2.800$ for a $20^3$ system
	exactly the same spin-flip history was generated with the two methods.
\bibitem{note2} In previous papers
	\protect\cite{Sethna,Dahmen1,Dahmen2}, the scaling
	function for the avalanche size distribution was defined as
	$D_{\pm}(S|r|^{1/\sigma},h/|r|^{\beta\delta})$. Since this function is
	not analytic in mean field for small $S|r|^{1/\sigma}$
	(see appendix B), we use
	in this paper the scaling function
	${\bar D_{\pm}}(S^\sigma |r|,h/|r|^{\beta\delta})$, which in mean field
	is analytic for small $S^\sigma |r|$. Note
	that the scaling function ${\bar {\cal D}_{+}^{(int)}}(S^{\sigma} |r|)$
	in reference\ 
	\protect\cite{Perkovic} has been mistakenly written as
	${\cal D}_{int}(S^\sigma |r|)$.
\bibitem{note4} In $3$ and $4$ dimensions, below the critical field $H_c$,
	we have not found any spanning avalanches. In $5$ dimensions, spanning
	avalanches appeared below the critical field $H_c=1.175$. This is
	due to the small system size ($30^5$) used for the $5$ dimensional
	simulation. We
	however, do not expect substantial corrections to the exponents
	derived from that data.
\bibitem{note5} For systems with short-range interactions, the
        critical behavior of a system depends only on the
        dimensionality $d$ and the symmetry group of the hamiltonian
        ${\cal H}$, and not on the microscopic details of the
        system. Two systems are in the
        same universality ``class'' if they have the same dimensionality
        and their hamiltonian is in the same symmetry group.
\end{references}
\end{document}